\documentclass[showpacs,showkeys,preprintnumbers,amsmath,amssymb]{revtex4}%twocolumn,
\usepackage{amsmath}
\usepackage{amsfonts}
\usepackage{amssymb}
\usepackage[colorlinks,linkcolor=blue,citecolor=red]{hyperref}
\usepackage{graphicx}
\usepackage{hyperref}
\usepackage{subfigure}
\usepackage{graphicx}% Include figure files
\usepackage{dcolumn}% Align table columns on decimal point
\usepackage{bm}% bold math

\usepackage{lscape} %landscape pages support
\def\dfrac{\frac}
\def\rt{$r_{+}$}

\begin{document}

%\preprint{APS/123-QED}

\title{Thermodynamic geometry of static and rotating regular black holes in conformal massive gravity}% Force line breaks with \\

\author{Saheb Soroushfar$^{a}$}
\email{soroush@yu.ac.ir}
\affiliation{$^{a}$Faculty of Technology and Mining, Yasouj University, Choram 75761-59836, Iran}

\author{Reza Saffari$^{b}$}
\email{rsk@guilan.ac.ir}
\affiliation{$^{b}$Department of Physics, University of Guilan, 41335-1914, Rasht, Iran}

\author{Amare Abebe$^{c}$}
 \email{amare.abbebe@gmail.com }
\affiliation{$^{c}$Center for Space Research, North-West University, Mahikeng 2745, South Africa}

\author{Haidar Sheikhahmadi$^{c,d}$}
 \email{h.sh.ahmadi@gmail.com;h.sheikhahmadi@ipm.ir}

\affiliation{$^{c}$ Center for Space Research, North-West University, Mahikeng 2745, South Africa,}
\affiliation{$^{d}$ School of Astronomy, Institute for Research in Fundamental Sciences (IPM),  P. O. Box 19395-5531, Tehran, Iran}

%\author{Authors$^{\star}$}
% \email{hhhhh@gmail.com}

%\affiliation{$^\star$ GGGGG}


%\affiliation{

\date{\today}% It is always \today, today,

\begin{abstract}
A version of massive gravity, namely conformal massive gravity, is employed to study the behavior of thermodynamic geometry for both the static and the rotating regular black holes. Whereas in thermodynamic geometry singularity can be interpreted as the second phase transition, seeing such phenomena as heat capacity behavior for both cases is investigated in detail. In addition, other thermodynamic quantities like the entropy and the temperature are analyzed as well. Another important property of the black holes is the so-called stability, in which utilizing the first phase transition of the heat capacity is detailed, for both cases, say the static and the spinning black holes.  It is also interestingly figured out that, the behavior of temperature of the uncharged black holes in the presence of hair parameter, $\lambda$, can mimic the charged ones. {The effects of scalar charge, $Q$, and hair parameters with both positive and negative signs and how it affects key parameters in the study of black holes are investigated as well.}
To investigate the properties of the black holes both regular thermodynamics and thermodynamic geometry approaches are considered. Then one can observe that aiming to realize the physics of the black holes, many complicated phenomena can be analyzed dramatically easier by considering the latter approach.
\end{abstract}

%Use showkeys class option if keyword
\maketitle

\tableofcontents

%=================================================================
%============== Introduction =====================================
%=================================================================

%One interesting proposal to probe the nature of DE is holographic DE (HDE) model \cite{Li:2004rb,Huang:2004ai,Pavon:2005yx,Setare:2007eq,Saaidi:2013yfa}.
\textcolor[rgb]{0.00,0.00,1.00}{\section{Introduction}}\label{Introduction00}
Among the most interesting and important real objects in nature are the black holes \cite{Barish:1999vh,TheLIGOScientific:2016agk,Abbott:2016izl,Lovelace:2016uwp,Calloni:2017whl,Akiyama:2019cqa,Akiyama:2019brx,Akiyama:2019sww,Akiyama:2019bqs,Akiyama:2019fyp,Akiyama:2019eap}.
In $1915$, the theory of general relativity, GR, was formulated by A. Einstein and he showed that gravity can affect light beams, and consequently the light trajectory \cite{1912a,1912b,1914a,1915a,1915b,1915c,1915d,1916a}. Some months later, K. Schwarzschild found the first exact solution for the vacuum Einstein's GR equations \cite{Schw1916}; this solution was devoted to a spherical symmetric static point mass.
It worth to be noted that after the Schwarzschild solution, independently J. Droste, found the same solution which contains more details about compact point masses \cite{Johannes1917}. For a review about these solutions and dealing with the concept of singularities in the early 20th century we refer the reader to \cite{Edington1920,Edington1924,tHoof2009,Throne1994}. Besides this static uncharged solution, one can mention the so-called Reissner-Nordstr\"{o}m, R-N, solution, discovered by  H. Reissner \cite{Reissner(1916)}, H. Weyl \cite{Weyl(1917)}, G. Nordstr\"{o}m \cite{Nordstron(1918)} and G. B. Jeffery \cite{Jeffery(1921)}.
 It was in $1963$ that successfully a more general and exact solution for a rotating black hole was found by R. Kerr \cite{Kerr:1963ud}. Extending the Kerr solution to an axisymmetric and charged black hole, E. Newman and his collaborators found such a solution two years later in $1965$ \cite{Newman:1965my}.
Based on the pioneering attempts by W. Israel \cite{Israel:1967wq}, B. Carter \cite{Carter:1971zc} and D. Robinson \cite{Robinson:1975bv} the no-hair conjecture was constructed, which means that one can completely describe a stationary black hole by virtue of the three parameters of the Kerr-Newman metric: mass, $M$, {electric charge, $q$,} and angular momentum, $J$ \cite{Chrusciel:2012jk}. We should emphasise here, despite the idea of  V. Belinsky, I. Khalatnikov, and E. Lifshitz \cite{Belinskii1969,Belinsky:1970ew} who tried to show that  singularities would not appear in any generic solution,  R. Penrose \cite{Penrose:1964wq} and S. Hawking \cite{Hawking1973} proved that singularities can appear generically, by means of the so-called global technique \cite{Ford:2003qt}; for more information about the properties of the singularities one can see \cite{Hawking:1969sw}. By considering the Kerr-Newman metric as the most general stationary black hole, one can obtain an inequality between those three aforementioned macrostates, i.e. {mass, electric charge, and angular momentum, appeared as $q^2+(J/M)^2\leq M^2$. }Black holes which satisfy this inequality with a minimum mass are known as extremal ones. Although for a configuration that violates this inequality a solution for Einstein's equation still does exist, such a solution will not pass the cosmic censorship conjecture \cite{Penrose:1969pc}, which means there is no event horizon surrounding the nominate black hole and it is called a naked singularity \cite{Christodoulou1999}.\\
Let us turn our attention here to the thermal properties of black holes which led to the well-known four  laws in thermodynamics of black holes, see \cite{Bekenstein:1972tm,Bardeen:1973gs}. For the Schwarzschild case whereas Hawking temperature \cite{Hawking:1974rv} is proportional to the inverse of the black hole mass, heat capacity , i.e. $C$, gets negative values and therefore such a system always is unstable \cite{Davies:1978zz, Hut:1977zx}. For the Reissner-Nordstr\"{o}m and Kerr-Newman black holes, for some intervals the heat capacity can get positive values while for other regions it stays negative; and, interestingly, in the middle region it will diverge and this is called \emph{second-order phase transition}, for the first time proposed by Paul C. W. Davies, after receiving D. W. Sciama's suggestion, \cite{Davies:1978zz}. For more details of the heat capacity  and turn-off point, that is divergence point, one can see \cite{Sokolowski:1980uva,Cai:1997cs}.
During these two past decades, after the introducing of the anti-de Sitter/conformal field theory, AdS/CFT, correspondence \cite{Maldacena:1997re,Maldacena:1997zz},  the investigations of the thermal properties of the black holes, that is the thermodynamics of the black holes, has attracted huge interest \cite{Aharony:1999ti,Shen:2005nu}.  E. Witten interestingly showed that the thermodynamics of AdS space can be explained by virtue of the dual CFT existing on the boundary of the AdS space \cite{Witten:1998zw}. The thermodynamical properties of phase transition for different circumstances were  investigated in the literature, for instance the Hawking-Page phase transition \cite{Hawking:1982dh}, R-N metric in the AdS space \cite{Chamblin:1999tk}, etc. Also one can see interesting results about the concept of holography and the relation between thermodynamics in the presence of gravitation, for a black hole for instance,  and  in the absence of gravitation, i.e. for its dual in the  flat space time with albeit one fewer dimension \cite{tHooft:1993dmi,Susskind:1994vu,Sheikhahmadi:2021jkz}.
Fortunately, the thermodynamic properties of a classical system can completely be reformulated based on the geometry \cite{Ruppeiner:1995zz,Ruppeiner:1983zz}. In fact for the first time it was F. Weinhold who introduced the geometrical approach to the thermodynamics \cite{Weinhold,Weinhold1975zx,Weinhold1975zya,Weinhold1976zxt}. In his leading papers, Weinhold successfully constructed a relation between a type of Riemannian geometry and the second derivative of the internal energy against entropy or other extensive quantities in the thermal system \cite{Weinhold,Weinhold1976zxt}. Such a configuration and its interpretation was physically meaningless and very soon after Weinhold's theorem, G.Ruppeiner took into account entropy and its second derivative against the internal energy, or other extensive quantities of the thermodynamic system, and have introduced a metric which was physically meaningful \cite{Ruppeiner}. Interestingly, metrics in \cite{Weinhold,Weinhold1975zx} and \cite{Ruppeiner} conformally were related and the conformal factor is the inverse of the temperature. Besides these leading proposals, we can consider some other  useful metrics as H. Quevedo metric \cite{Quevedo:2006xk} and Hendi- Panahiyan-Eslam Panah-Momennia, HPEM, metric \cite{Hendi:2015rja} for our analysis. In the former one,  a relation between differential geometry and the properties of thermodynamic system in the framework of the classical thermodynamic equilibrium has been supposed. And in the latter metric, a geometrical phase space due to  the extensive thermodynamic quantities was constructed successfully.  Briefly if one wanted to make  a relation between the properties of the equilibrium manifold to the macroscopic properties, they can be summarised as follows:
\begin{description}
	 \item[$\bullet$]  Curvature $\longleftrightarrow$ Thermodynamic Interactions,
  \item[$\bullet$] Singularity $\longleftrightarrow$ Curvature Phase Transition,
  \item[$\bullet$] Thermodynamic Geodesics $\longleftrightarrow$ Quasi-Static Processes.
\end{description}
As discussed in the aforementioned paragraphs the nature of the singularity located at the heart of the black holes has not yet been understood completely and to realize the physics of this weird phenomena we have no any complete formulated remedy up to now. On the other hand, to cope with the well-known hierarchy problem one way is making a  quantum field theory model of the massive gravity due to the presence of the spin-2 particles, that is gravitons. The first studies of massive gravity dates back to the 1939 paper of M. Fierz and W. Pauli   \cite{Fierz:1939ix,Pauli:1939xp}, in which they introduced a free massive spin-2 action in flat space time. More investigations showed that in the presence of coupling the Fierz-Pauli theory will not tend to the linearised general relativity in the massless limit \cite{vanDam:1970vg,Zakharov:1970cc}. This phenomena is well known as the van Dam-Veltman-Zakharov, vDVZ, discontinuity. To resolve this problem A. I. Vainshtein suggested and showed that by adding nonlinear terms to the Fierz-Pauli action the vDVZ discontinuity will disappear \cite{Vainshtein:1972sx}. This remedy itself faced a problem with its extra degree of freedom known as the Boulware-Deser, BD, ghost \cite{Boulware:1973my}. Although recently some papers came out to {cure the BD ghost problem}, by using the extension in the Fierz-Pauli massive gravity,  known as de Rham, Gabadadze and Tolley, dRGT, { remedy} \cite{deRham:2010ik,deRham:2010kj}, their solution {faces} some other problems like local causality  \cite{Deser:2013eua}. This problem means that for the very high energy scales, e.g. Planck scales, it is not consistent and therefore one cannot quantise it. To overcome this drawback one viable way may be considering gravity invariant conformal  transformations, and one can expect it passes the quantisation criterion  \cite{tHooft:2011aa,Faria:2013hxa}. Recently Bebronne and Tinyakov have obtained a spherical vacuum solution in a massive gravity framework \cite{Bebronne:2009mz,Bebronne:2007qh}. In \cite{Capela:2011mh} the authors examined the validity of the four black hole thermodynamic laws in massive gravity. For more details about massive gravity, its thermodynamics etc.,  we refer the reader to \cite{Faria:2020kbv,deRham:2020yet,Kenna-Allison:2019tbu,Cayuso:2019ieu,Jusufi:2019caq,Dubovsky:2004,EslamPanah:2018ums,Kumara:2019xgt,AhmedRizwan:2019yxk,Gunasekaran:2012dq,Hoseini:2016nzw,Hoseini:2016tvu,Hoseini:2016ztk}.\\
The aforementioned interesting topics motivated us to investigate the so-called  conformal massive black holes by means of a geothermodynamic approach. In this approach, the second phase transition and its properties help us to{ realize} the physics of singularities { in both cases the static and the rotating black holes}. Interestingly, this technique will be useful to understand the behaviour of temperature, entropy and other extensive thermodynamic quantities too.

{This work} is organised as follows: In Sec.\,\ref{section1} both the physical and mathematical properties of a static black hole in conformal massive gravity besides thermodynamic behaviour in both classical and geothermodynamics frameworks will be discussed. {In} Sec.\, \ref{section3} the metric of a rotating regular black hole in the conformal massive gravity and thermodynamic properties to see the behaviour of second phase transition and temperature evolutions will be investigated. Ultimately Sec.\, \ref{section4} is devoted to the discussion and conclusions of the work.\\
\\

%=================================================================
%============== Static regular black hole in conformal massive gravity=====================================
%=================================================================

%=================================================================
%============== Static regular black hole in conformal massive gravity=====================================
%=================================================================

\textcolor[rgb]{0.00,0.00,1.00}{\section{Static regular black hole in conformal massive gravity}\label{section1}}
In this section, we want to supply a brief introduction to the metric of a static regular black hole in conformal massive gravity (CMG). The massive gravity theory usually could be described by the following action \cite{Dubovsky:2004,Jusufi:2019caq}
\begin{equation}\label{LagrangianMG}
\mathcal{S}_{MG}=\int{d^{4}x\sqrt{-g}} \bigg[\frac{R}{16\pi}+\Lambda^{4}\mathfrak{F}(\mathcal{X},W^{ij})\bigg],
\end{equation}
where
\begin{eqnarray}\label{GoldstoneFields}
&&W^{ij}=\frac{{{\partial ^\mu }{\psi ^i}{\partial _\mu }{\psi ^j}}}{\Lambda^{4}} - \frac{{{\partial ^\mu }{\psi ^i}{\partial _\mu }{\psi ^0}{\partial ^\upsilon }{\psi ^j}{\partial _\upsilon }{\psi ^0}}}{{{\Lambda ^4}\mathcal{X}}},\\
&&\mathcal{X}=\frac{{{\partial ^0}{\psi ^i}{\partial _0}{\psi ^i}}}{{{\Lambda ^4}}},
\end{eqnarray}
 { and $ \mathfrak{F}(\mathcal{X},W^{ij}) $ is considered as a function of the four scalar Goldstone fields $ \psi ^i $ and $ \psi ^0 $, which are minimally coupled to gravity by derivative interactions. As carefully discussed in  \cite{Dubovsky:2004,Bebronne:2009mz} this model is generally covariant and can satisfy Lorentz symmetry criterion at the level of the action.  For instance in \cite{Bebronne:2009mz}, different solutions for the function of $\mathfrak{F}(\mathcal{X},W^{ij})$, almost analytical solutions, are introduced and among them one can mention
\begin{equation}\label{Fa}
\begin{aligned}
\mathfrak{F}(\mathcal{X},W^{ij}) =& c_{0}\left(\frac{1}{\mathcal{X}}+w_{1}\right) \\
&+c_{1}\left(w_{1}^{3}-3 w_{1} w_{2}-6 w_{1}+2 w_{3}-12\right)\,,
\end{aligned}
\end{equation}
where $w_{n}=\operatorname{Tr} W^{n}, n=1, 2, 3$, $c_0$ is a constant and $c_1=\pm 1$,
 or as another example the following expression
\begin{equation}\nonumber
\begin{aligned}
\mathfrak{F}(\mathcal{X},W^{ij}) =& c_{0}\left(z_{1}+2+\frac{z_{1}^{3}-6 z_{1} z_{2}+8 z_{3}}{3}\right) \\
&+2 c_{1}\left(z_{1}^{2}-2 z_{2}-1-2 \frac{z_{1}^{3}-6 z_{1} z_{2}+8 z_{3}}{3}\right)\,,
\end{aligned}
\end{equation}
where $z_{n}=\operatorname{Tr} Z^{n}, n=1, 2, 3$, and $Z\equiv \mathcal{X} W^{ij}$.
 With regard to studying the consequences of symmetry breaking one may consider a background solution as
\begin{equation}\nonumber
g_{\mu \nu}=\eta_{\mu \nu}, \quad \psi^{0}=\alpha t, \quad \psi^{i}=\beta x^{i}\,,
\end{equation}
 with zero energy momentum tensor for some constant $\alpha$ and $\beta$, in this model of massive gravity they can be considered as $\alpha=\beta=\Lambda^2$ .} In Eq.\eqref{LagrangianMG} the constant $ \Lambda =(m M_P)^{1/2}$   has the dimension of mass, here $m$ refers the graviton mass and Planck mass denoted by $M_P$,  and $ R $ stands for Ricci scalar. Then by introducing a conformal transformation, and using \eqref{Fa} that appeared in \cite{Bebronne:2009mz}, Kimet Jusufi et al., \cite{Jusufi:2019caq} have introduced a static spherically symmetric black hole, in the presence of CMG correction as follows
\begin{align}\label{metric}
\begin{split}
ds^{2}=-\bigg(1+\frac{L^{2}}{r^{2}}\bigg)^{\frac{\left| \lambda  \right|}{2}+2}\bigg(1-\frac{2M}{r}-\frac{Q}{r^{\lambda}}\bigg)dt^{2} \\
+\bigg(1+\frac{L^{2}}{r^{2}}\bigg)^{\frac{\left| \lambda  \right|}{2}+2}\bigg(1-\frac{2M}{r}-\frac{Q}{r^{\lambda}}\bigg)^{-1} dr^{2}\\
+\bigg(1+\frac{L^{2}}{r^{2}}\bigg)^{\frac{\left| \lambda  \right|}{2}+2}r^2(d\theta^2 +\sin^2\theta d\phi^2 ),
\end{split}
\end{align}
where $ L $ is the CMG parameter, introduced for dimensional reasons, $ \lambda $ is the hair parameter{and $Q$ is the scalar charge}. {As shown in the article \cite{Jusufi:2019caq}, and given the observational constraints of the galaxy $M87^*$, there is no particular reason that the $Q $ and $\lambda$ parameters accept only positive values, and the model evolution should be examined for negative values as well.}  For more details of derivation of this metric we refer the reader to \cite{Bebronne:2009mz}. In the next sections, we are going to investigate the geothermodynamics of such a metric risen from CMG. Before {going }deeply through such an investigation, we will try to obtain thermodynamic key parameters like temperature and heat capacity.

\begin{figure}[]
	\centering
\subfigure[$\lambda=0.5$]{\includegraphics[width=7cm]{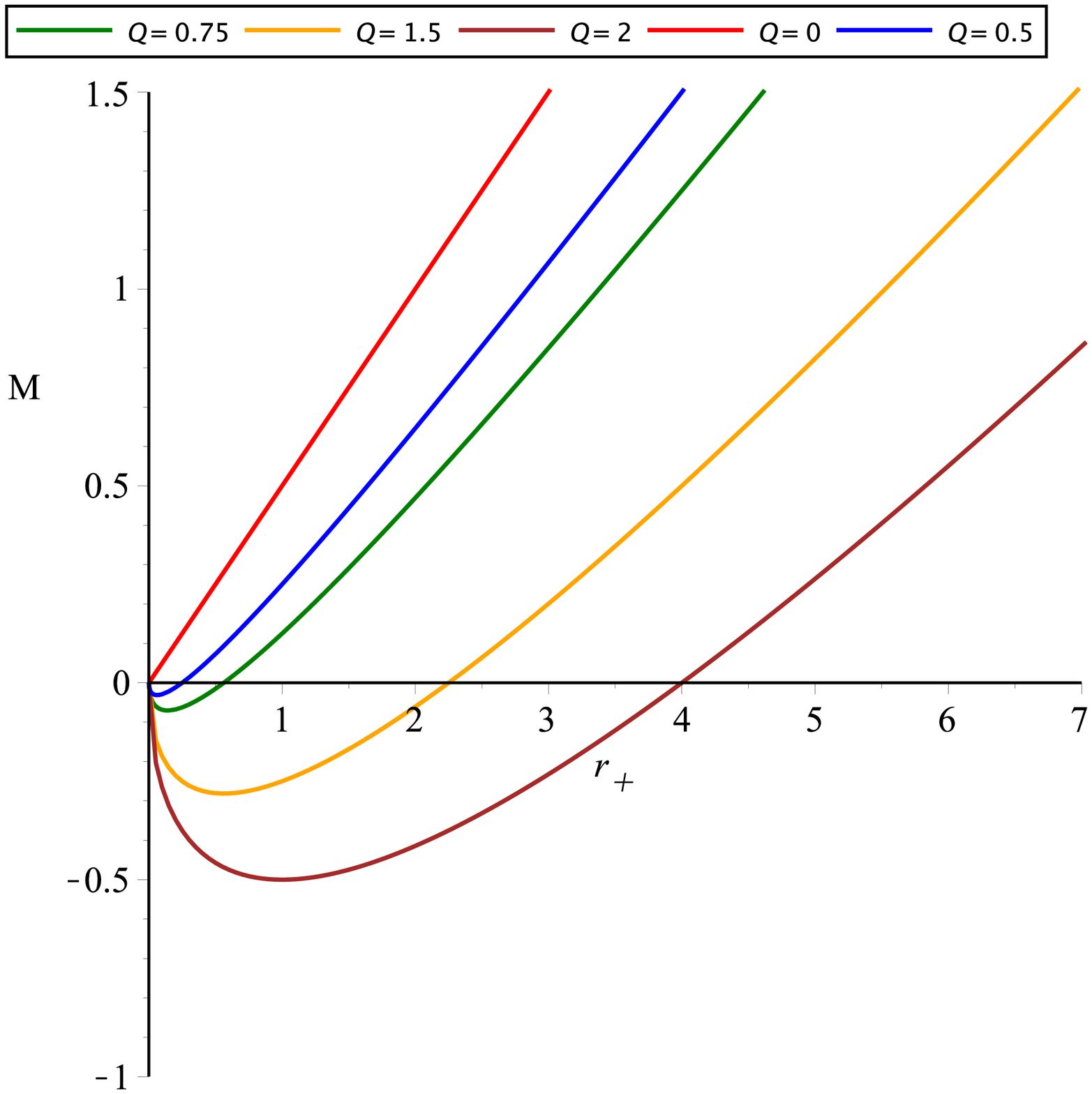}}
\subfigure[$Q=0.5$]{	\includegraphics[width=7cm]{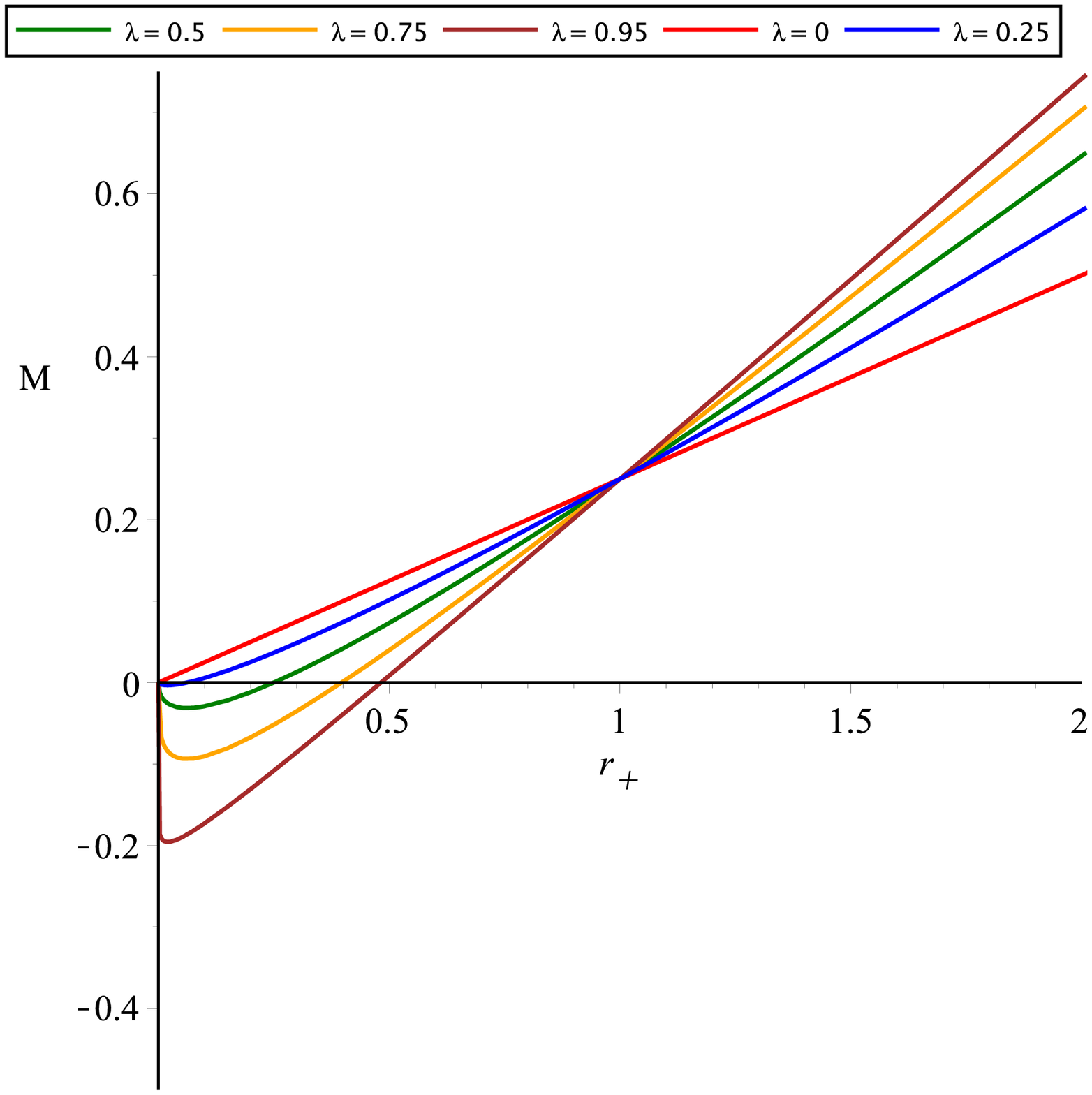}}
	\caption{ {The Variation of mass function  in terms of horizon radius $ r_{+}$, for a static regular black hole in the presence of  CMG corrections is illustrated. Obviously plot (a) focuses on the changes in amount of $Q$ when the parameter $\lambda$ gets a fixed amount. In figure (b)  the changes are related to the parameter  $\lambda$ when  $Q$ is a fixed constant. In figure (a),  when $Q=0$, that is a Schwarzschild-like solution, mass function gets only positive values, but by increasing the size of the black hole and for R-N black holes in the presence of the CMG corrections mass function accepts smaller negative values then enters to the positive regions. For figure (b), when hair parameter, $\lambda=0$  and $Q\neq0$,  mass function in the presence of a CMG correction shows the same behaviour, as figure (a), when $Q$ tends zero. There  may be a physical interpretation for negative mass that appears in this model with the help of quantum gravity, see \cite{Mann:1997jb} for more details. }  }
	\label{pic:M}
\end{figure}

\begin{figure}[]
	\centering
	\subfigure[]{
		\includegraphics[width=0.38\textwidth]{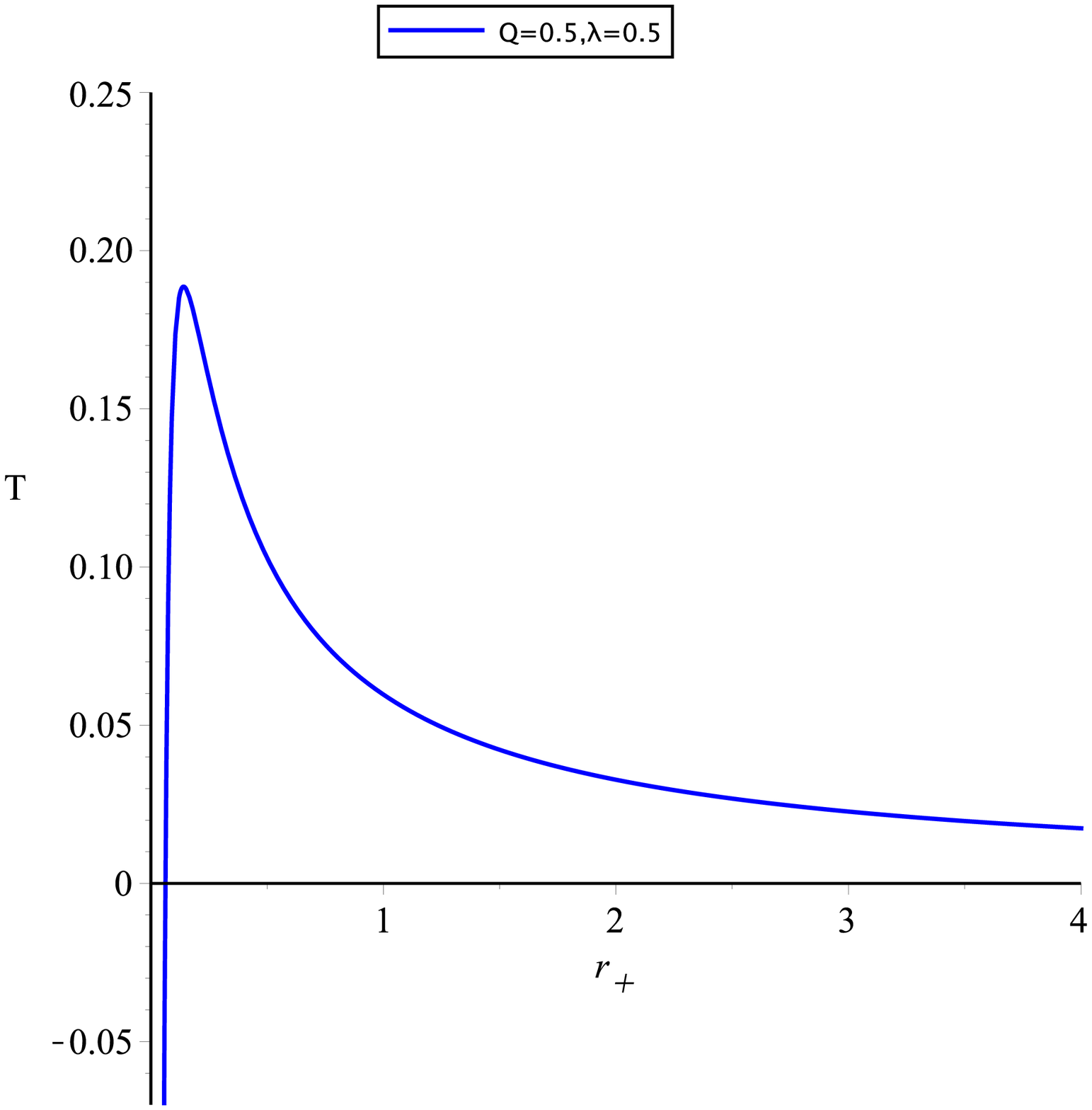}
	}
\subfigure[$\lambda=0.5$]{
	\includegraphics[width=0.38\textwidth]{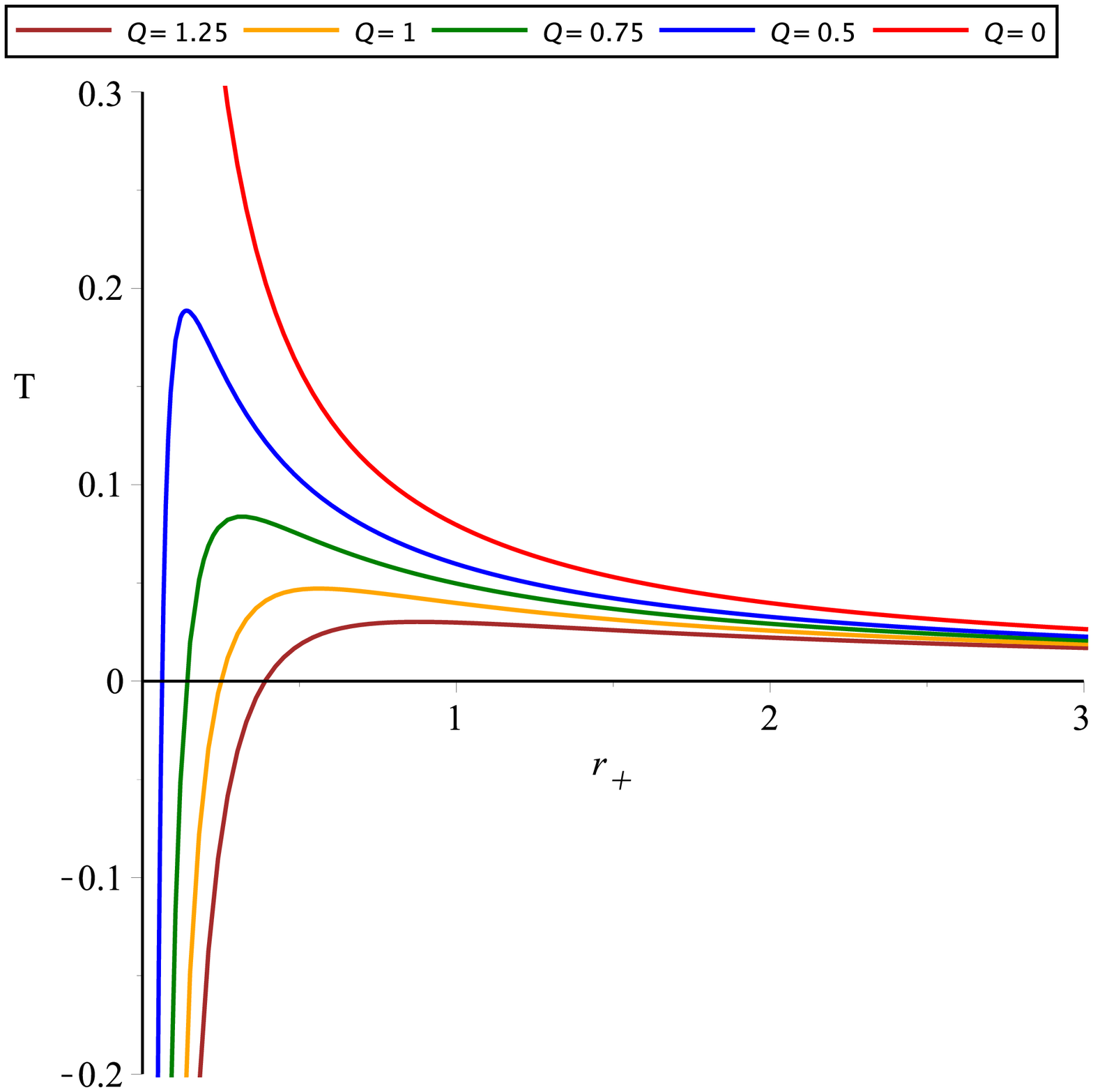}
}
	\subfigure[$Q=0.5 $]{
		\includegraphics[width=0.38\textwidth]{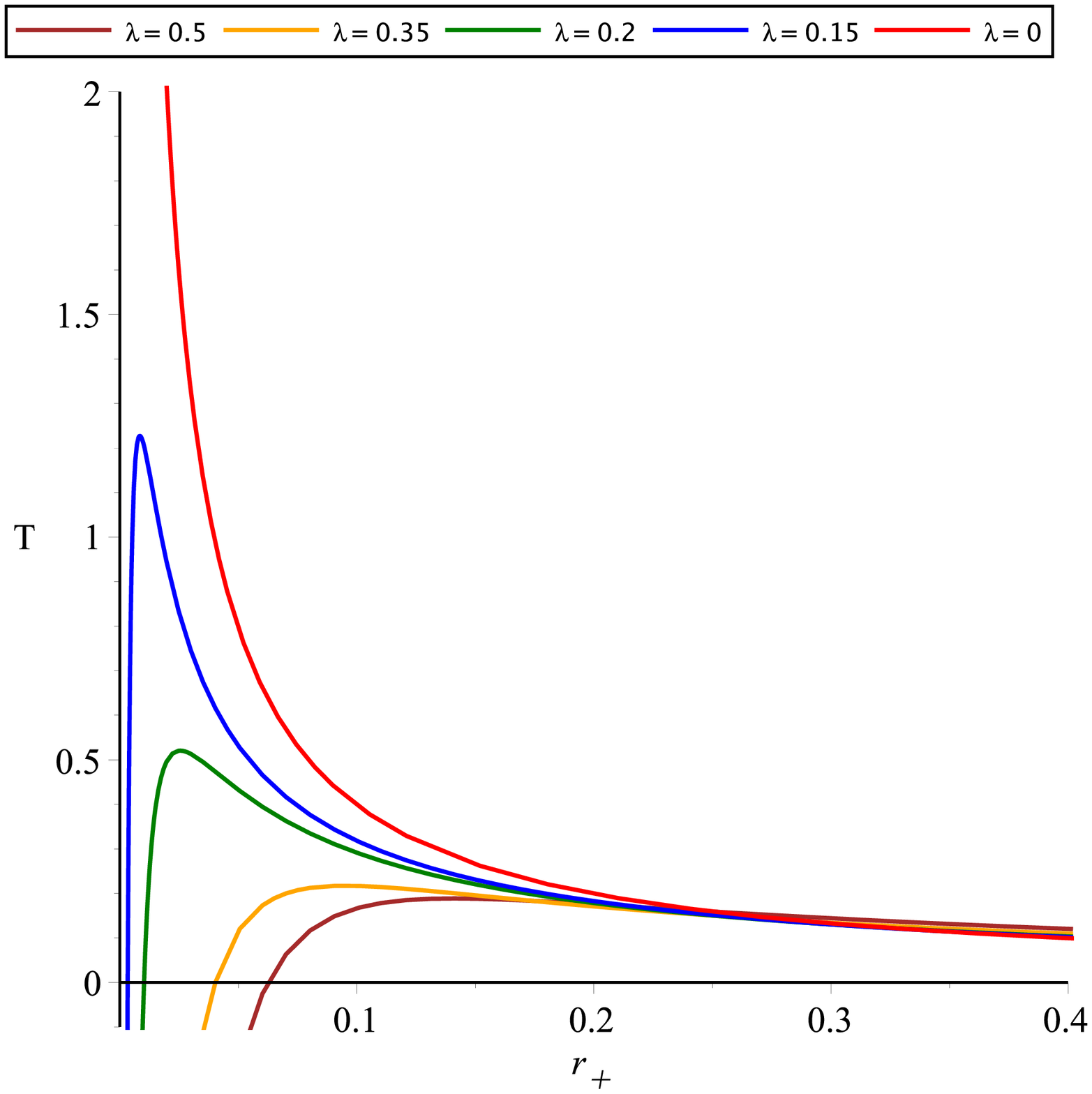}
	}
	\caption{{These figures show the evolution of temperature in terms of horizon radius $ r_{+}$, for a static regular black hole in CMG background. In figure (a) the behaviour of temperature when both the parameters $Q$ and $\lambda$ get nonzero values is plotted. The plot (b) is devoted to investigate a specific case when $Q$ tends to zero but $\lambda$ not, that is the well-known Schwarzschild-like case in the presence of CMG corrections. And if one wanted to see the R-N limits in the absence of CMG corrections,  one can follow figure (c).  Theses plots indicate that larger black holes will be colder compared to the smaller ones. }}
	\label{pic:T}
\end{figure}

\begin{figure}[]
	\centering
		\subfigure[]{
		\includegraphics[width=0.3\textwidth]{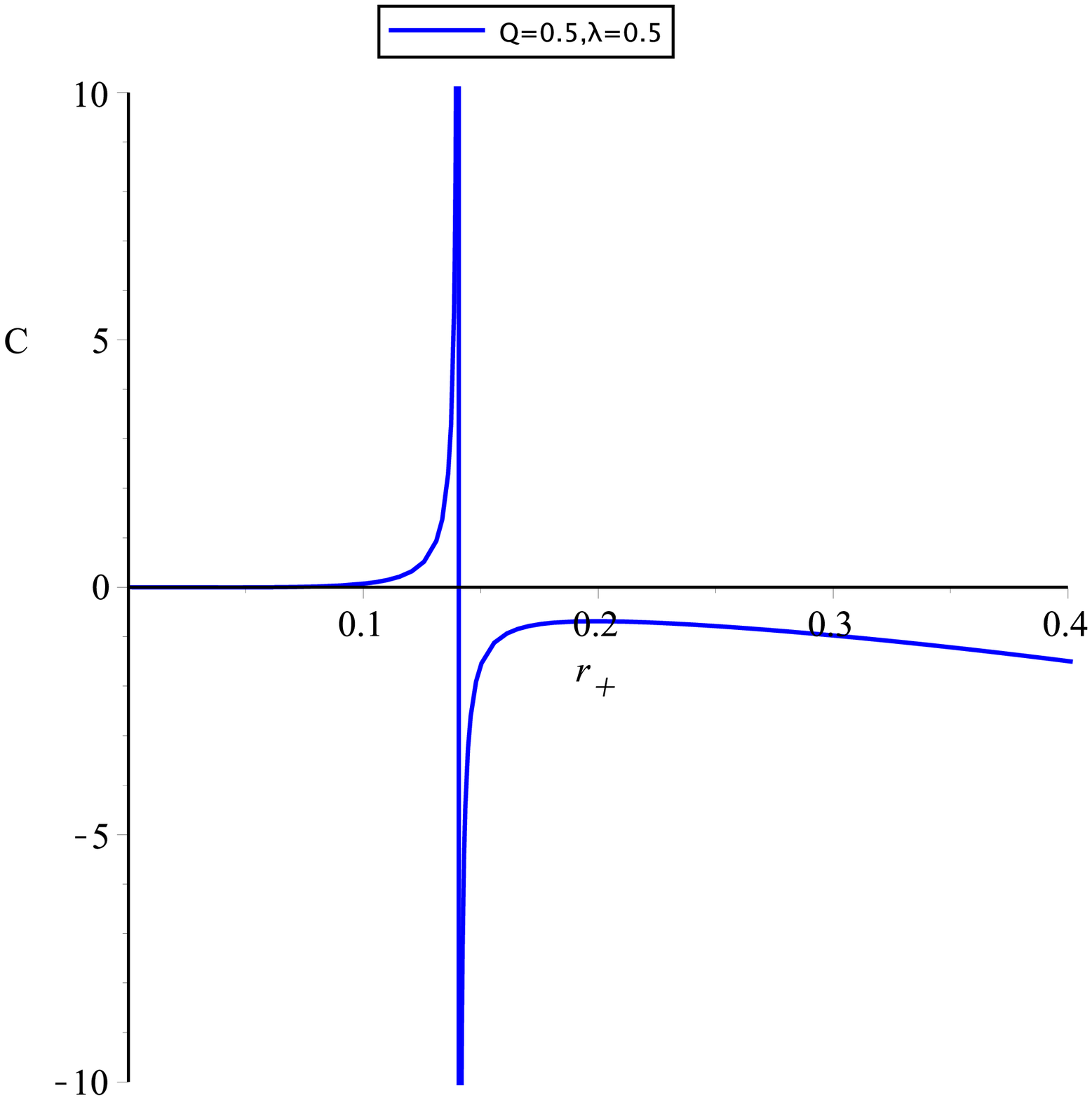}
	}
\subfigure[closeup of figure (a)]{
	\includegraphics[width=0.3\textwidth]{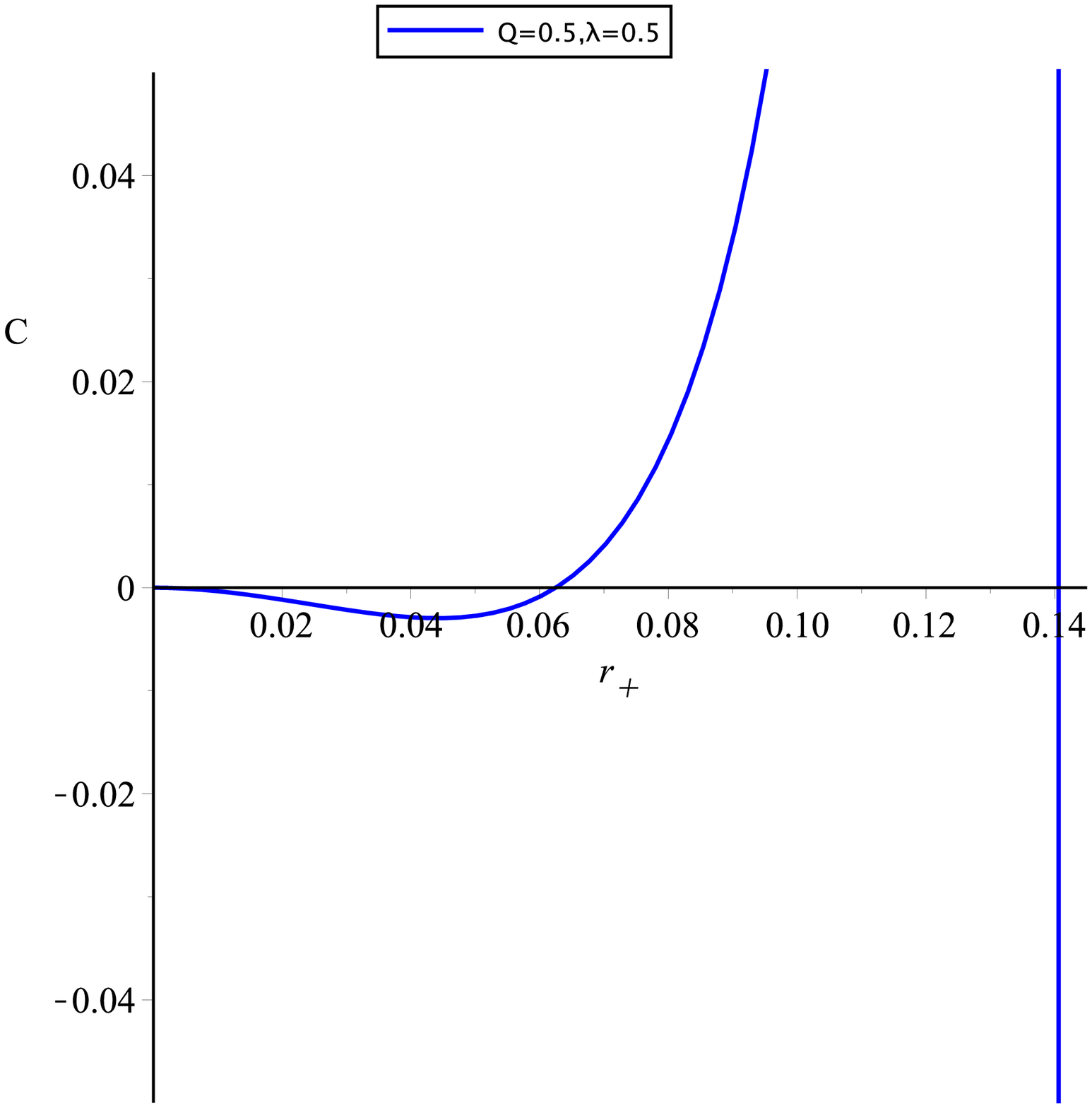}
}
	\subfigure[$\lambda=0.5$]{
		\includegraphics[width=0.3\textwidth]{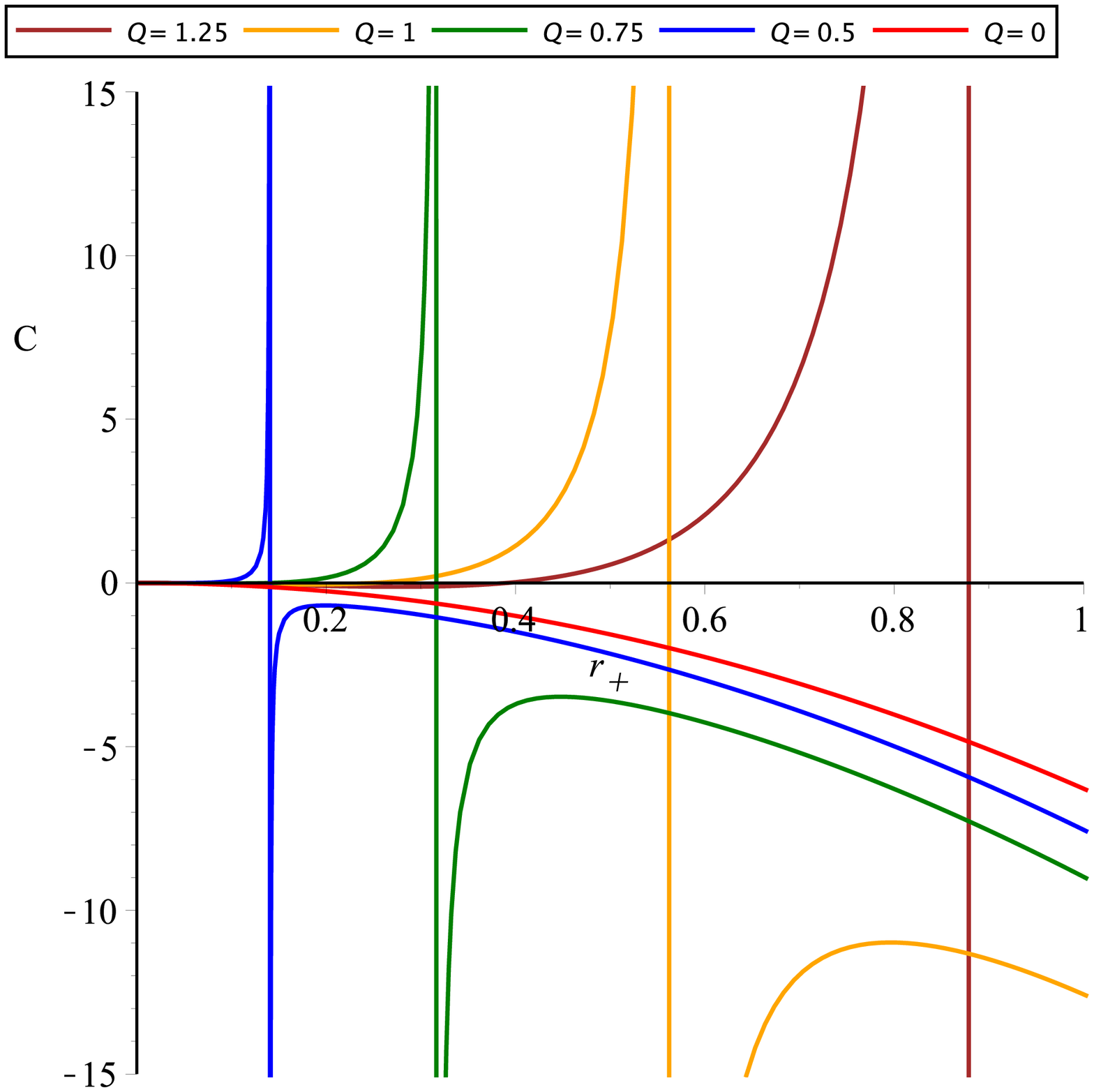}
	}
\subfigure[closeup of figure (c) ]{
	\includegraphics[width=0.3\textwidth]{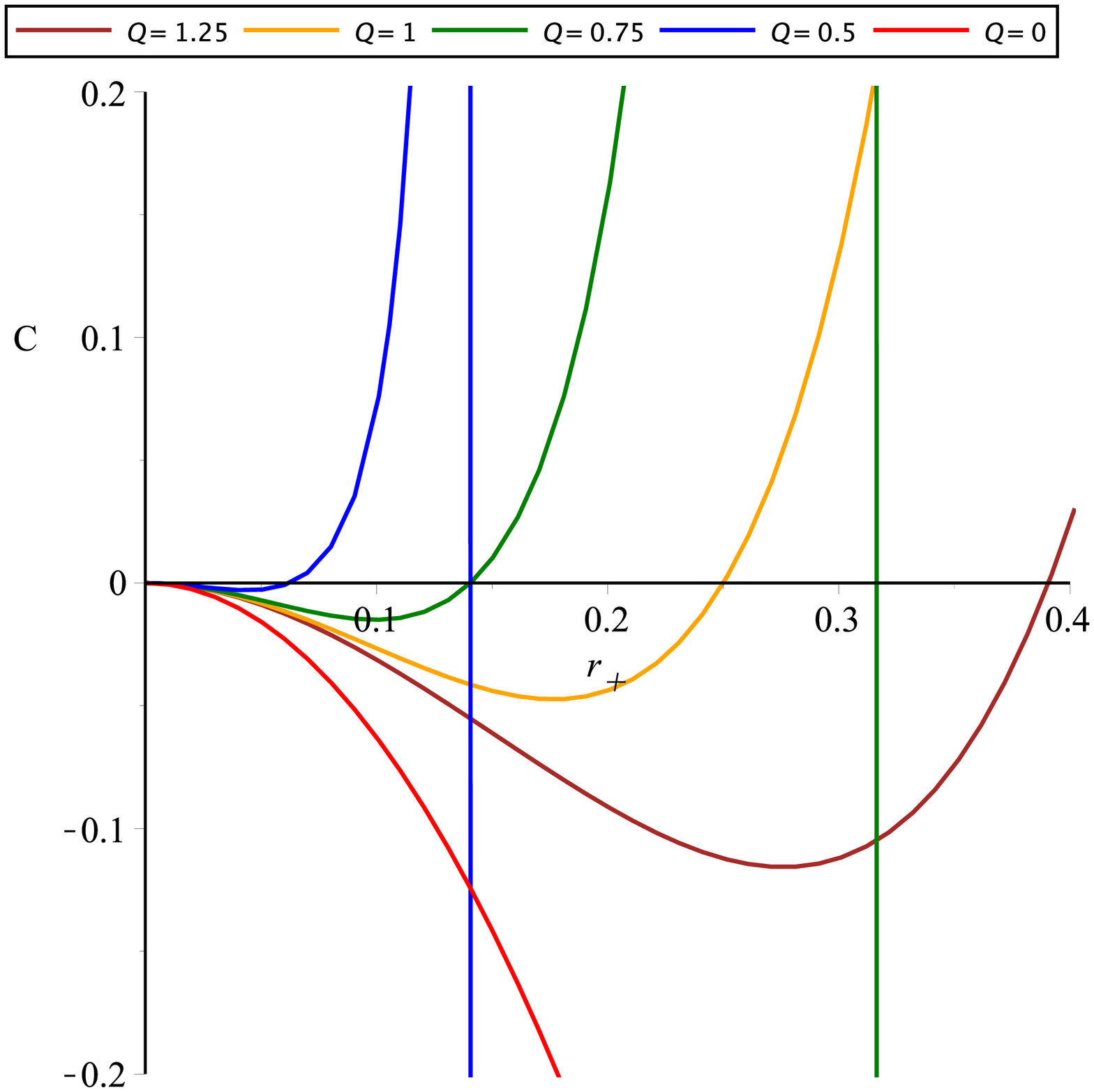}
}
	\subfigure[$ Q=0.5 $]{
		\includegraphics[width=0.3\textwidth]{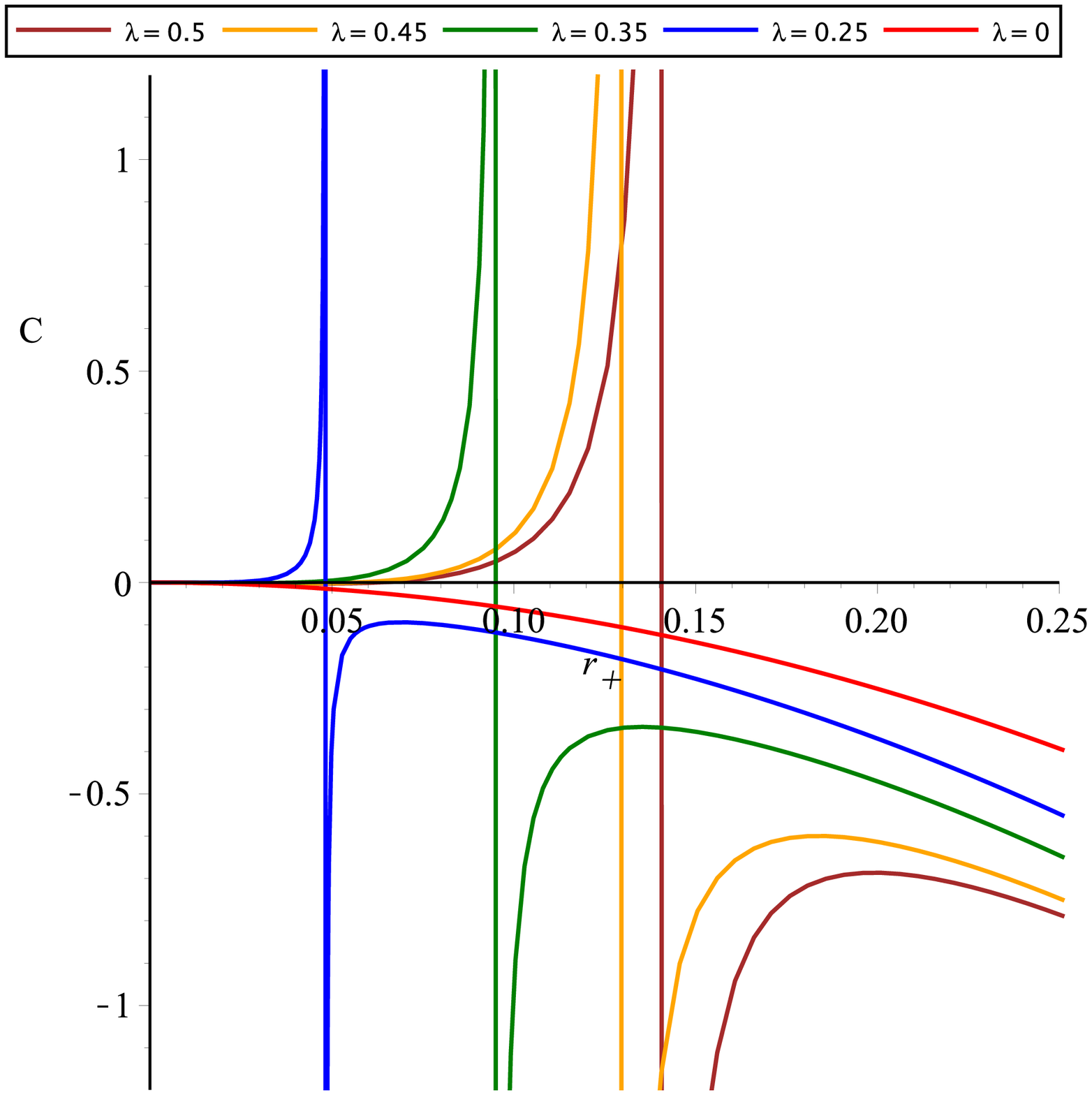}
	}
\subfigure[closeup of figure (e) ]{
	\includegraphics[width=0.3\textwidth]{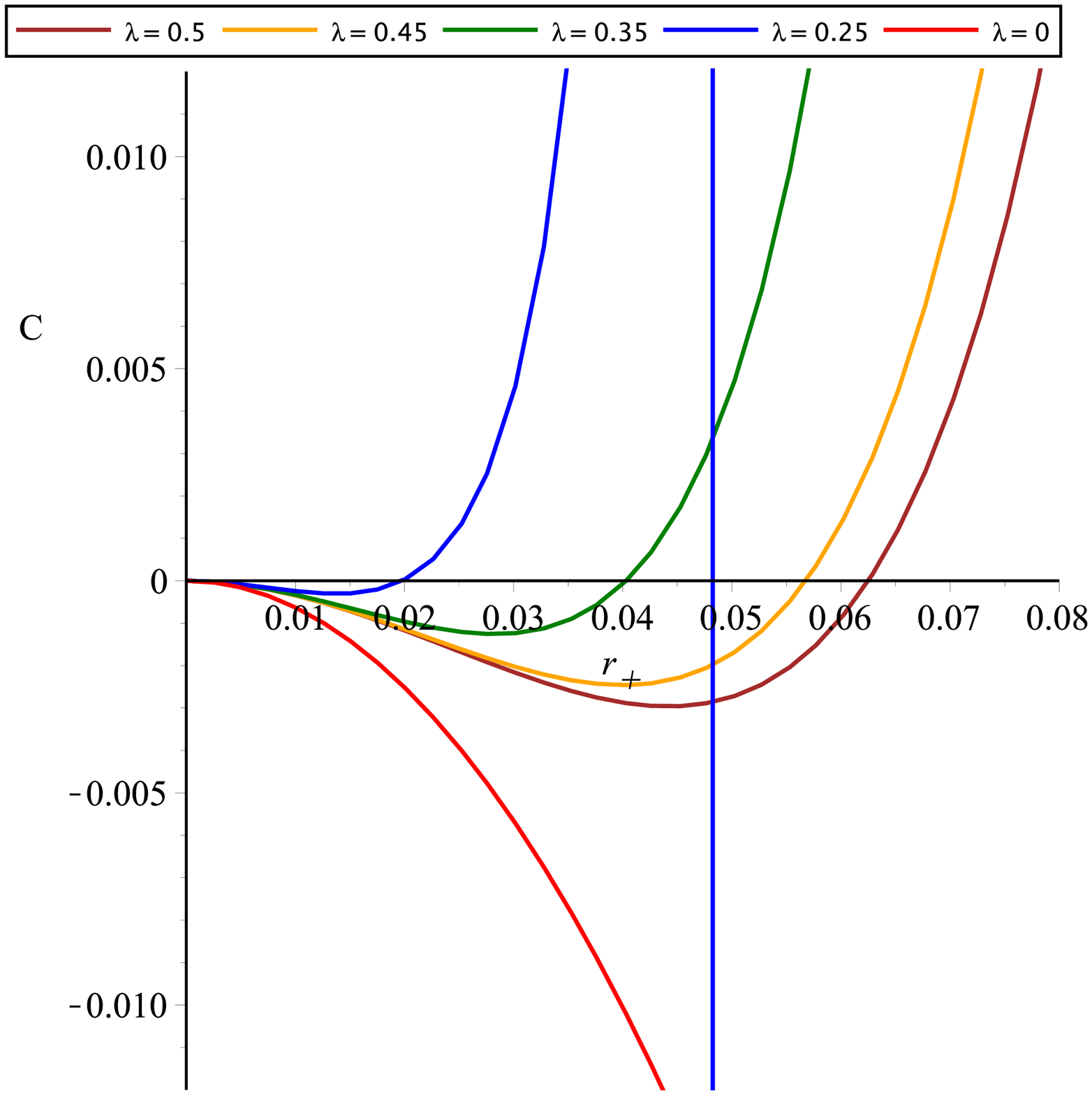}
}
\\
	\caption{{These figures indicate the variations of the heat capacity in terms of horizon radius $ r_{+}$, for a static regular black hole in the presence of the CMG corrections. As a general case, when both parameters $Q$ and $\lambda$ get nonzero  values then figures (a) and (b) result. Besides these,  plots (c) and (d)  contain a Schwarzschild-like solution, when $Q=0$, and R-N solutions in the presence of a fixed CMG correction. Finally, to investigate the effects of variation of the hair parameter on the R-N like black holes evolution figures (e) and (f) are plotted. }}
	\label{pic:C}
\end{figure}

%%%%%%%%%%%%%%%%%
\textcolor[rgb]{0.00,0.00,1.00}{\subsection{Thermodynamics}\label{sub1}}
Now, we are ready to calculate, some important, thermodynamics quantities of a regular static black hole in CMG. In doing so, first utilizing  the condition  $f(r_{+})=0$, at the event horizon $ r = r_{+}$, and the relation between  $S$  and event horizon radius, i.e.  $ (S=\pi r^{2}_{+})$ we {can calculate mass of the black hole} as a function of the extensive quantities entropy, $ S $, and  {scalar charge},$ Q $.
Therefore, mass function can be expressed as follows
\begin{equation}\label{mass}
M(S,Q)=-\frac{1}{2}\sqrt {\frac{S}{\pi }}\bigg(\frac{{ {{\bigg(\frac{S}{\pi } \bigg)}^\frac{\lambda}{2} } - Q}}{{{{\bigg(\frac{S}{\pi } \bigg)}^\frac{\lambda}{2} }}} \bigg)\,.
\end{equation}
Then, by considering the first law of black hole thermodynamics
\begin{equation}\label{RQlow}
	{\it dM}={\it TdS}+\varphi {\it dQ}\,,
\end{equation}
Hawking temperature and heat capacity can be expressed, respectively, as following

\begin{equation}\label{T}
	T=\bigg(\frac{\partial M}{\partial S}\bigg)= \frac{1}{4}\frac{{Q(\lambda  - 1)({\pi ^{\frac{\lambda }{2}}}{S^{1 - \frac{\lambda }{2}}}) + S}}{{{S^{\frac{3}{2}}}{\pi ^{\frac{1}{2}}}}}\,,
\end{equation}
and
\begin{equation}\label{C}
C= T\bigg(\frac{\partial S}{\partial T}\bigg)=\dfrac{\bigg(\frac{\partial M}{\partial S}\bigg)}{\bigg(\frac{\partial^2 M}{\partial S^2}\bigg)}=-\frac{{ 2S\bigg(({\pi ^{\frac{\lambda }{2}}}{S^{\frac{\lambda }{2}}})Q(\lambda  - 1) + 1\bigg)}}{{{\pi ^{\frac{\lambda }{2}}}{S^{\frac{\lambda }{2}}}Q({\lambda ^2} - 1) + 1}}\,.
\end{equation}
{where, in Eq.(\ref{RQlow}), $\varphi=({\partial M}/{\partial Q})$ is the {scalar  potential, in similarity to the electrostatic potential \cite{Pacilio:2018gom}}}. {To impose some constraints on the model, aiming to distinguish physical or non-physical results, one can these equations useful. }In other words, they express that the root of heat capacity (i.e. $ C=T=0 $) is indicating a boundary between physical  ($T > 0$) and non-physical ($ {T < 0} $) black holes, which can be called a \emph{physical limitation point}. Due to this property, {usually one can expect that} the thermodynamic system shows a change in the sign of the heat capacity. Also it is well-known, the divergence of the heat capacity usually indicates a phase transition {critical point} of a black hole. Therefore, both of these points can be expressed as the following equations \cite{EslamPanah:2018ums}.
\begin{equation}
\left\{\begin{array}{ll}
T=\left(\frac{\partial M}{\partial S}\right)=0\,, &  { physical~ limitation~points } \\
\left(\frac{\partial^{2} M}{\partial S^{2}}\right)=0\,, \quad & { phase~ transition~ critical ~points }
\end{array}\right.
\end{equation}
To illustrate mentioned above properties, the thermodynamic parameters including mass, temperature, and heat capacity, i.e. $M$, $T$ and $C$ are plotted versus {event} horizon radius, $r_+$, aiming to investigate their behaviors in more detail, see Figs.~\ref{pic:M}, \ref{pic:T} and \ref{pic:C}.
Fig.~\ref{pic:M} (a), (b) shows the behavior of mass for different values of $Q$ and $\lambda$ versus $r_+$. It can be illustrated, from these figures, the mass of this black hole has a minimum point, after that it starts growing  by  increasing in the amount of $r_+$. In addition, clearly  from Fig.~\ref{pic:M} (a), (b), it will be observed that by increasing the value of $Q$ and $\lambda$, the value of the minimum point starts to decrease. {In Fig.~\ref{pic:M} (a),  when $Q=0$, a Schwarzschild-like solution, mass function gets only positive values, but by increasing the size of the black hole and for R-N black holes in the presence of the CMG corrections mass function accepts smaller negative values then enters to the positive regions. For Fig.~\ref{pic:M} (b), when hair parameter, $\lambda=0$  and $Q\neq0$,  mass function in the presence of a CMG correction shows the same behaviour, as figure (a), when $Q$ tends to zero! Another interesting behavior of the mass function that appeared in Fig.~\ref{pic:M} (b), not in Fig.~\ref{pic:M} (a), is converging this function at $r_{+}\approx1 $  to a fixed amount for $M$. It means for a R-N black hole with a fixed amount of $Q$, when hair parameter varies then mass function is almost independent of variation of parameter $\lambda$ at $r_{+}\approx 1.$  }
{  Now let's investigate the negative mass that appeared in the mass function versus $r_+$. In one side, the possibility of a negative mass actually is one of the cases that has been studied in detail in Literature \cite{Bondi:1957zz,Belletete:2013nqa}. There are some examples including classical physics, quantum mechanics, general relativity, quantum field theory, Hawking radiation, and even observational experiments that the concept of negative mass can be investigated, see  \cite{Bondi:1957zz,Belletete:2013nqa,Castelvecchi:2016gbf}.
 In \cite{Bondi:1957zz} the possibility of a negative mass, especially from a gravitational point of view, is well discussed. That is, arguments such as what exists for a positive mass can also be applied to a negative mass. Although it is relatively impossible to prove this claim in our day-to-day experiences, it may be necessary to justify some important phenomena in physics, such as Hawking radiation. Also recently in \cite{Khamehchi:2017xzz} a claim that a negative mass can play the role of dark matter and dark energy at the same time was proposed.
As it has been argued in \cite{Farnes:2017gbf}, the negative mass is a potential candidate for dark matter because they do not combine through gravity and the structures do not form naturally and it does not allow nuclear fusion, for instance, and consequently electromagnetic radiation.
On the other hand, as mentioned, the traces of this concept can also be traced back to Hawking radiation \cite{Castelvecchi:2016gbf}.
Because according to the mechanism governing this radiation, negative energy can be assigned to the antiparticle that is absorbed by the black hole at its horizon, and by considering Einstein's mass-energy equivalence, it may be related to a negative mass and this causes the so-called black hole evaporation.
However, linking such a claim to our work, which is a semi-classical study, maybe hasty, and for this claim to be true, it is necessary to examine carefully the concepts of quantum gravity or at least quantum field theory in curved space-time. But in its raw form it can be interpreted that for black holes where the mass parameter becomes negative for values of the event horizon, the black hole is losing mass and a kind of Hawking radiation, which may be related to the presence of CMG corrections, as can be seen from Fig.~\ref{pic:M} for example.}
{On the other side, there may be another physical interpretation for negative mass appeared in this model with the help of  quantum gravity. For instance, Robert Mann in \cite{Mann:1997jb} showed that by utilizing quantum fluctuations some regions with negative energy can exist that under some certain conditions will undergo gravitational collapse to form a black hole!. The exterior of such a black hole definitely has negative mass with a non-trivial topology.}

%%%%%%%%%%%%%%%%%%%%%%
\begin{figure}[]
	\centering
	\subfigure[]{
		\includegraphics[width=0.38\textwidth]{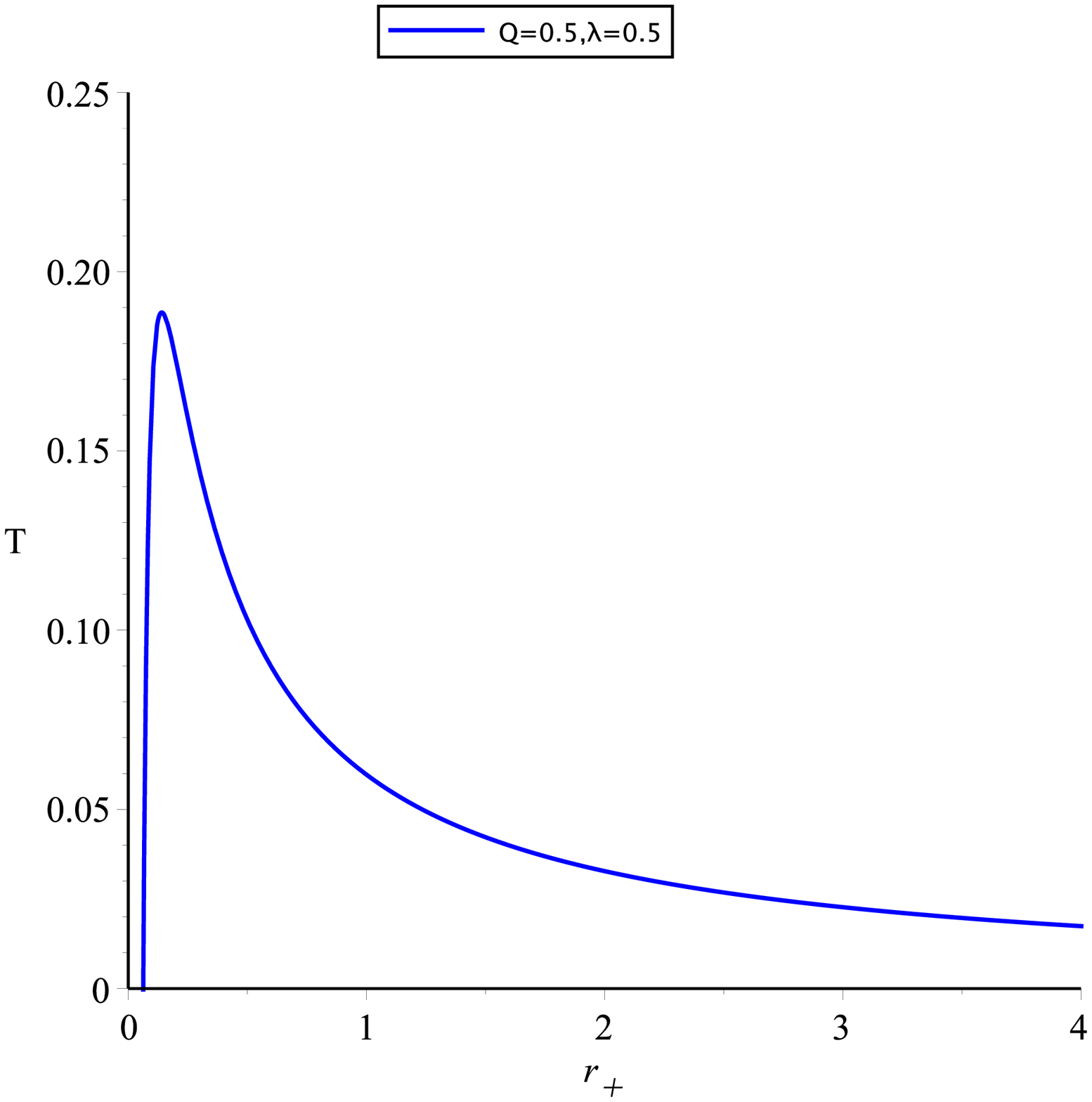}
	}
\subfigure[$\lambda=0.5$]{
	\includegraphics[width=0.38\textwidth]{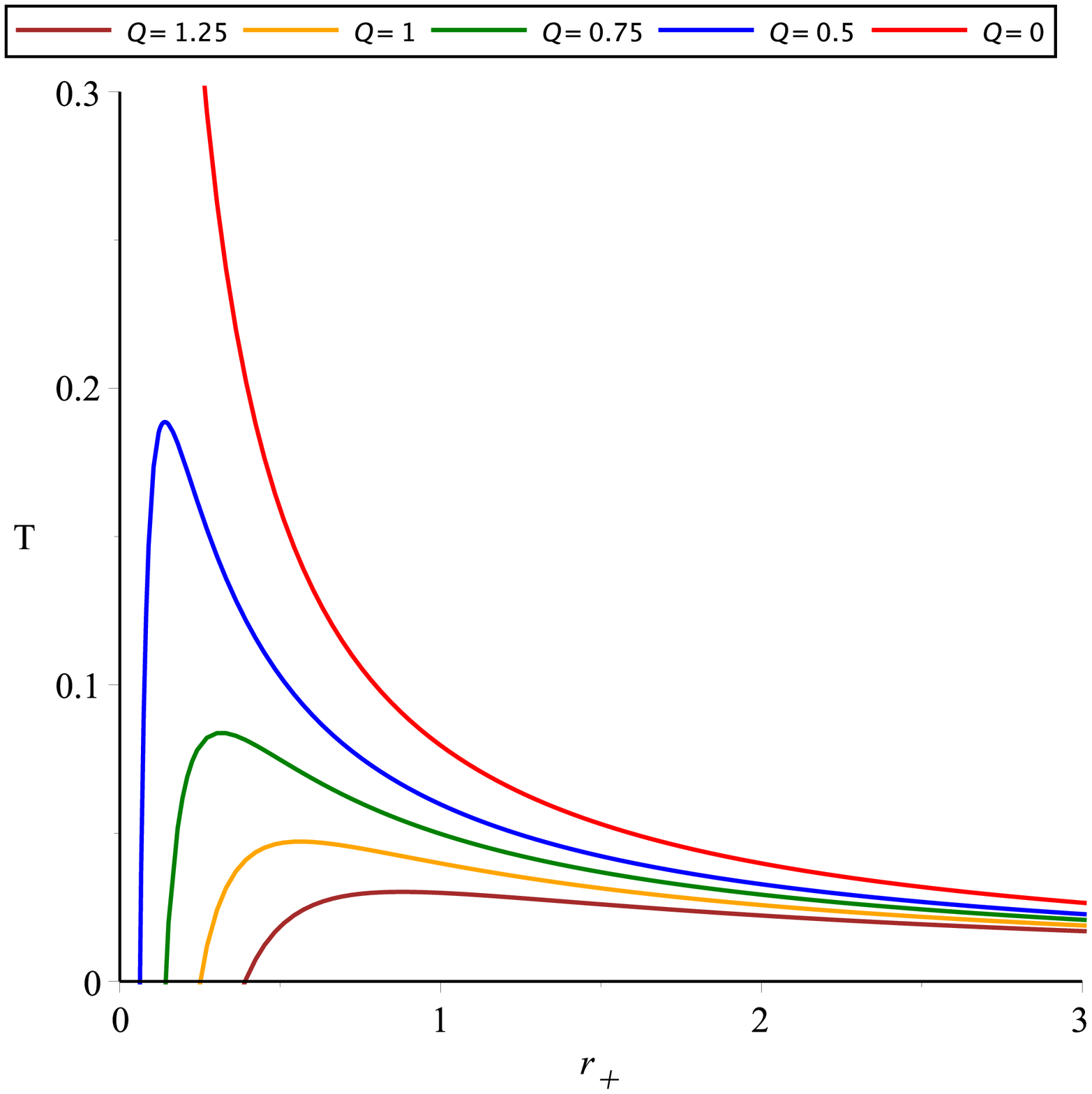}
}
	\subfigure[$Q=0.5 $]{
		\includegraphics[width=0.38\textwidth]{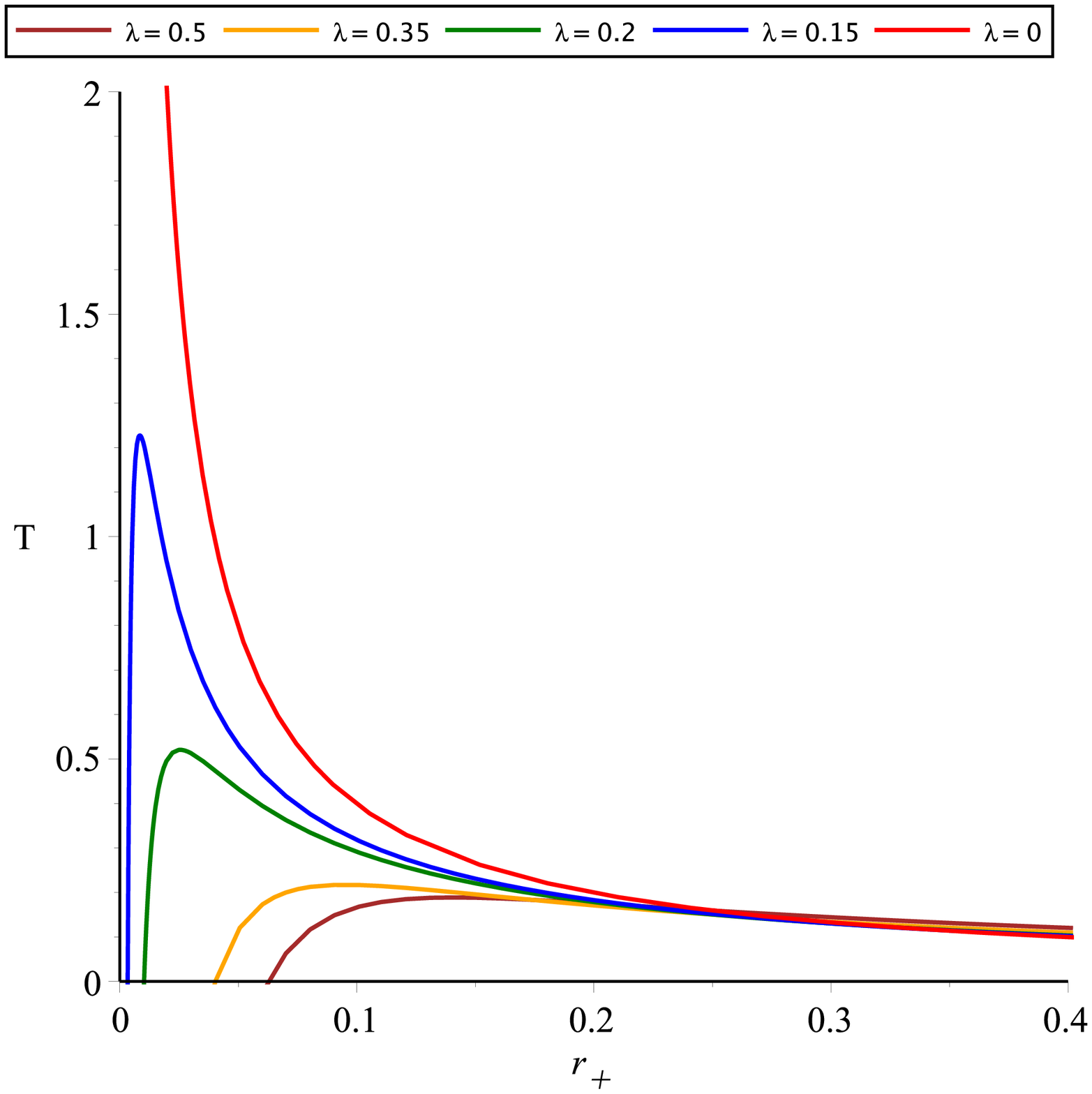}
	}
	\caption{{These figures show the evolution of temperature in terms of horizon radius $ r_{+}$, for a static regular black hole in CMG background. The behaviour of temperature here is as same as Fig.~\eqref{pic:T}  except that a cut-off length is provided here to avoid any conflict with quantum gravity. }}
	\label{cut-off:T-St}
\end{figure}

\begin{figure}[]
	\centering
\subfigure[]{
	\includegraphics[width=0.4\textwidth]{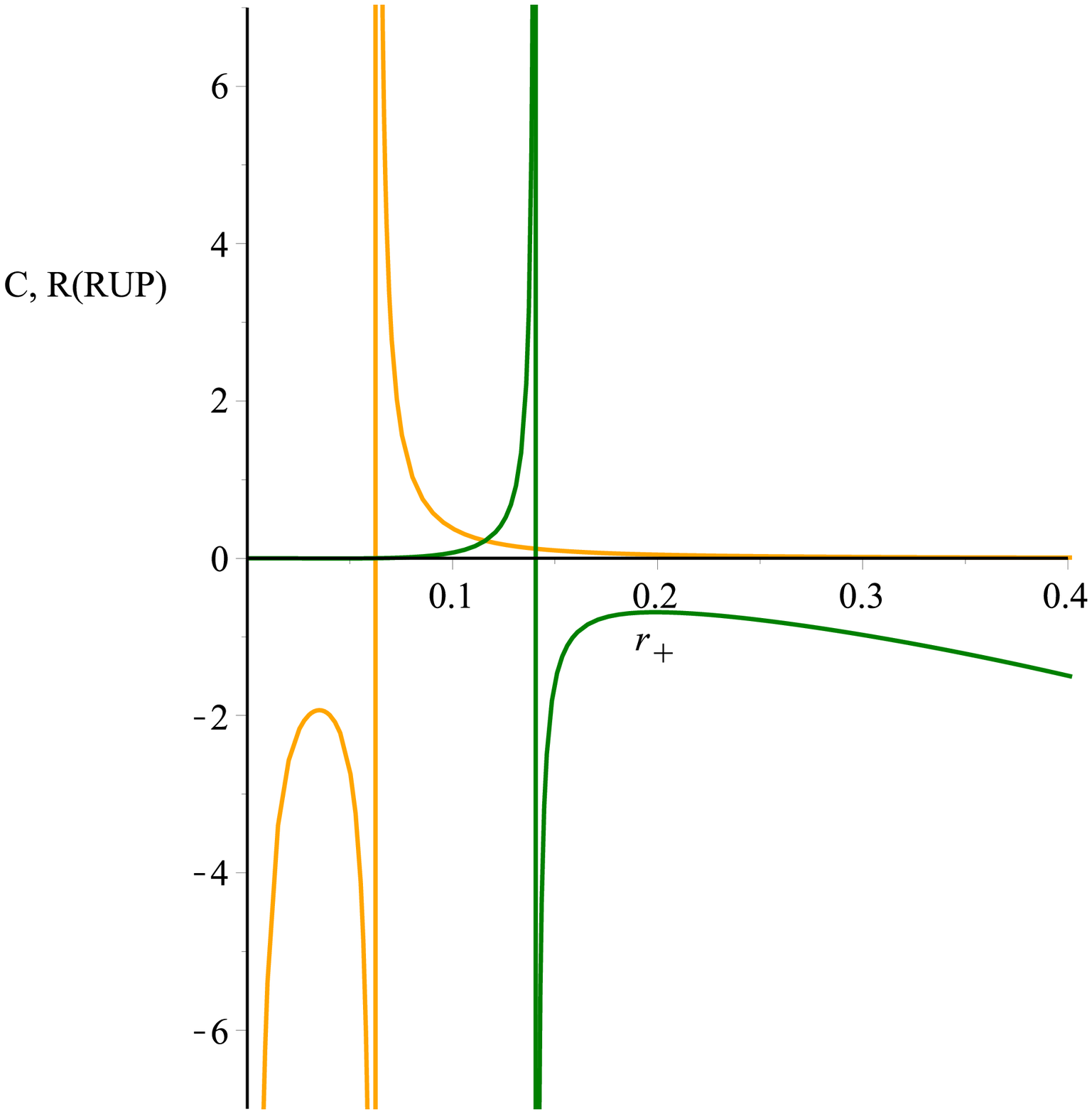}
}
	\subfigure[Closeup of figure (a)]{
		\includegraphics[width=0.4\textwidth]{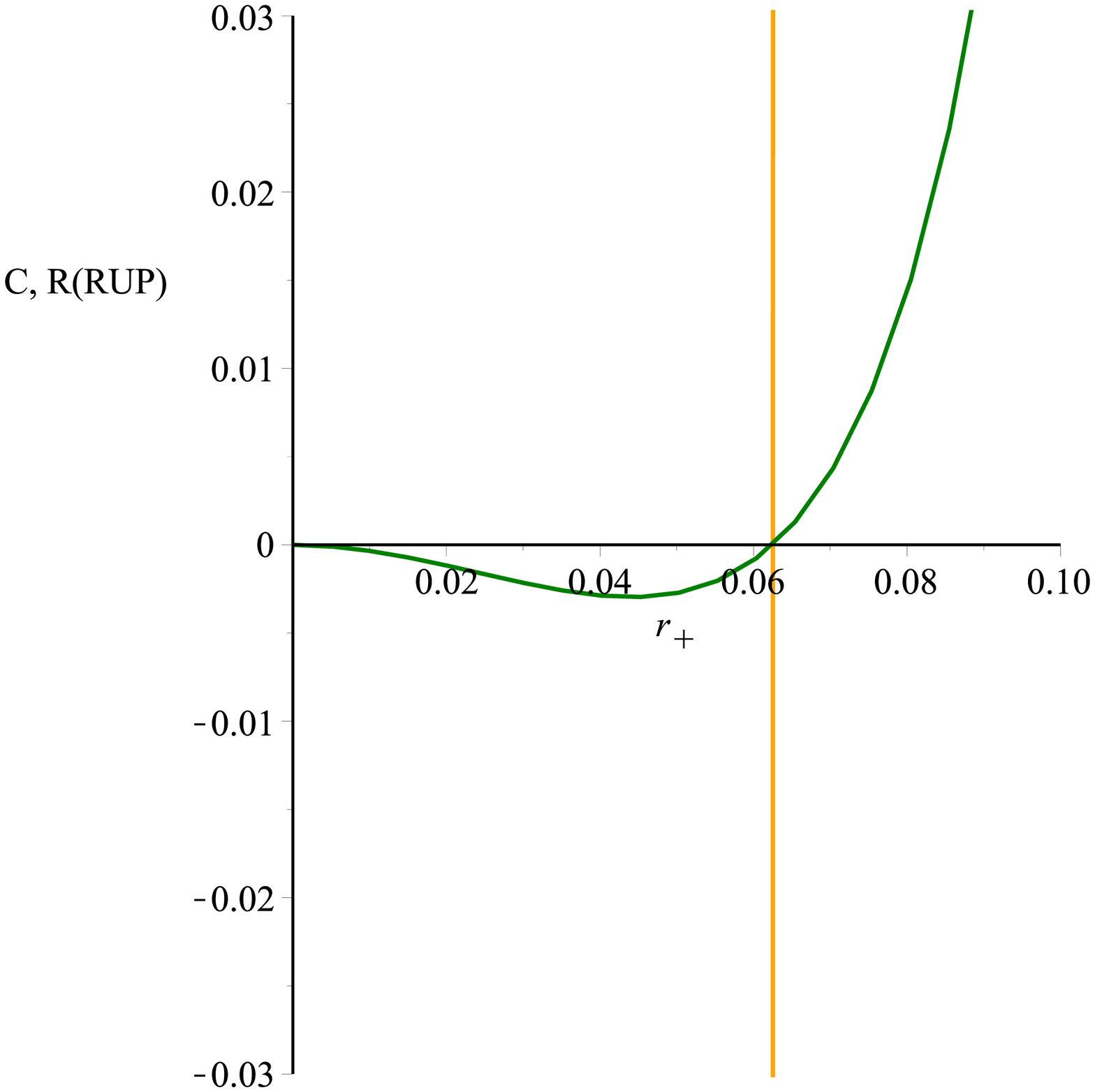}
	}
\subfigure[]{
	\includegraphics[width=0.4\textwidth]{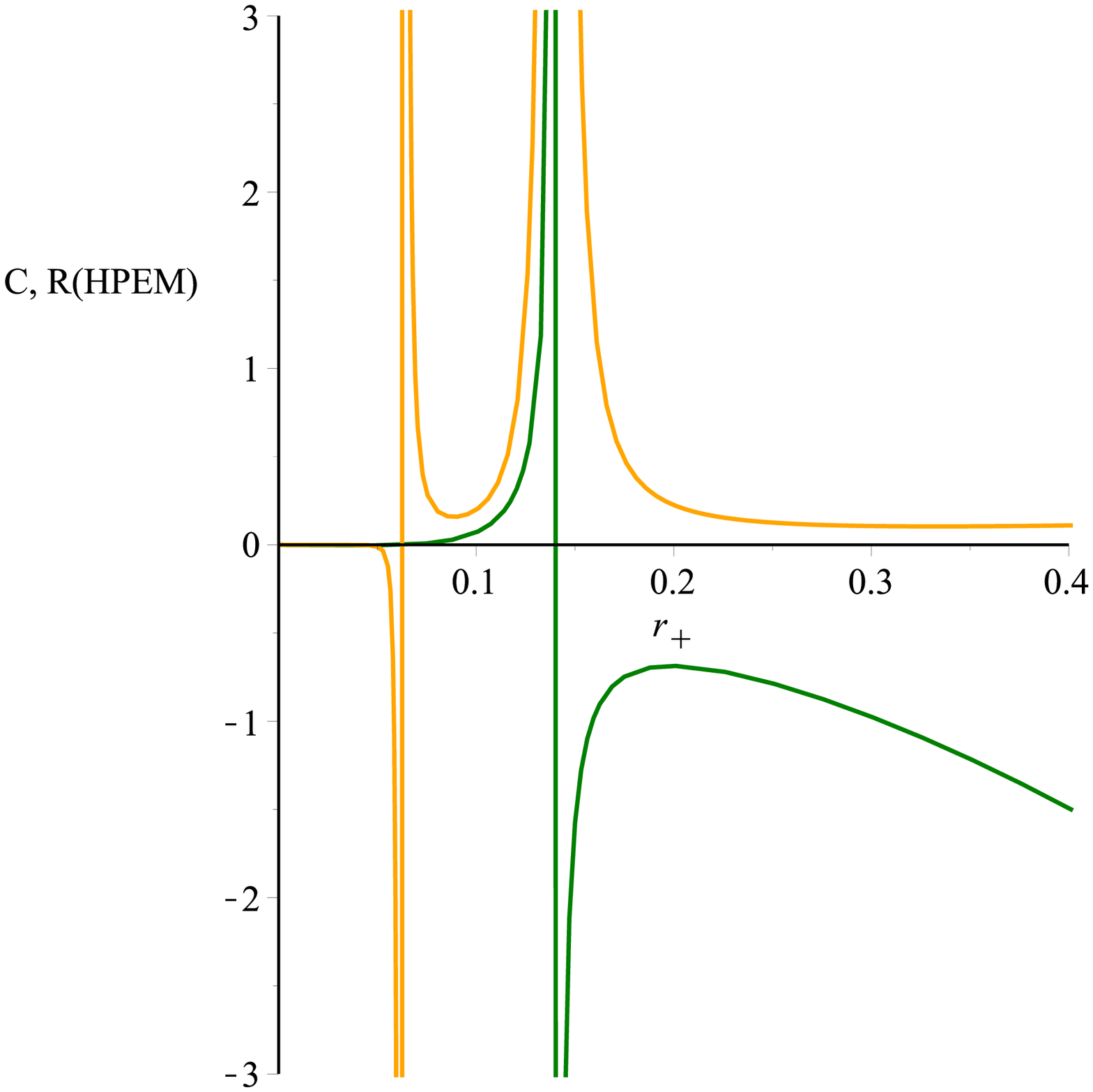}
}
\subfigure[Closeup of figure (c)]{
	\includegraphics[width=0.4\textwidth]{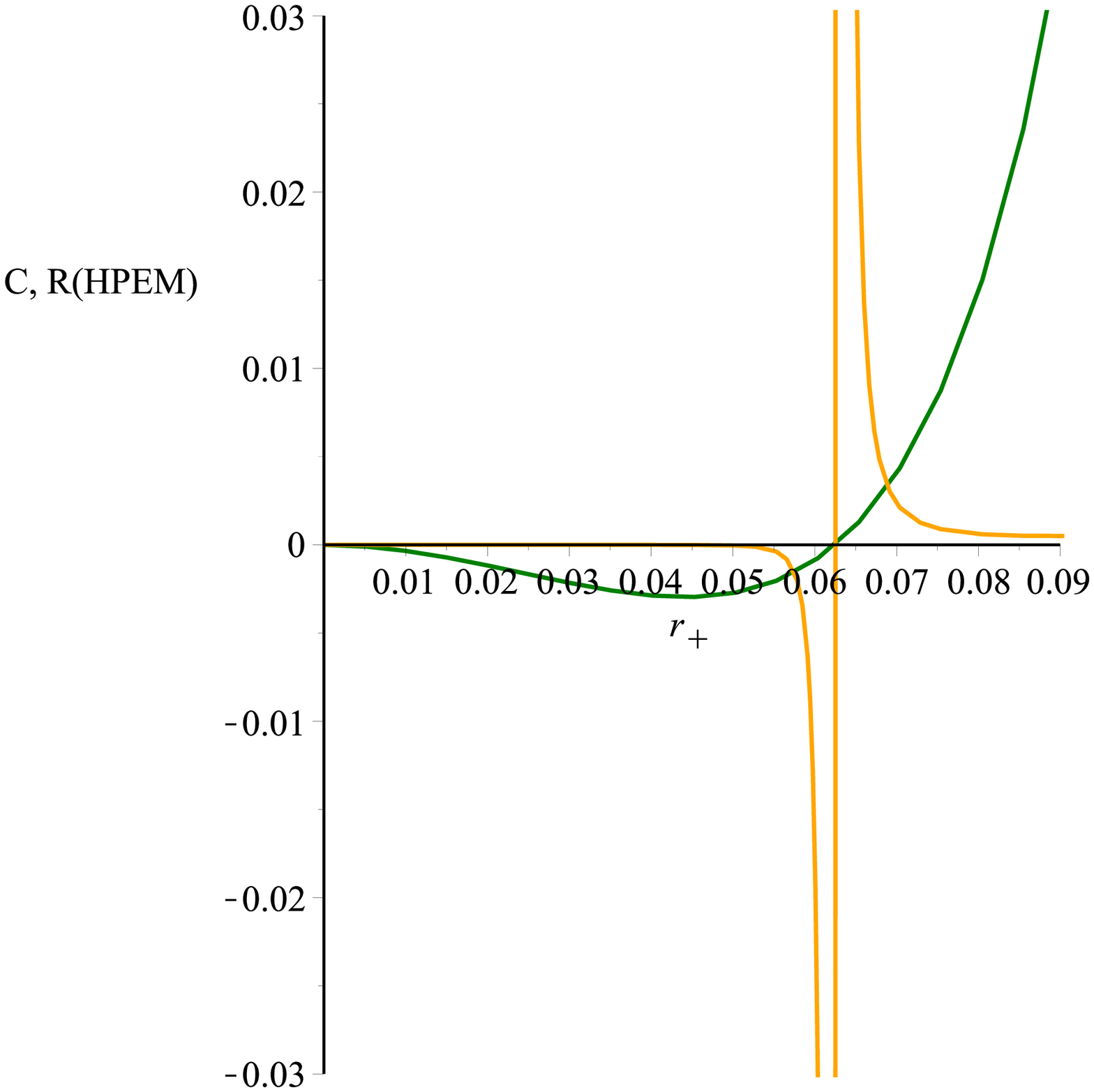}
}
	\caption{{In these figures, (a) and (c), one observes the curvature scalar variation of Ruppeiner and HPEM metrics respectively, distinguished by orange lines, besides heat capacity variation, showed by Green lines, in terms of horizon radii for parameters $ Q=\lambda=0.5 $. Interestingly, the divergence points of the Ricci scalar of the HPEM metric is in good concordance with both the zero point (physical limitation point) and the divergence point (transition critical point) of heat capacity, see figure (c). In other words, the divergence points of the Ricci scalar of the HPEM metric capable with both the first and the second phase transitions.}}
	\label{pic:CRupHPEM}
\end{figure}

Moreover, the behavior of temperature for different values of $Q$ and $\lambda$ is brought in Fig.~\ref{pic:T}.  {From Fig.~\ref{pic:T} (a) when both the parameters $Q$ and $\lambda$ get certain values, we see at first that temperature is in the negative regions at a particular range of $r_+$, then it tends to zero. After that,  temperature { }{enters} the positive regions and { }{increases} to reach a maximum point, then  it decreases by increasing in amount of $r_+$. In Fig.~\ref{pic:T} (b), temperature is illustrated when $\lambda$ gets a fixed amount but $Q$ can vary. When $Q=0$, that is a Schwarzschild-like black hole with CMG corrections, there is no phase transition for such a configuration. By considering nonzero values for $Q$ , the behavior of temperature for larger black holes interestingly  shows a convergence to almost zero temperature which means as long as black holes become larger then they will be colder as well. Then by considering a fixed vale for the parameter $Q$, we examined the impacts of the variation of the hair parameter, $\lambda$ on the evolution of temperature in Fig.~\ref{pic:T} (c). Interestingly, when $\lambda=0,$ the behavior of temperature is as same as Fig.~\ref{pic:T} (b), when metric is Schwarzschild-like, say when $Q=0$ in the presence of the CMG corrections. It obviously means that for some particular cases, hair parameter, $\lambda,$ can act like a charge parameter aiming to investigate the evolution of  temperature of the black hole. The analysis of  Fig.~\ref{pic:T} (c) can approve our aforementioned claims in case (b) of Fig.~\ref{pic:T}. Some other interesting physical properties can be realized from Fig.~\ref{pic:T} (c), they are related to the peak of the Hawking temperature and the converging point of the temperature for larger black holes. By comparing Figs.~\ref{pic:T} (c) and (b) one immediately realizes that by decreasing the values of parameter $\lambda$ maximum points get larger amounts and also for larger black holes temperature will be larger than cases appeared in Fig.~\ref{pic:T} (b) . In addition to the above mentioned properties,  these plots show that as long as the values of $Q$, and $\lambda$ are  increased, the peak value of temperature becomes smaller in value.} {From the quantum gravity phenomenology, it should be noticed that if the black hole evaporates then the quantum effects near the Planck regime should force the evaporation of the black hole to stop thereby leaving behind a black hole remnant. Therefore, the existence of negative regions in the temperature diagram has no physical justification and can be eliminated by introducing the cut-off length so that it does not conflict with quantum gravity, see Fig.~\eqref{cut-off:T-St}. }
%%%%%%%%%%%%%%%%%%%%%%%%%%%%%%
%%%%%%%%%%%%%%%%%%%%%%%%%%%%
Figs.~\ref{pic:C} {(a)-(f), are} devoted to investigate the behavior of heat capacity of {a static black hole in the presence of CMG corrections} for different values of $Q$, and $\lambda$. From these plots  it is {figured out} that the heat capacity contains a root point besides a divergence one, which correspond to physical limitation and phase transition critical points respectively. {As a general case, in Fig.~\ref{pic:C} (a), both the parameters $Q$ and $\lambda$ get nonzero values and it shows up some interesting physical properties. To illustrate these properties clearly a close-up diagram of Fig.~\ref{pic:C} (a) is  shown as Fig.~\ref{pic:C} (b), which shows how and where heat capacity do the first and the second phase transitions. For ${r_1} \simeq 0.01\leq{r_ + }\leq{r_2} \simeq 0.06$ heat capacity gets negative values and it indicates that black hole will be in an unstable phase, and then for ${r_ + }> 0.06$  it enters positive regions and the black hole will be in its stable phase. {And for its second phase transition one can look at Fig.~\ref{pic:C} (a) again,  in which it appears  at ${r_ + }\simeq 0.14$ indicating the phase transition critical point of the black hole.} Now let us turn our attention to the evolution of heat capacity by emphasising on the effects of the  parameters $Q$ and $\lambda$ when they are varied separately. In Figs.~\ref{pic:C} (c), (d), immediately one can conclude that, when you have a Schwarzschild-like black hole, i.e. when $Q=0$, whereas heat capacity only gets negative values then it  completely is in an unstable phase.   By introducing nonzero values of $Q$ to the problem in question it behaves similarly to Figs.~\ref{pic:C} (a), (b). In a similar procedure when $\lambda$ tends to zero, Figs.~\ref{pic:C} (e), (f) follow the discussed behaviour of Figs.~\ref{pic:C} (c), (d). Again one can observe that why the parameter $\lambda$ is considered as a hair of the black hole!.
In addition to these, from whole the Figs. \ref{pic:C} (c)-(f), one can see that by increasing the values of $Q$, and $\lambda$, the phase transition points of the system are shifting along horizon radius $r_+$. In this regard and for a comparison we refer the reader to \cite{Sanchez:2016ger,Fernando:2016qhq,CruzNeto:2020lqb} and references therein.}
%%%%%%%%%%%%%%%%%%%%%%
\\
\textcolor[rgb]{0.00,0.00,1.00}{\subsection{Thermodynamic geometry}\label{sub2}}
{In this section, we are going to study the thermodynamic geometry of a static regular black hole with CMG corrections. In order to do such an investigation, by means of  the so-called geothermodynamic approach, various methods  such as Weinhold, Ruppeiner, Quevedo, and HPEM formalisms, based on their advantages and disadvantages, can be employed.}
The Weinhold geometry and its line element  are respectively expressed by~\cite{Weinhold,Soroushfar:2016nbu}
\begin{equation}\label{Weinhold}
	g^{W}_{i j}=\partial _{i}\partial _{j}M(S, Q)\,,
\end{equation}
and
\begin{eqnarray}
	ds^{2}_{W}&=&M_{SS}dS^{2}+M_{QQ}dQ^{2} + 2M_{SQ}dSdQ \,,
\end{eqnarray}
where $M_{SS}\equiv\partial^2M/\partial^2S,$ $M_{QQ}\equiv\partial^2M/\partial^2Q,$ and $M_{SQ}\equiv\partial^2M/\partial S\partial Q$.
Hence, the relevant {Hessian }matrix reads
\begin{equation}
g^{W}=\left[\begin{array}{ll}
M_{S S} & M_{S Q} \\
M_{Q S} & 0
\end{array}\right]\,.
\end{equation}
Since the curvature scalar of the Weinhold metric is zero, i.e. $ R^{W}=0 $, obviously one cannot get any physical information from this method! { Indeed for this metric, whereas the curvature scalar tends to zero there is no any point which becomes singular and therefore this system does not contain any interaction thermodynamics. To overcome this drawback there are some proposals which we will deal with one by one. }
{To start one can consider  the Ruppeiner formalism for instance, which its metric is  defined as }~\cite{Ruppeiner,Salamon,Mrugala:1984}
\begin{equation}
	ds^{2}_{R}=\frac{1}{T}ds^{2}_{W},
\end{equation}
and the relevant matrix for this black hole, is expressed by
\begin{equation}
	g^{R}=\bigg(\frac{{{4 S^{\frac{3}{2}}}{\pi ^{\frac{1}{2}}}}}{{Q(\lambda  - 1)({\pi ^{\frac{\lambda }{2}}}{S^{1 - \frac{\lambda }{2}}}) + S}}\bigg)
g^{W}
\end{equation}
The curvature scalar of the Ruppeiner metric reads
\begin{equation}
	R^{Rup}=\frac{1}{2}\frac{\lambda }{{S({\pi ^{\frac{\lambda }{2}}}{S^{\frac{{ - \lambda }}{2}}}Q(\lambda  - 1) + 1)}}\,,
\end{equation}
which obviously is nonzero. To illustrate  thermodynamic phase transition and physics of the black hole in geothermodynamic approach,  the resulting curvature scalar of Ruppeiner metric is plotted versus horizon radius $ r_{+} $, see Figs.~\ref{pic:CRupHPEM} (a), (b). It can be seen that the divergence of the curvature scalar of Ruppeiner metric coincides  with zero points of the heat capacity (physical limitation point).
%and has not any point corresponding to divergence point of the heat capacity (phase transition critical points).
{In Fig.~\ref{pic:CRupHPEM} (a), interestingly the upper branch of the curvature scalar in two points crosses the heat capacity diagram. One of them is located in a region in which the black hole is in its stable phase, and the second crossing point is at the phase transition critical points. This plot therefore interestingly indicates interaction thermodynamics.}

\begin{figure}[]
	\centering
\subfigure[$\lambda=+0.5$]{\includegraphics[width=5cm]{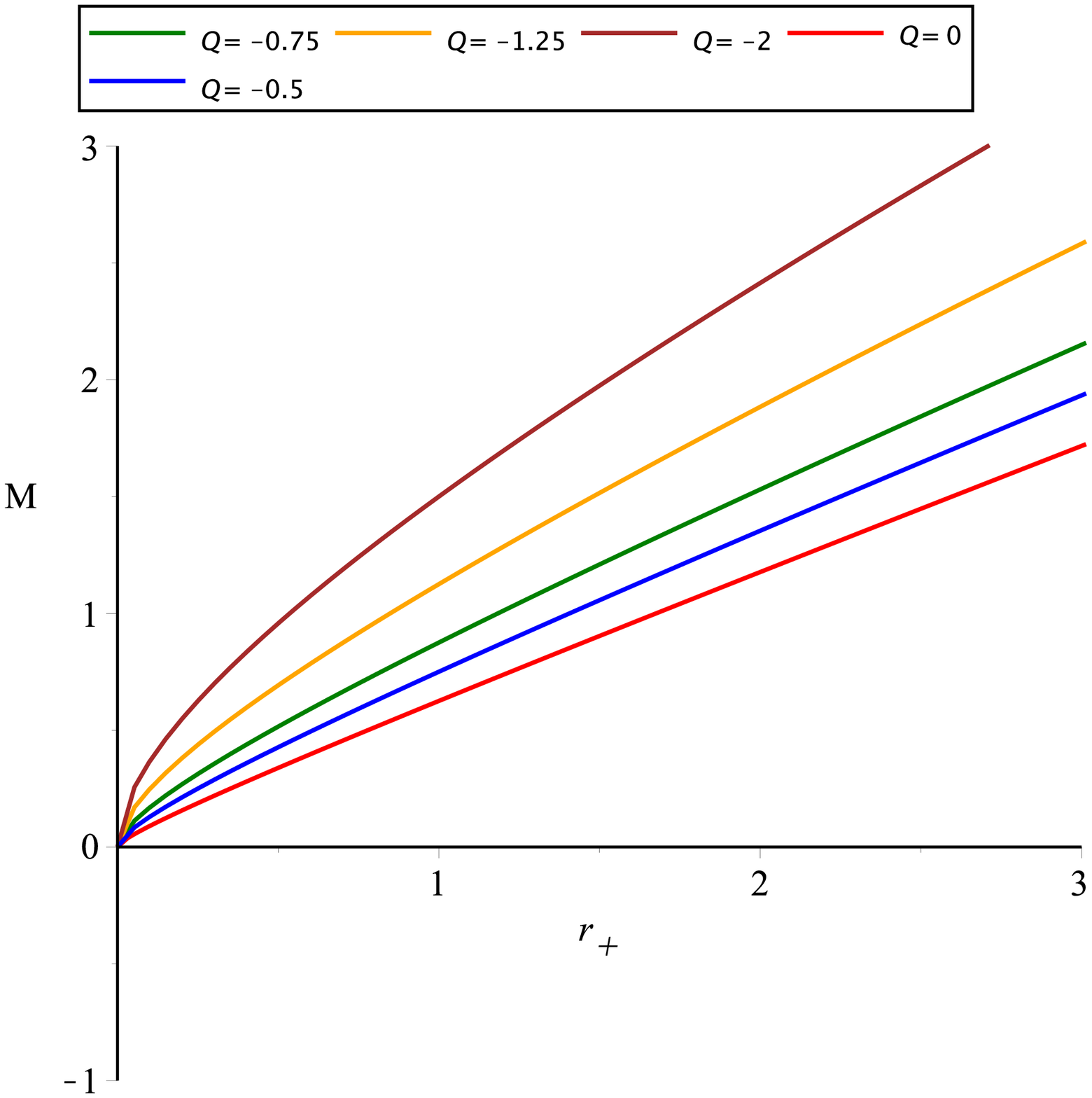}}
\subfigure[$Q=+0.5$]{	\includegraphics[width=5cm]{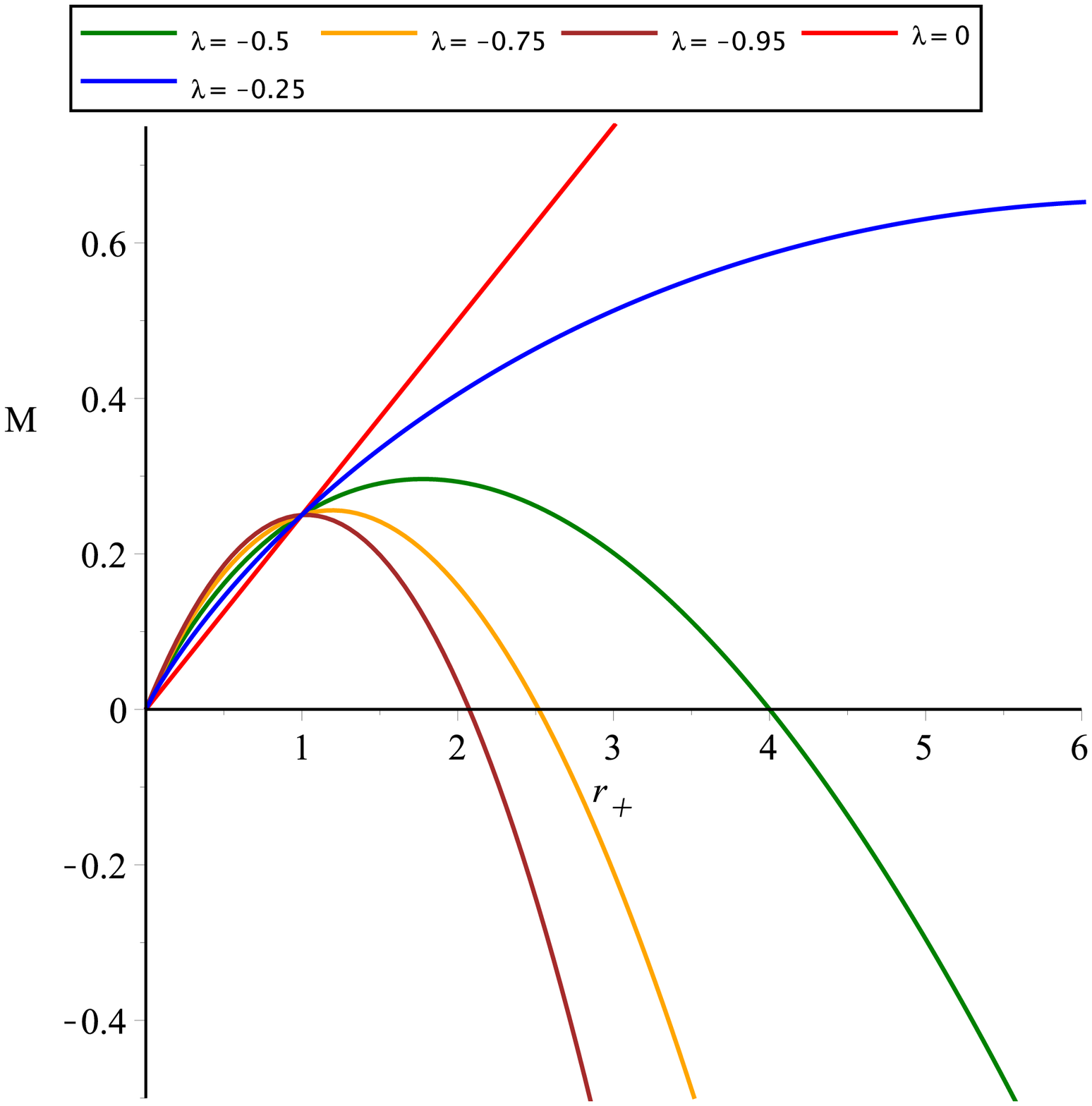}}
\subfigure[$\lambda=-0.5$]{\includegraphics[width=5cm]{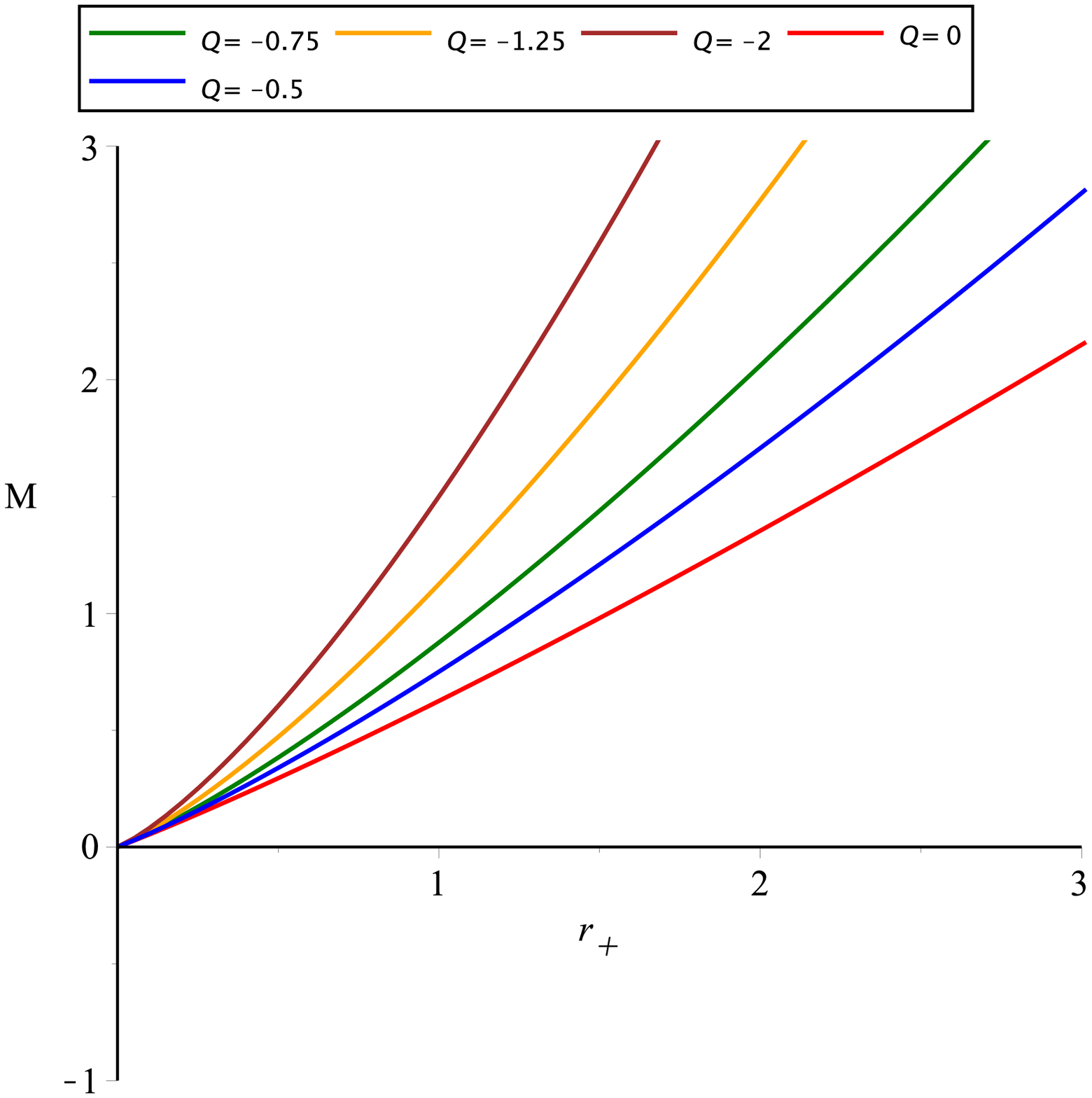}}
\subfigure[$Q=-0.5$]{	\includegraphics[width=5cm]{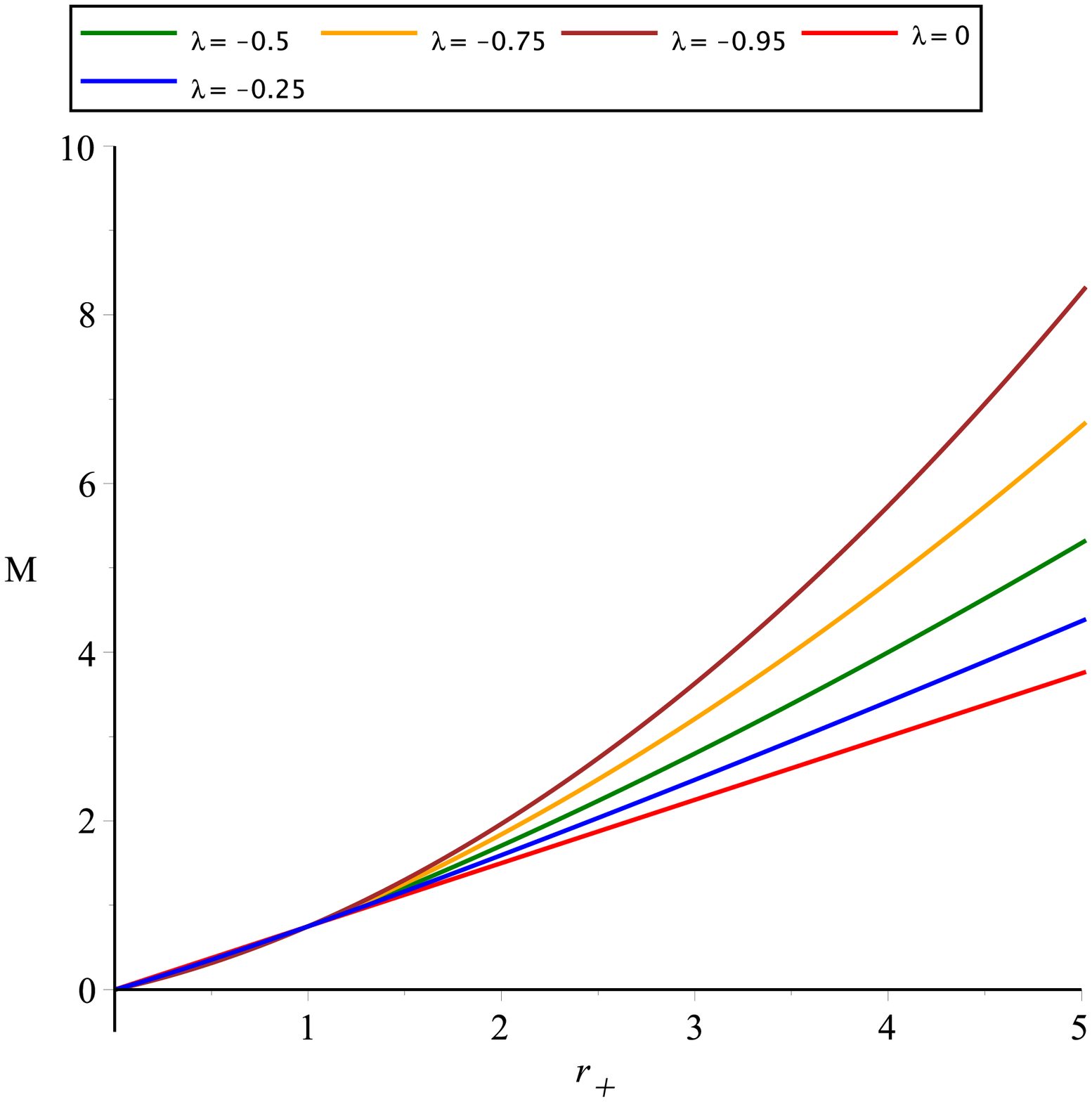}}
	\caption{ {The Variation of mass function  in terms of horizon radius $ r_{+}$, for a static regular black hole in the presence of  CMG corrections is illustrated. In plot (a) the focus is on the changes in amount of $Q<0$ when the parameter $\lambda$ gets a fixed positive amount. In figure (b)  the changes are related to the parameter  $\lambda <0 $ when  $Q$ gets a fixed positive constant. Then in figure (c),  mass function varies for $Q<0$ when $\lambda$ gets a fixed negative value is plotted. And finally figure (d) shows the behaviour of the mass function $\lambda<0$ when $Q$ is a fixed negative constant.  For figures (a), (c) and (d) mass function gets only positive values and only plot (b) in some cases gets negative values for the mass function.}  }
	\label{pic:MNS}
\end{figure}

\begin{figure}[]
	\centering
	\subfigure[$Q=+0.5$]{
		\includegraphics[width=0.38\textwidth]{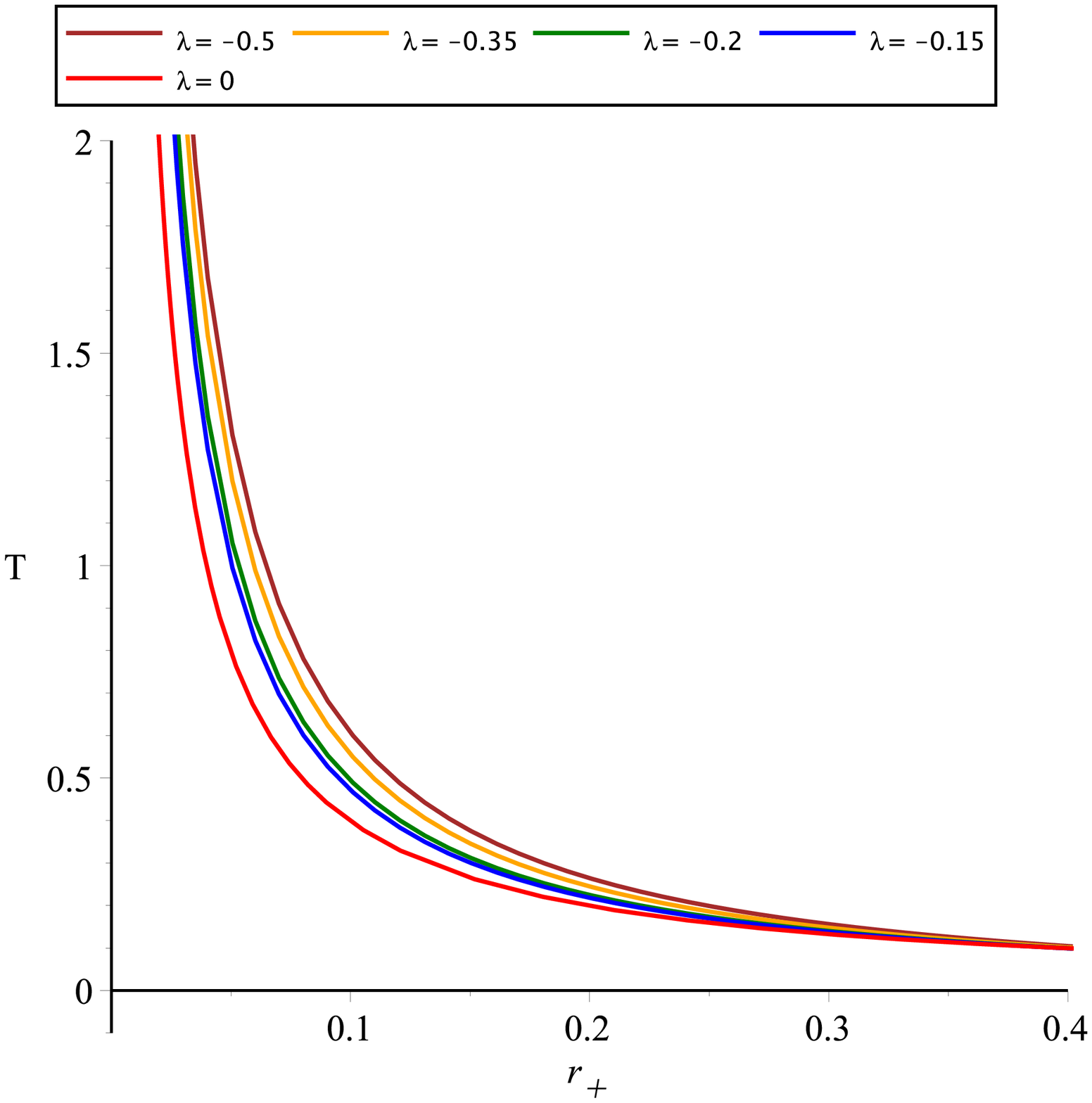}
	}
\subfigure[$\lambda=+0.5$]{
	\includegraphics[width=0.38\textwidth]{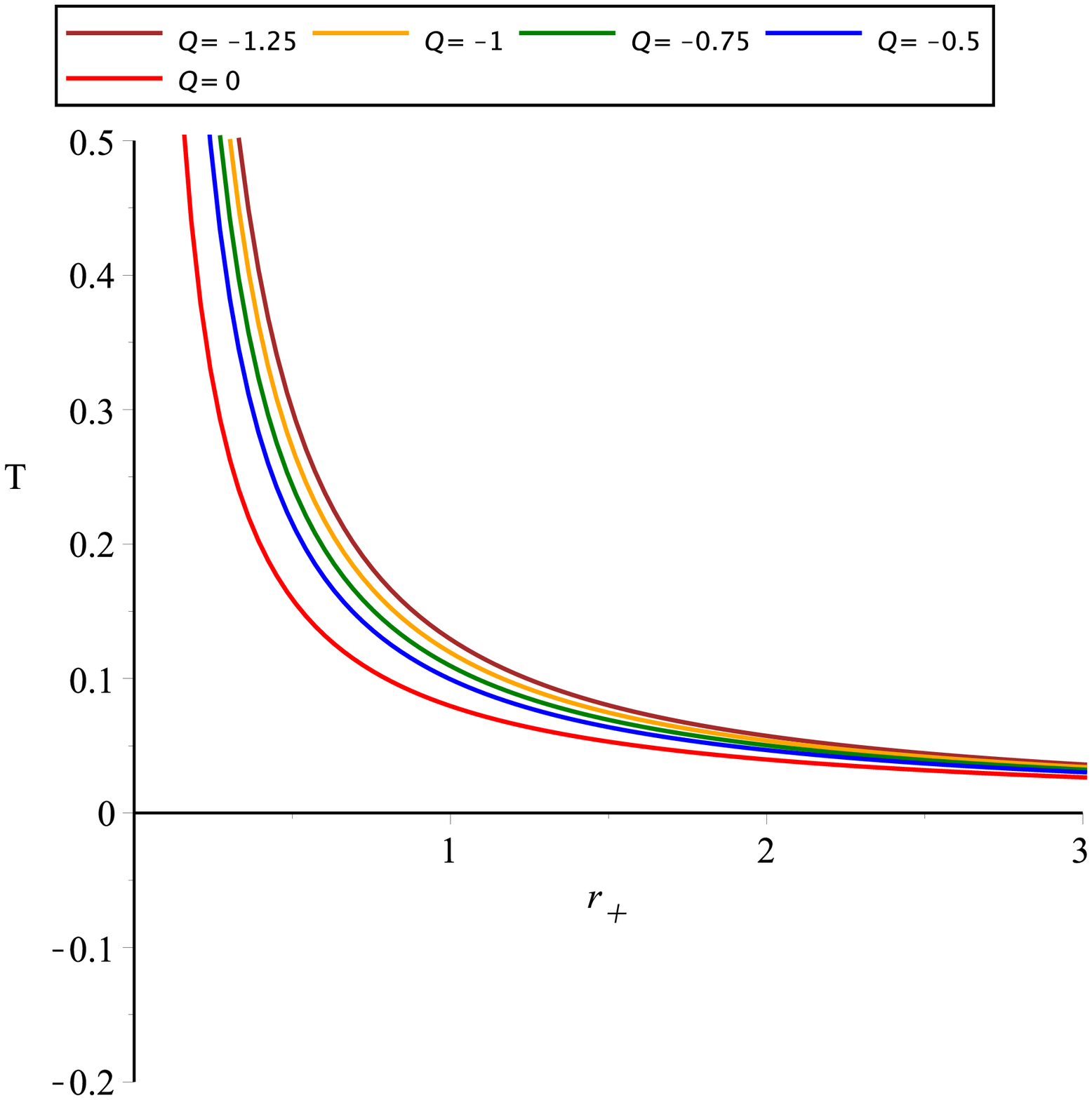}
}
	\subfigure[$Q=-0.5$]{
		\includegraphics[width=0.38\textwidth]{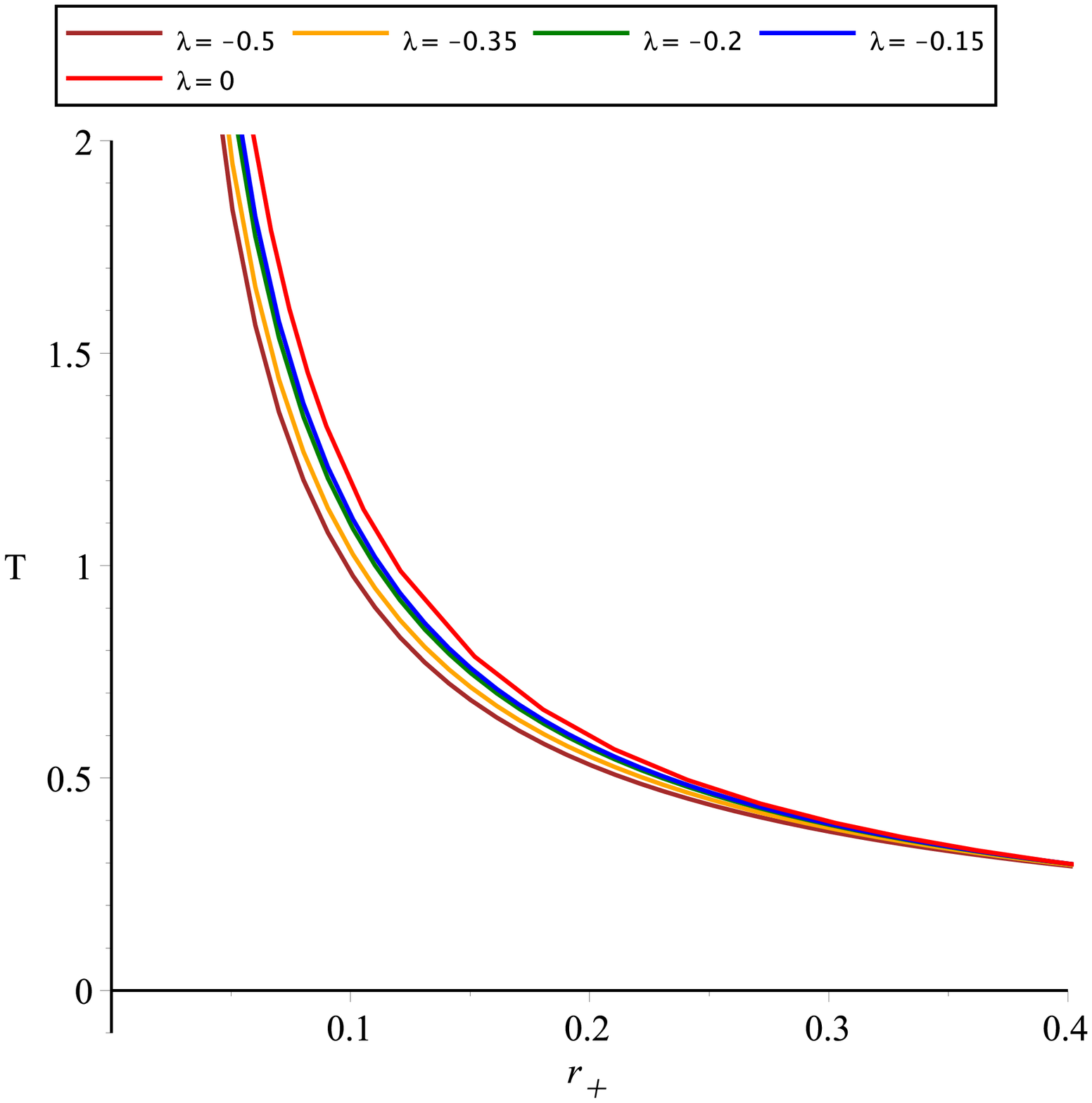}
}
	\subfigure[$\lambda=-0.5$]{
		\includegraphics[width=0.38\textwidth]{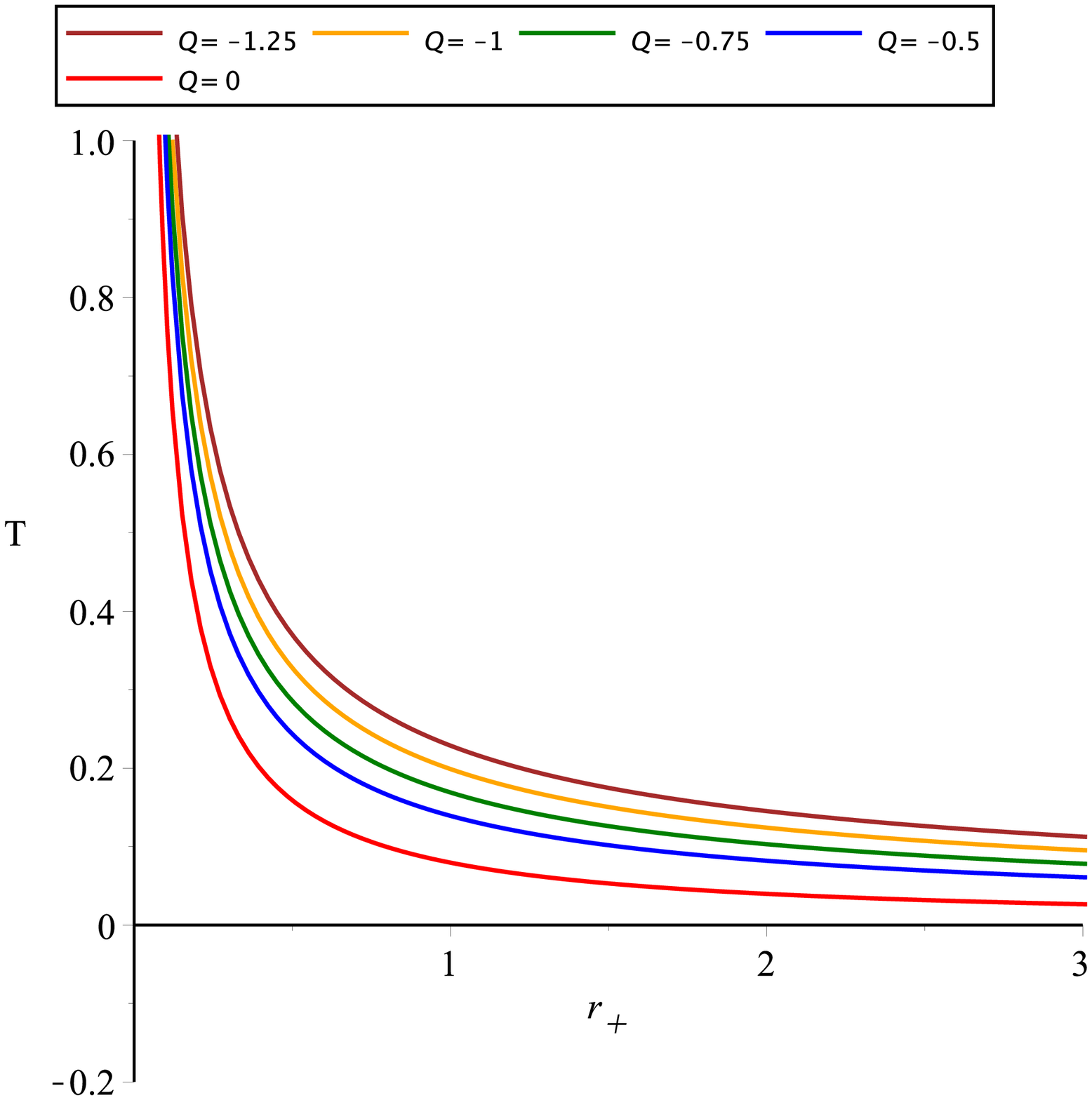}
	}
	\caption{{These figures show the evolution of temperature in terms of horizon radius $ r_{+}$, for a static regular black hole in CMG background. In figure (a) the behaviour of temperature when  $\lambda$ gets nonzero values and $Q=0.5$ is plotted. The plot (b) is illustrated when $Q<0$ varies and $\lambda $ is a positive constant. In plot (c) and (d) both the parameters $Q$ and $\lambda$ are negative. Interestingly for these plots the non-physical region does not appear. }}
	\label{pic:TNS}
\end{figure}

\begin{figure}[]
	\centering
\subfigure[$\lambda=+0.5$ ]{
	\includegraphics[width=5cm]{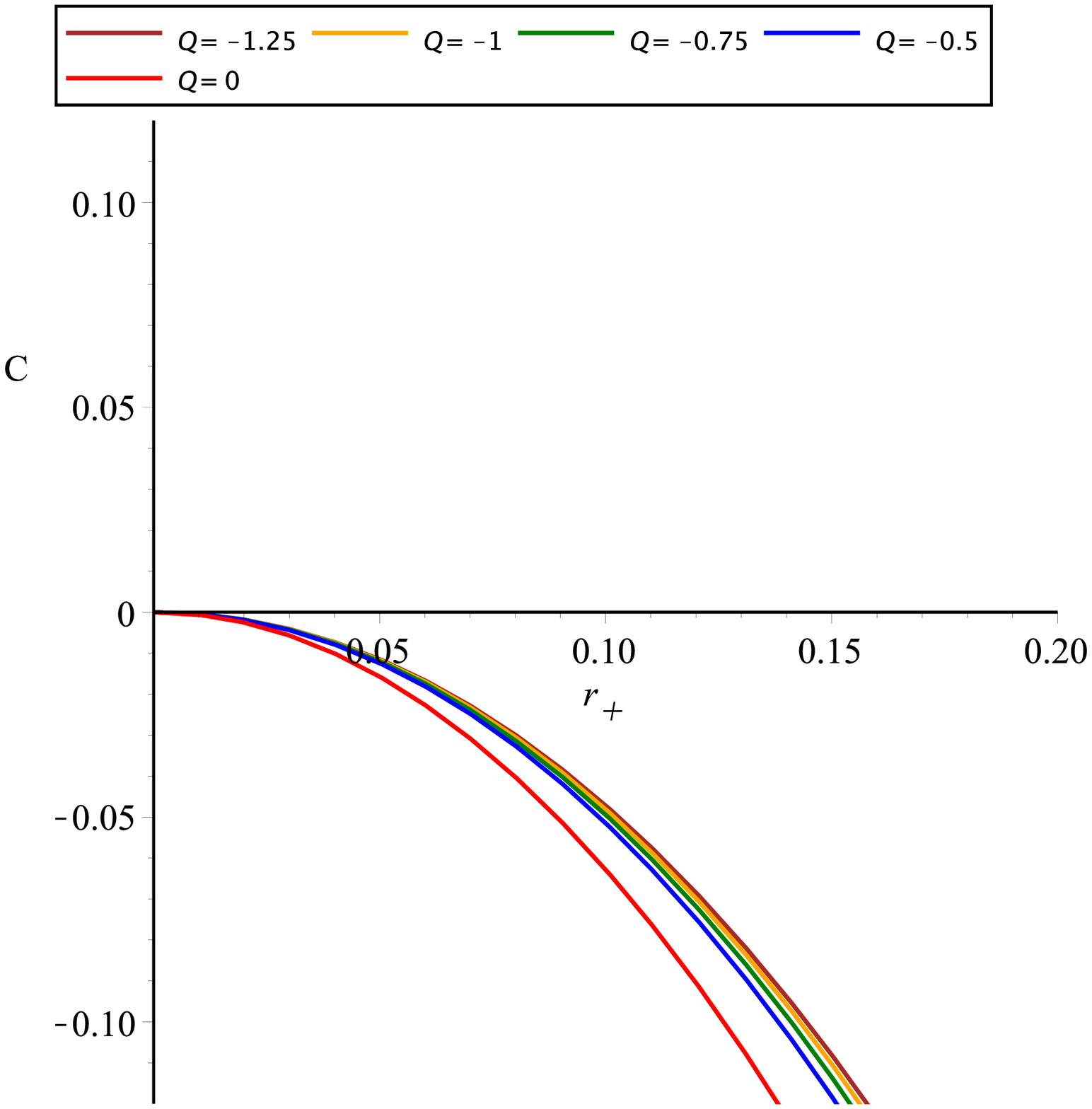}
}
	\subfigure[$Q=+0.5$]{
		\includegraphics[width=5cm]{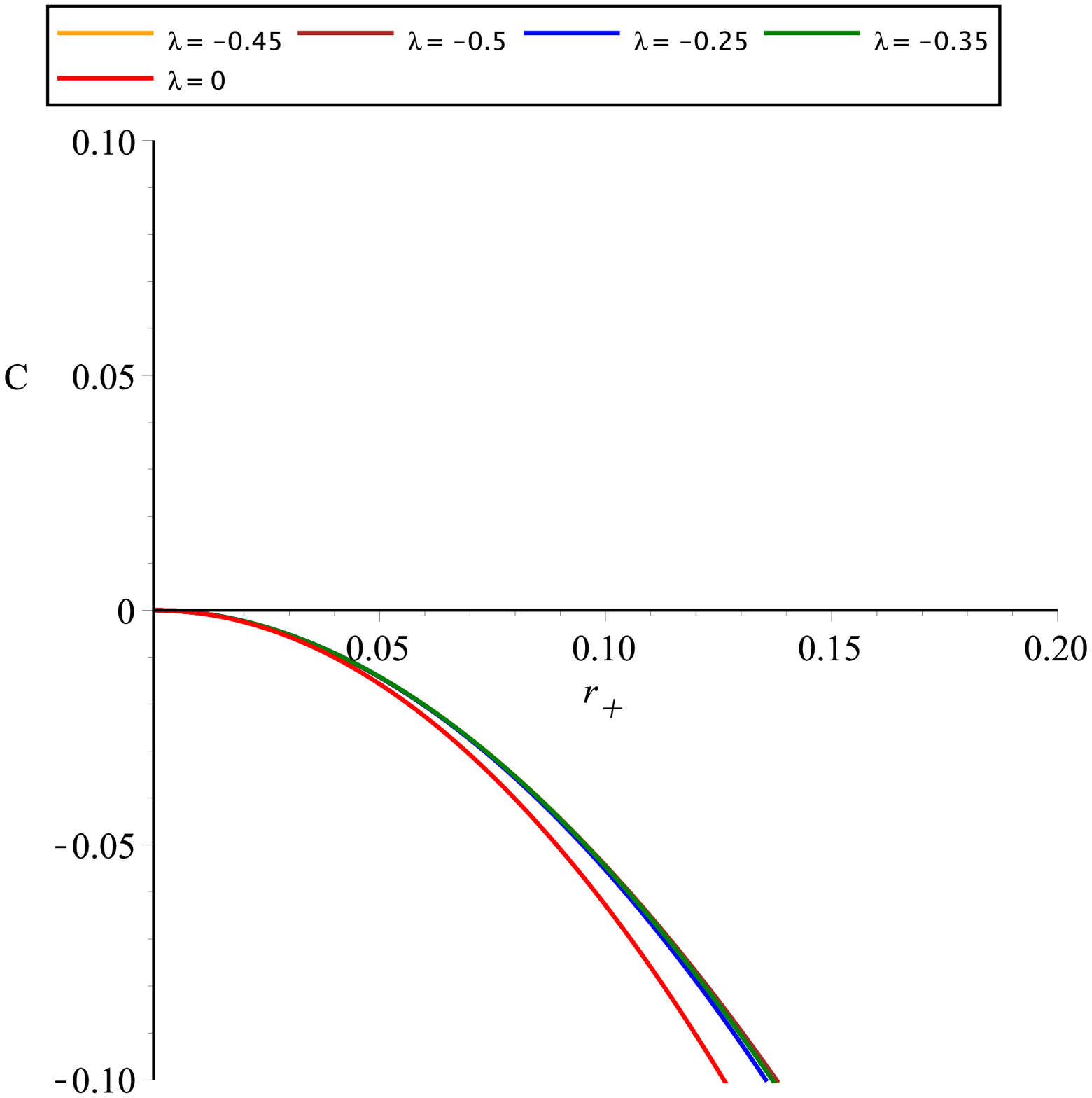}
	}
\subfigure[ $\lambda=-0.5$]{
	\includegraphics[width=5cm]{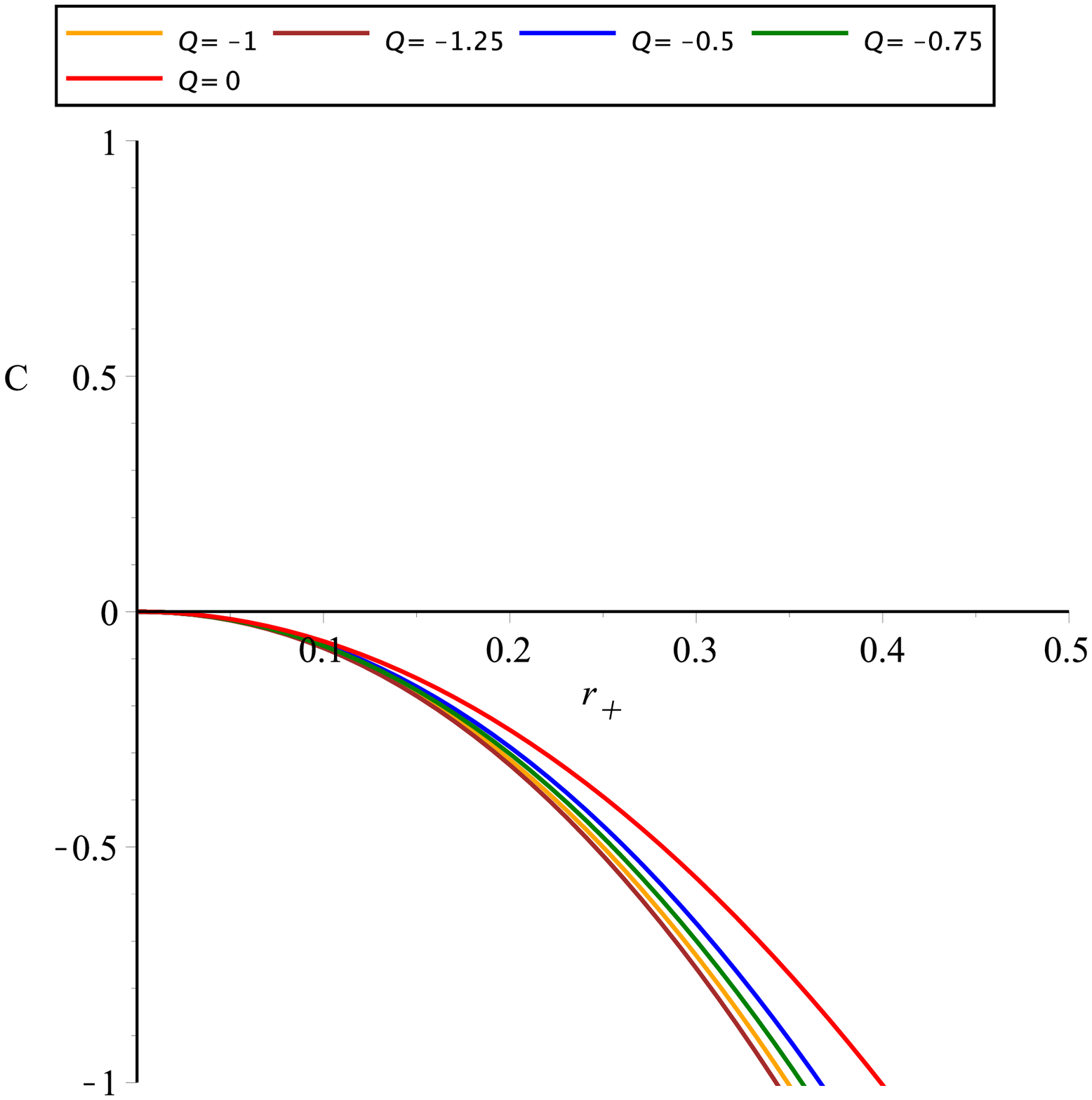}
}
\subfigure[$Q=-0.5$]{
	\includegraphics[width=5cm]{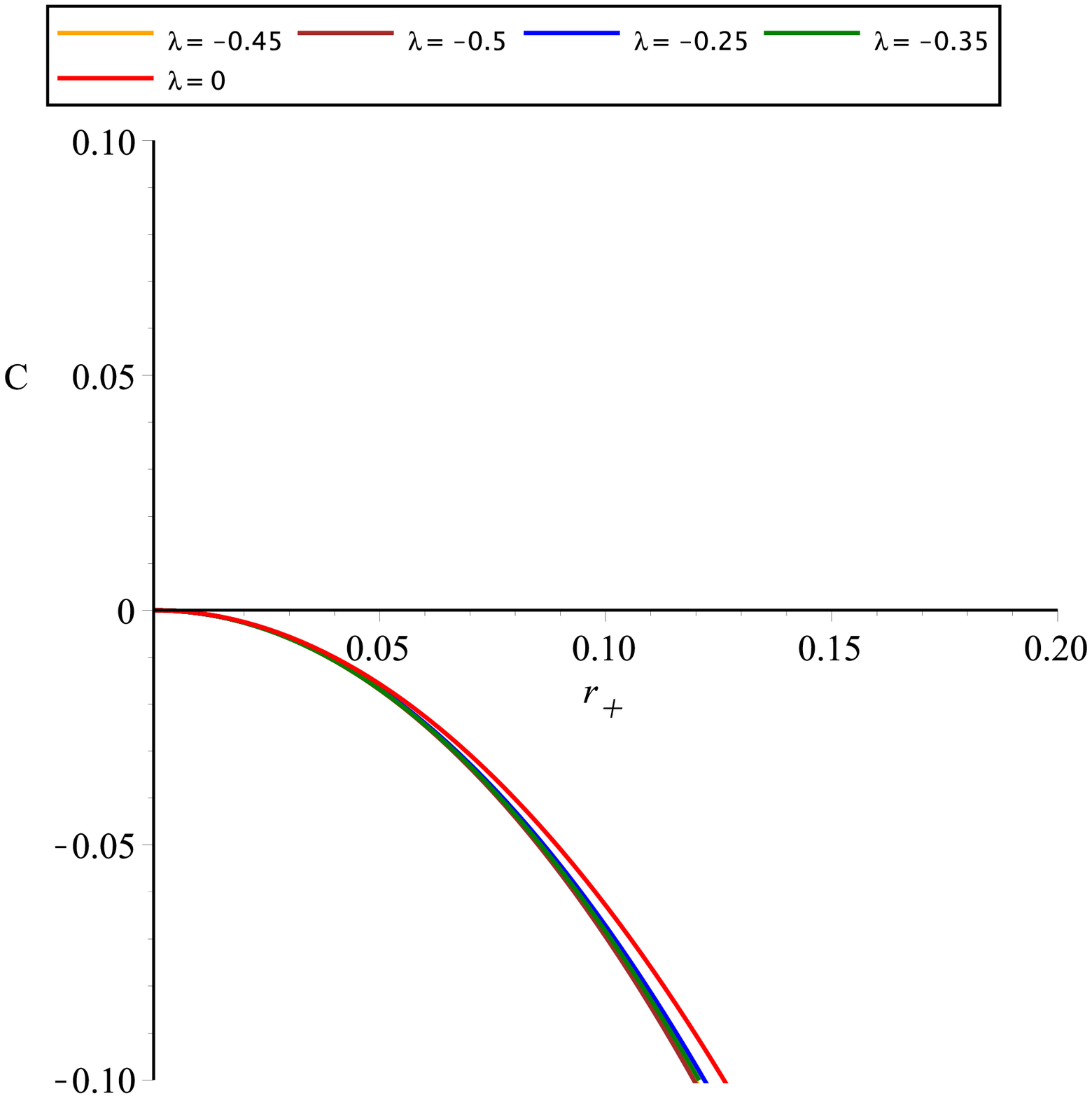}
}
\\
	\caption{{These figures indicate the variations of the heat capacity in terms of horizon radius $ r_{+}$, for a static regular black hole in the presence of the CMG corrections. In plot (a) the behaviour of heat capacity when  $Q$ gets negative values and $\lambda>0$ is demonstrated. The plot (b) is illustrated when $\lambda<0$ varies and $Q$  is a positive constant. In plot (c) and (d) both the parameters $Q$ and $\lambda$ are negative in which for (c) parameter $Q$ varies and $\lambda$ is a negative constant and vice versa for the diagram (d). }}
	\label{pic:CNS}
\end{figure}

{Now let us turn our attention to other geothermodynamic attempts whit much more fruitful physical results. } In this step, we are going to employ  the Quevedo and HPEM metrics to study the thermodynamic geometry of a static regular black hole in the presence of the CMG corrections. The Quevedo metric is expressed by
~\cite{Quevedo:2006xk}:

\begin{equation}\label{GTD}
	g=\left(E^{c}\frac{\partial\Phi}{\partial E^{c}}\right)\left(\eta_{ab}\delta^{bc}\frac{\partial^{2}\Phi}{\partial E^{c}\partial E^{d}}dE^{a}dE^{d}\right),
\end{equation}
in which
\begin{equation}
	\frac{\partial\Phi}{\partial E^{c}}=\delta_{cb}I^{b},
\end{equation}
where $ E^{a} $ and $ I^{b} $ are the extensive and intensive thermodynamic variables, and $ \Phi $ is the thermodynamic potential.
In addition to the above case,  the HPEM metric with $ n $ extensive variables ($ n\ge 2 $) is given by \cite{Hendi:2015rja,EslamPanah:2018ums,Hendi:2015xya,Soroushfar:2019ihn}
\begin{equation}
	ds_{HPEM}^{2}= \dfrac{SM_{S}}{\left(\prod_{i=2}^n \frac{\partial^{2} M}{\partial \chi_{i}^{2}}\right)^{3}}\left(-M_{SS}dS^{2}+\sum_{i=2}^n \left(\frac{\partial^{2} M}{\partial \chi_{i}^{2}}\right)d\chi_{i}^{2}\right) ,
\end{equation}
 where $ \chi_{i} (\chi_{i}\neq S) $, $ M_{S}={\partial M}/{\partial S} $ and $ M_{SS} ={\partial^{2} M}/{\partial S^{2}} $  are {extensive} parameters.
{Therefore}, these two metrics can be rewritten as \cite{Hendi:2015rja,EslamPanah:2018ums,Hendi:2015xya,Soroushfar:2019ihn}
\begin{equation}
d s^{2}=\left\{\begin{array}{ll}
\left(S M_{S}+Q M_{Q}\right)\left(-M_{S S} d S^{2}+M_{Q Q} d Q^{2}\right) &  \text{ Quevedo~ Case~ I } \\
\\
S M_{S}\left(-M_{S S} d S^{2}+M_{Q Q} d Q^{2}\right) &  \text{ Quevedo~ Case~ II } \\
\\
\frac{S M_{S}}{\left(M_{Q Q}\right)^{3}}\left(-M_{S S} d S^{2}+M_{Q Q} d Q^{2}\right) & \text{ HPEM }
\end{array}\right.
\end{equation}
and the denominator for their Ricci {scalar }could be appeared in the following form \cite{EslamPanah:2018ums,Hendi:2015xya,Soroushfar:2019ihn}
\begin{equation}
{denom}(R)=\left\{\begin{array}{ll}
2 M_{S S}^{2} M_{Q Q}^{2}\left(S M_{S}+Q M_{Q}\right)^{3} &  \text{ Quevedo ~Case~ I } \\
\\
2 S^{3} M_{S S}^{2} M_{Q Q}^{2} M_{S}^{3} & \text{ Quevedo ~Case ~II } \\
\\
2 S^{3} M_{S S}^{2} M_{S}^{3} & \text{ HPEM }
\end{array}\right.
\end{equation}
{Investigating the above equations shows that the curvature scalar of the Quevedo formalism does not give any physical data about the system. But, as can be seen from Figs.~\ref{pic:CRupHPEM} (c), (d), intriguingly the divergence points of the Ricci scalar of the HPEM metric, is in concordance with both the zero point (physical limitation point) and the divergence point (transition critical point) of heat capacity. In other words, the divergence points of the Ricci scalar of the HPEM metric capable with both types of phase transitions of the heat capacity, say the first and the second phase transitions. Therefore, it is clear that we can obtain more physical information considering HPEM formalism instead of Ruppeiner or other above mentioned metrics.}{In the next section, by means of both the usual and thermodynamic geometry techniques we will go through  investigating the rotating regular black holes in the presence of CMG corrections.}
%%%%%%%%%%%%%%%%%%%%%%%%%%%%%%%%
%%%%%%%%%%%%%%%%%%%%%%%%%%%%%%%%
%%%%%%%%%%%%%%%%%%%%%%%%%%%%%%%
\textcolor[rgb]{0.00,0.00,1.00}{\subsection{Investigating the static black hole for $Q<0$ and $\lambda<0$}\label{substatic2}}
{In this subsection, we want to investigate the effects of $Q<0$ and  $\lambda<0$ on the evolution of the parameters  mass, temperature and heat capacity.
For this purpose the behaviour of the mass function is depicted in Fig.~ \eqref{pic:MNS} and  it can be realized that for plots (a), (c) and (d) mass function gets only positive values and only plot (b) in some cases obtains negative values. In addition to these, the behaviour of temperature and heat capacity are illustrated in Fig.~\eqref{pic:TNS} and Fig.~\eqref{pic:CNS}. In all plots (a) to (d) of  Fig.~\eqref{pic:TNS}, temperature stays in physical region. For a static regular black hole in the presence of CMG correction when one of the parameters $Q$ and $\lambda$ or both of them get negative values heat capacity does not show any phase transition.}

{We will do such studies for rotating  black holes in the following sections to better understand the effects of sign change of the scalar charge and the  hair parameter $\lambda$ on the evolution of the system.}

\clearpage

\textcolor[rgb]{0.00,0.00,1.00}{\section{Rotating regular black hole in conformal massive gravity}\label{section3}}
{Following the previous section devoted to investigating the static black holes in the presence of CMG corrections, in this section we are going to consider a more general case that is a  rotating charged regular black hole, Kerr-Newman metric, with CMG corrections. Such a metric can be expressed as follows} \cite{Jusufi:2019caq}

\begin{eqnarray}
ds^{2}&&=-\bigg(1+\frac{L^{2}}{\rho^{2}}\bigg)^{\frac{\left| \lambda  \right|}{2}+2}\bigg(1-\dfrac{2 M r+Q r^{2-\lambda}}{\rho^2}\bigg)dt^2 \\\nonumber
&&+\bigg(1+\frac{L^{2}}{\rho^{2}}\bigg)^{\frac{\left| \lambda  \right|}{2}+2}\dfrac{\rho^2}{\Delta}dr^2-2 \bigg(1+\frac{L^{2}}{\rho^{2}}\bigg)^{\frac{\left| \lambda  \right|}{2}+2} a \sin^2 {\theta}\bigg(\dfrac{2 M r+Q r^{2-\lambda}}{\rho^2}\bigg)dt d\phi \\\nonumber
&&+\bigg(1+\frac{L^{2}}{\rho^{2}}\bigg)^{\frac{\left| \lambda  \right|}{2}+2}\sin^2 {\theta}\bigg[r^2+a^2+\sin^2 {\theta}\bigg(\dfrac{2 M r+Q r^{2-\lambda}}{\rho^2}\bigg)\bigg]d\phi^2 \,,
\end{eqnarray}
where
\begin{eqnarray}
&&\rho^2=r^2+a^2 \cos^2 \theta\,,\\
&&\Delta=r^2+a^2-2 M r-Q r^{2-\lambda}\,,
\end{eqnarray}
are introduced for brevity. In this metric by $a$ we mean $J/M$ where $J$ refers angular momentum and, as indicated before, parameters $L$ and  $\lambda$ are appeared due to CMG corrections.

%%%%%%%%%%%%%%%%%%%%%%%%%%%%%%%%%%%%%%%%%%%%%%%%%%%%%%%%

\begin{figure}[]
	\centering
	\subfigure[$\lambda=0.25$, $a=0.15 $]{
		\includegraphics[width=0.38\textwidth]{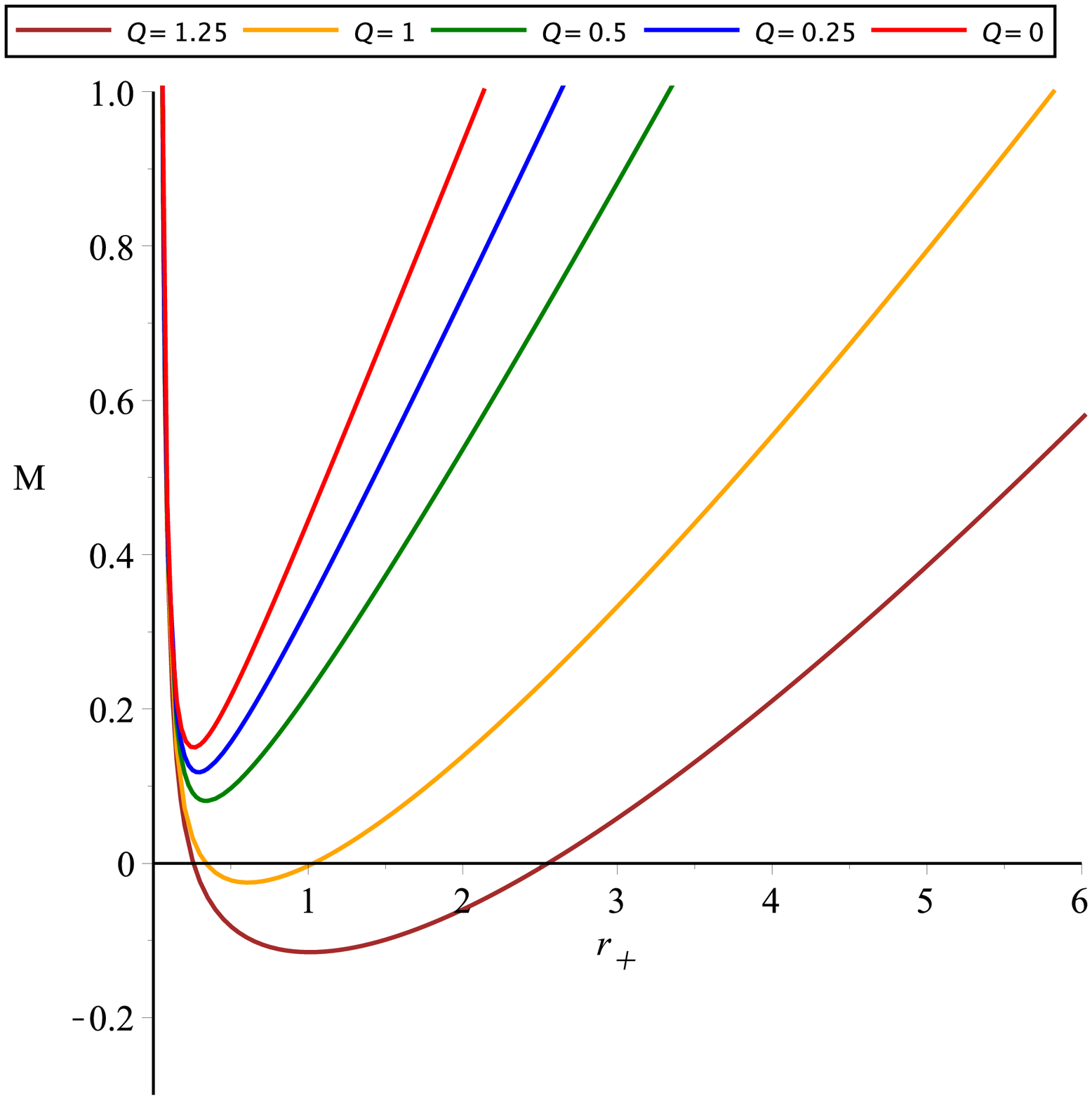}
	}
	\subfigure[$Q=0.25$, $a=0.15 $]{
		\includegraphics[width=0.38\textwidth]{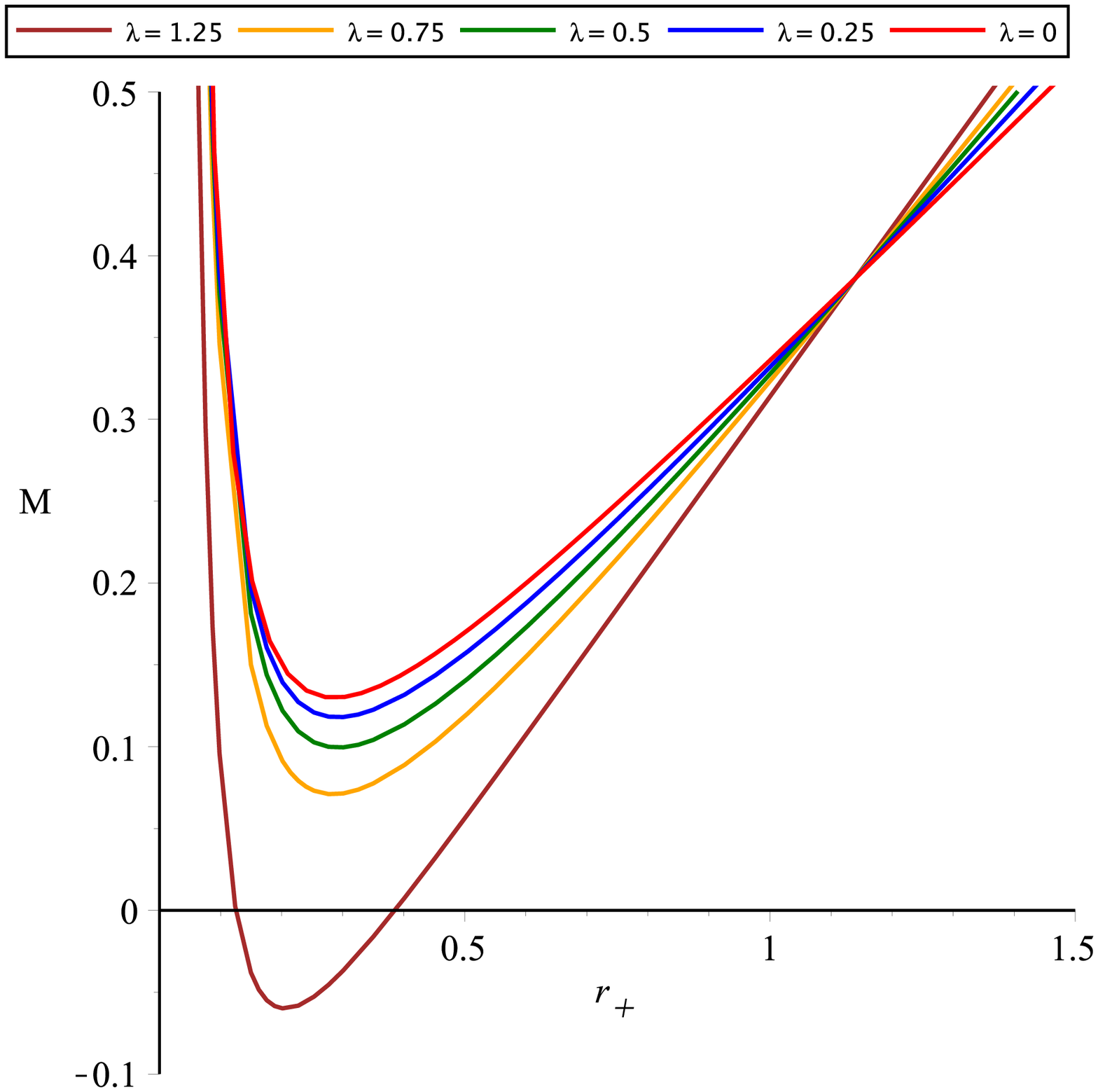}
	}
\subfigure[$Q=0.25$, $\lambda=0.25 $]{
	\includegraphics[width=0.38\textwidth]{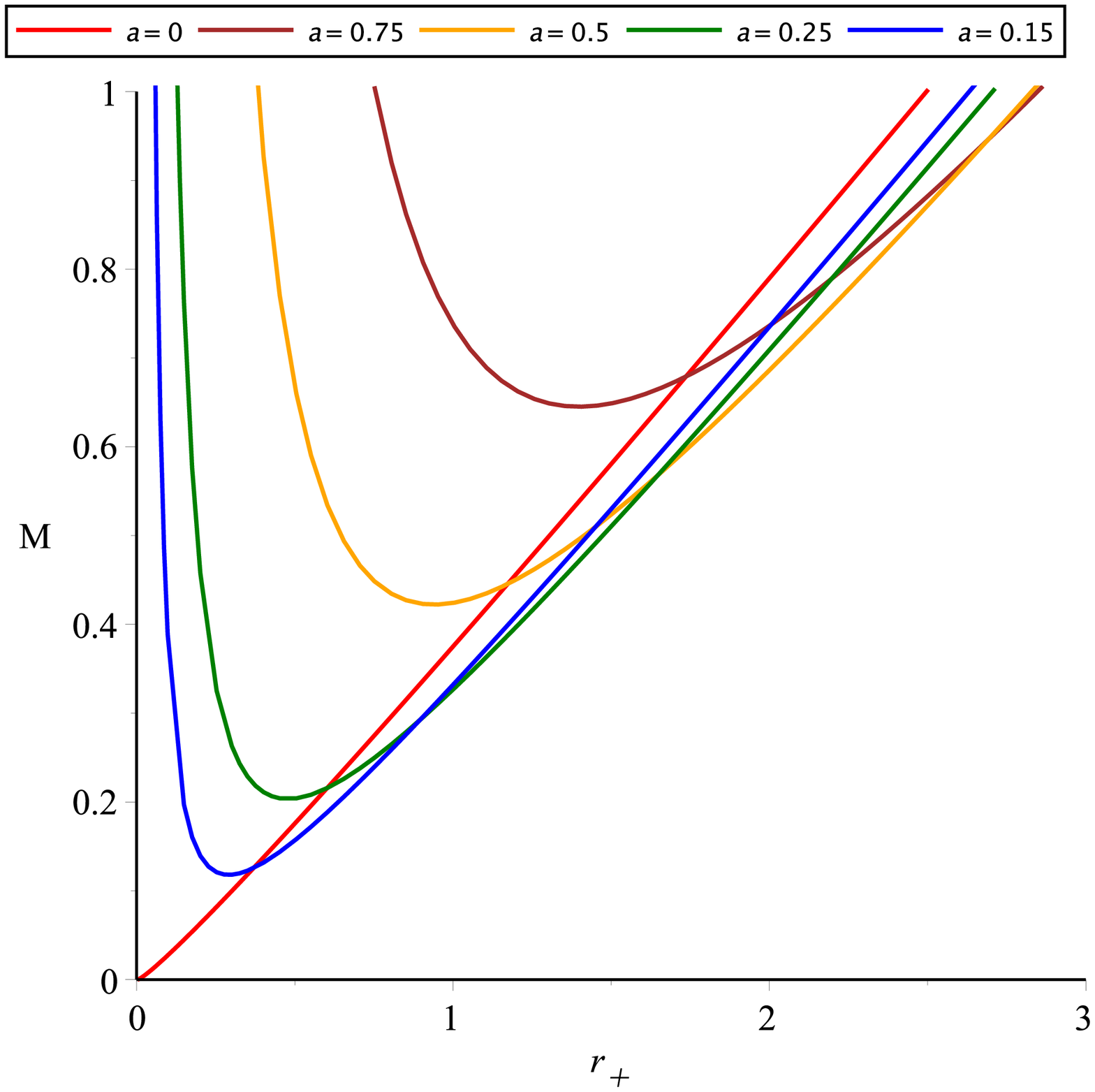}
}
	\caption{{ }{Variations of mass function in terms of horizon radius $ r_{+}$, for a rotating regular black hole with CMG corrections. In figure (a), different values of the parameter $0\leqslant Q\leqslant 1.25$ are employed, in which the solution with $Q=0$ is Kerr solution with CMG correction. In a similar procedure, plot (b) is shown` to investigate the impact of the variation of the  hair parameter $0 \leqslant \lambda \leqslant 1.25$. Then to see the effects of the rotation parameter $0\leqslant a\leqslant 0.15$, diagram (c) is plotted in which the critical solution $a=0$ goes back to non-rotating R-N solution with CMG correction. } }
	\label{pic:MR}
\end{figure}

\begin{figure}[]
	\centering
	%\subfigure[]{
		%\includegraphics[width=0.38\textwidth]{Rt/T.eps}
%	}
	\subfigure[$\lambda=0.25$, $a=0.15 $]{
		\includegraphics[width=0.38\textwidth]{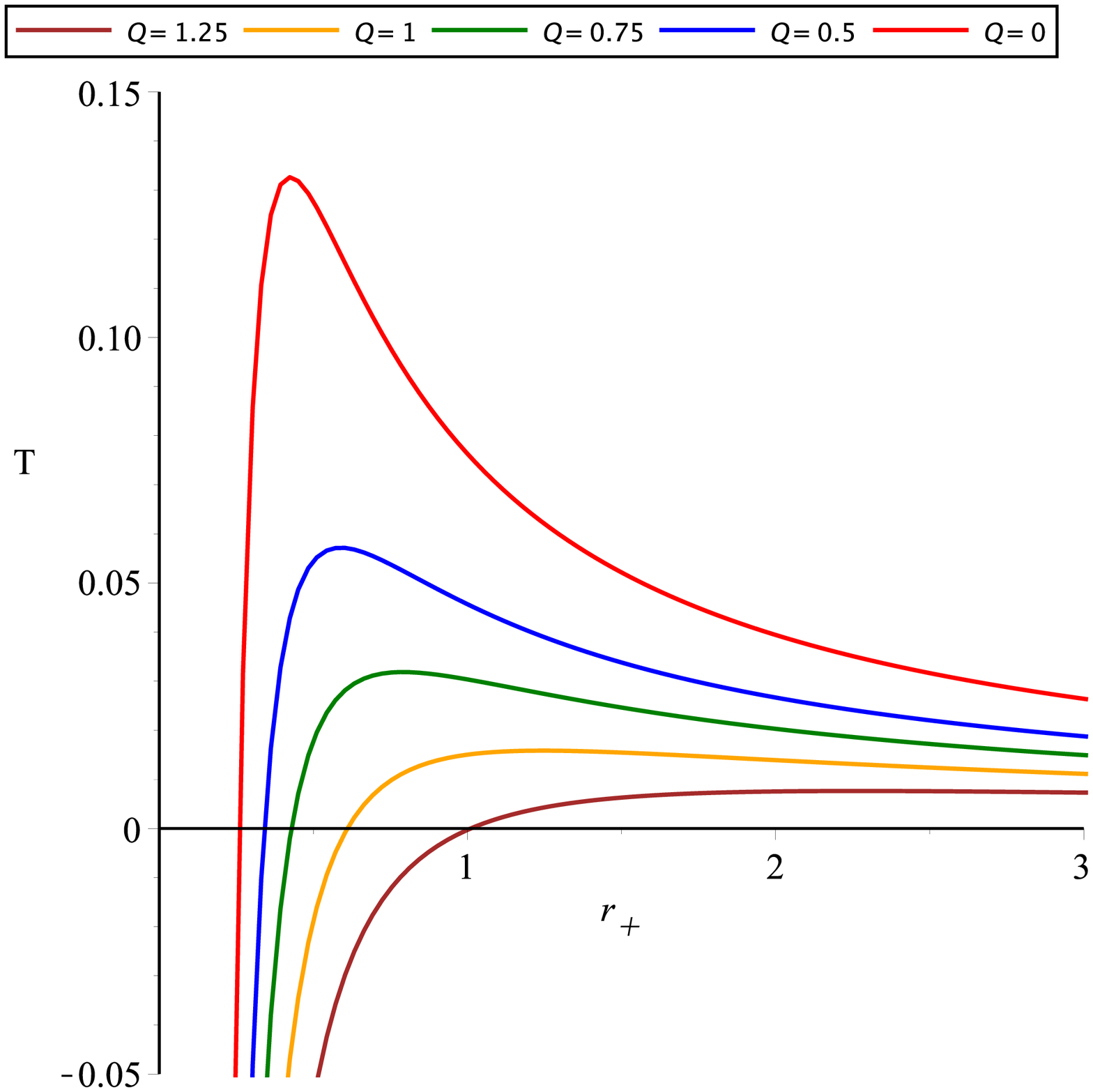}
	}
	\subfigure[$Q=0.25 $, $a=0.15 $]{
		\includegraphics[width=0.38\textwidth]{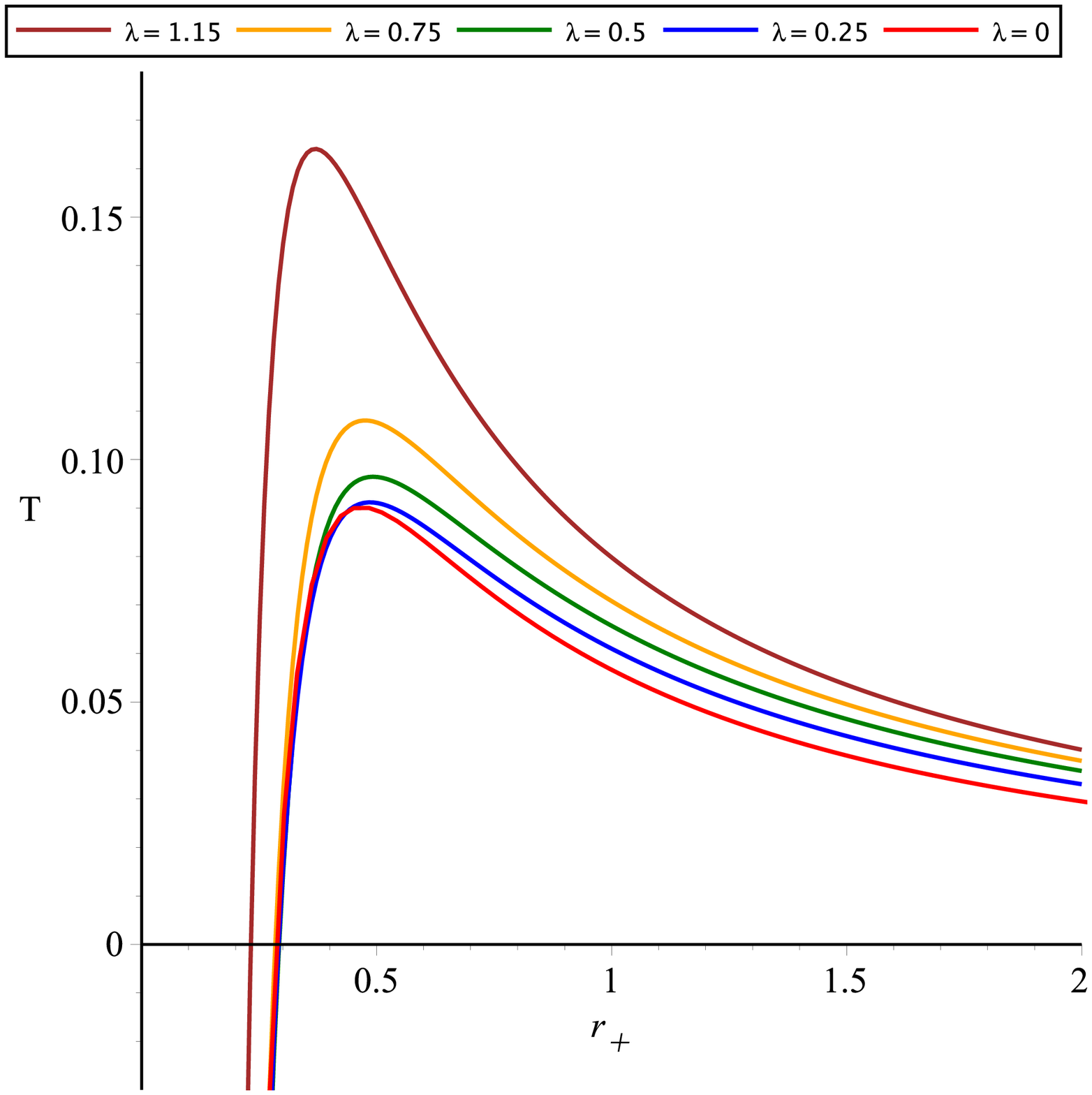}
	}
\subfigure[$Q=0.25$, $\lambda=0.25 $]{
	\includegraphics[width=0.38\textwidth]{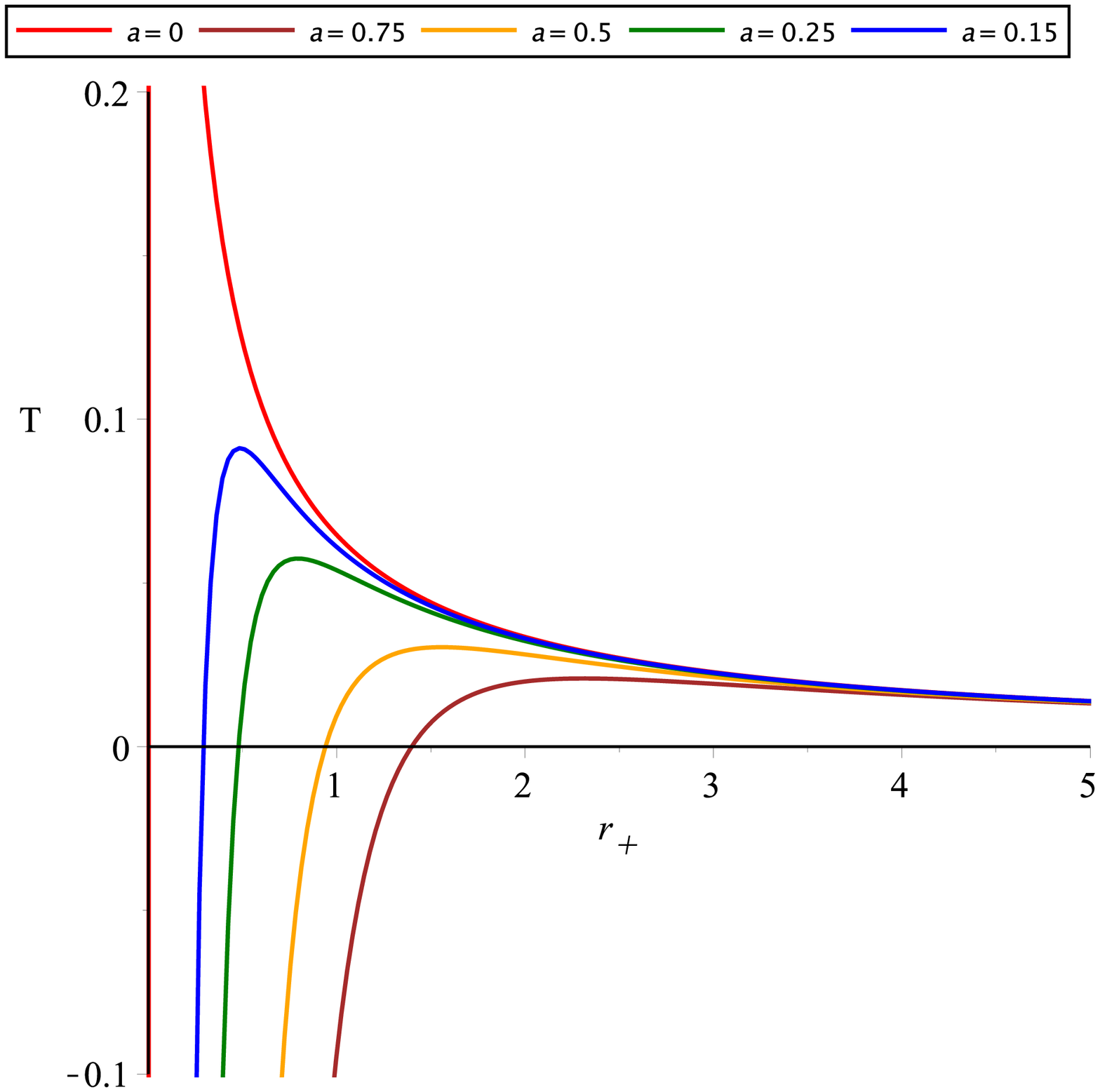}
}
	\caption{{ }{These plots show the evolution of Hawking temperature  in terms of horizon radius, for a rotating regular black hole with CMG corrections. In figure (a), different values of the parameter $0\leqslant Q\leqslant 1.25$ are considered, which the solution with $Q=0$ indicates a Kerr solution in the presence of the  CMG correction. Furthermore, plot (b) is devoted to investigate the effect of the variation of the  CMG parameter $0 \leqslant \lambda \leqslant 1.25$, in which $\lambda=0$ tends to a Kerr-Newman like solution. Then to see the effects of the variation of the rotation  parameter $0\leqslant a\leqslant 0.15$, diagram (c) is plotted in which the critical solution $a=0$ refers a  non-rotating R-N solution with CMG correction. }}
	\label{pic:TR}
\end{figure}

\begin{figure}[]
	\centering
	\subfigure[]{
		\includegraphics[width=0.3\textwidth]{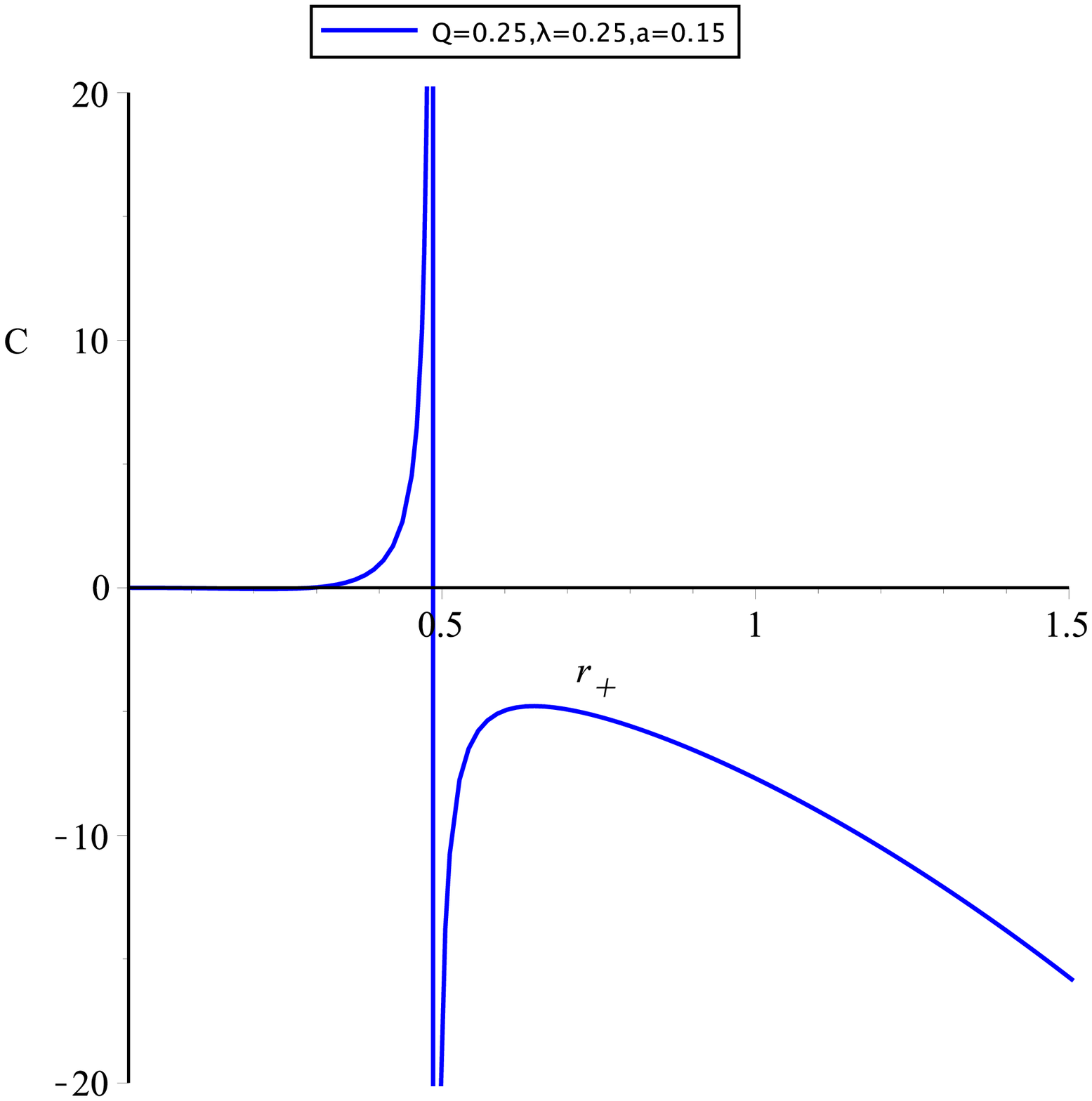}
	}
	\subfigure[closeup of figure (a)]{
		\includegraphics[width=0.3\textwidth]{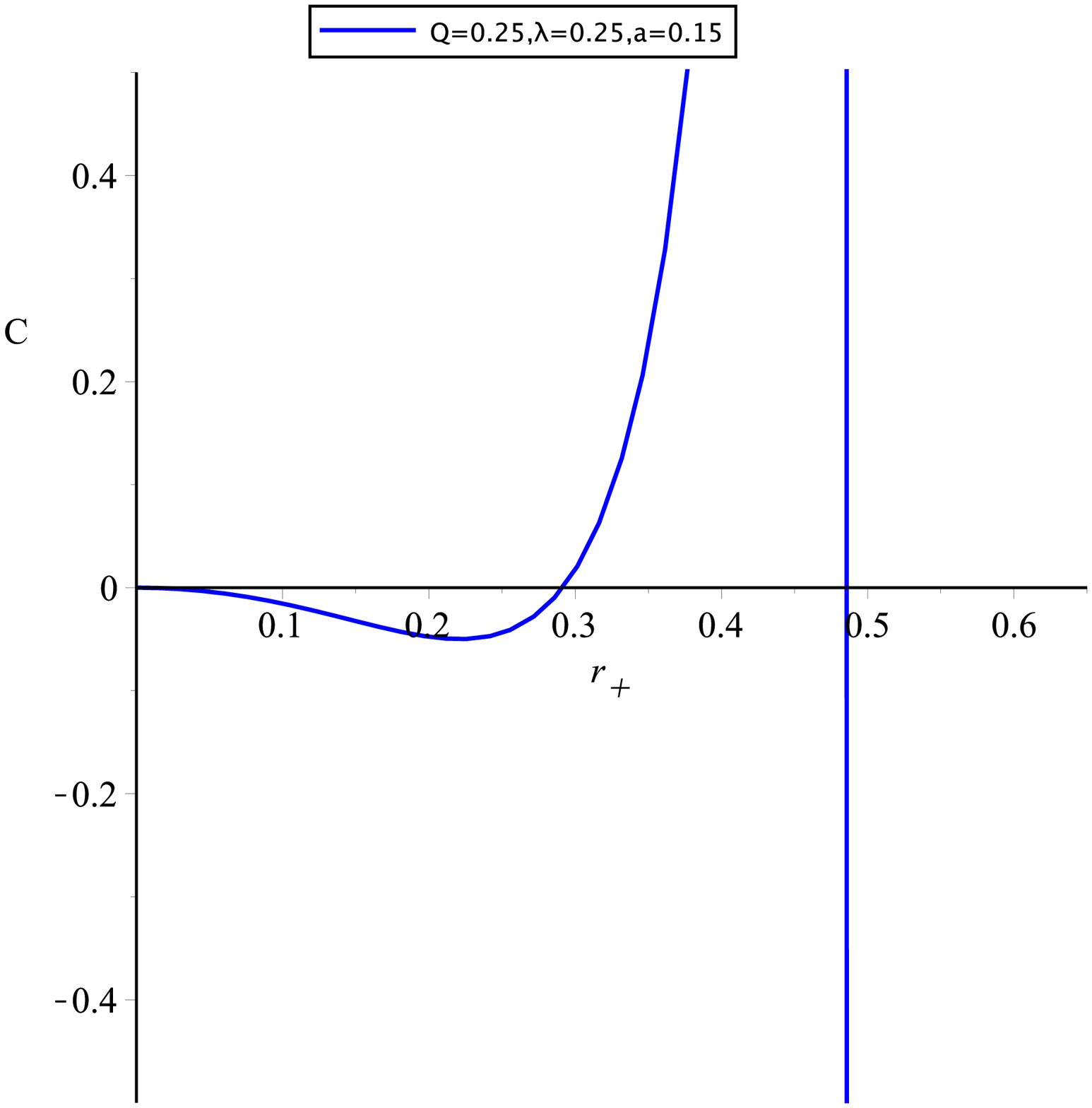}
	}
	\subfigure[$\lambda=0.25$, $a=0.15 $]{
		\includegraphics[width=0.3\textwidth]{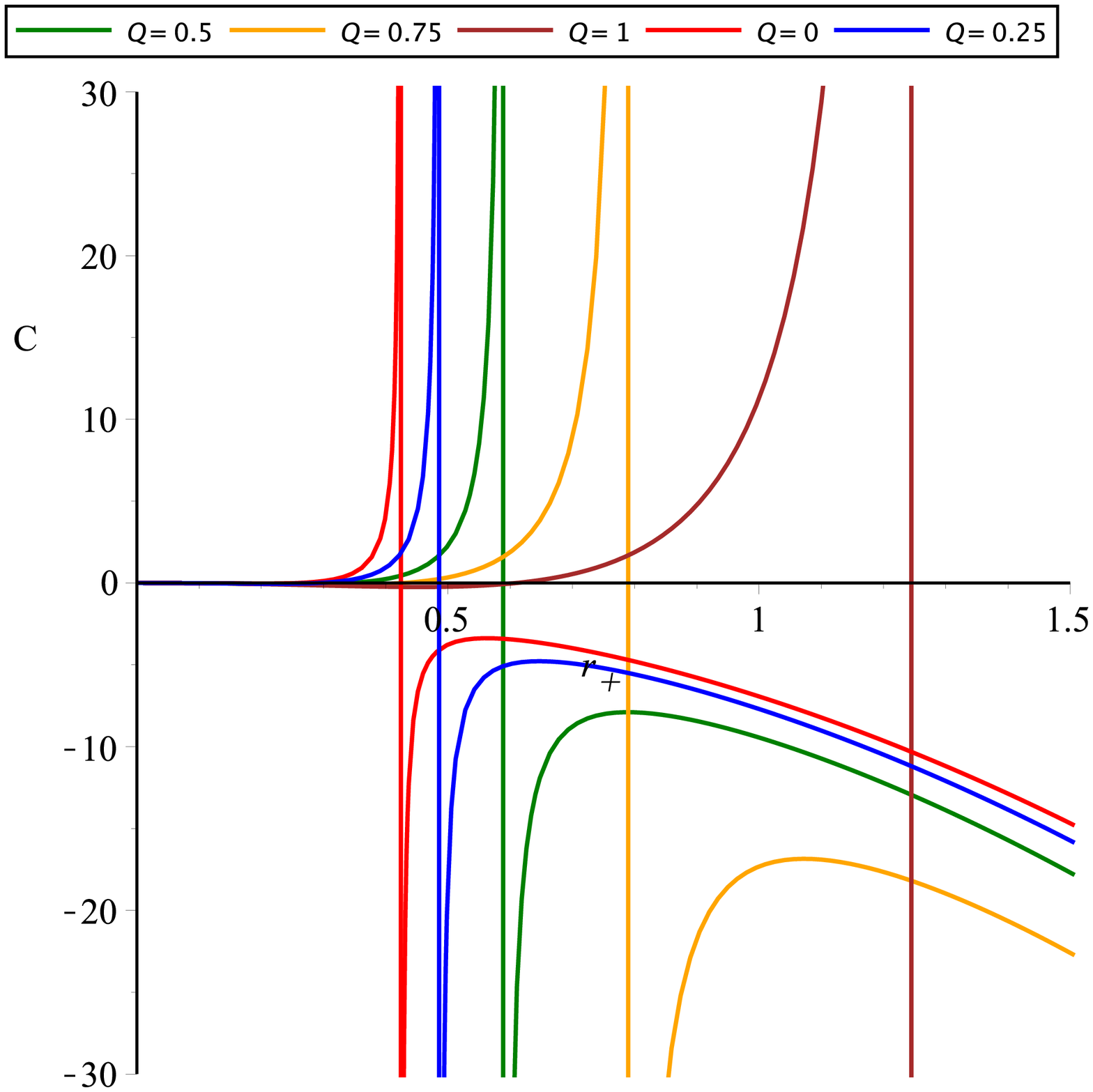}
	}
	\subfigure[closeup of figure (c) ]{
		\includegraphics[width=0.3\textwidth]{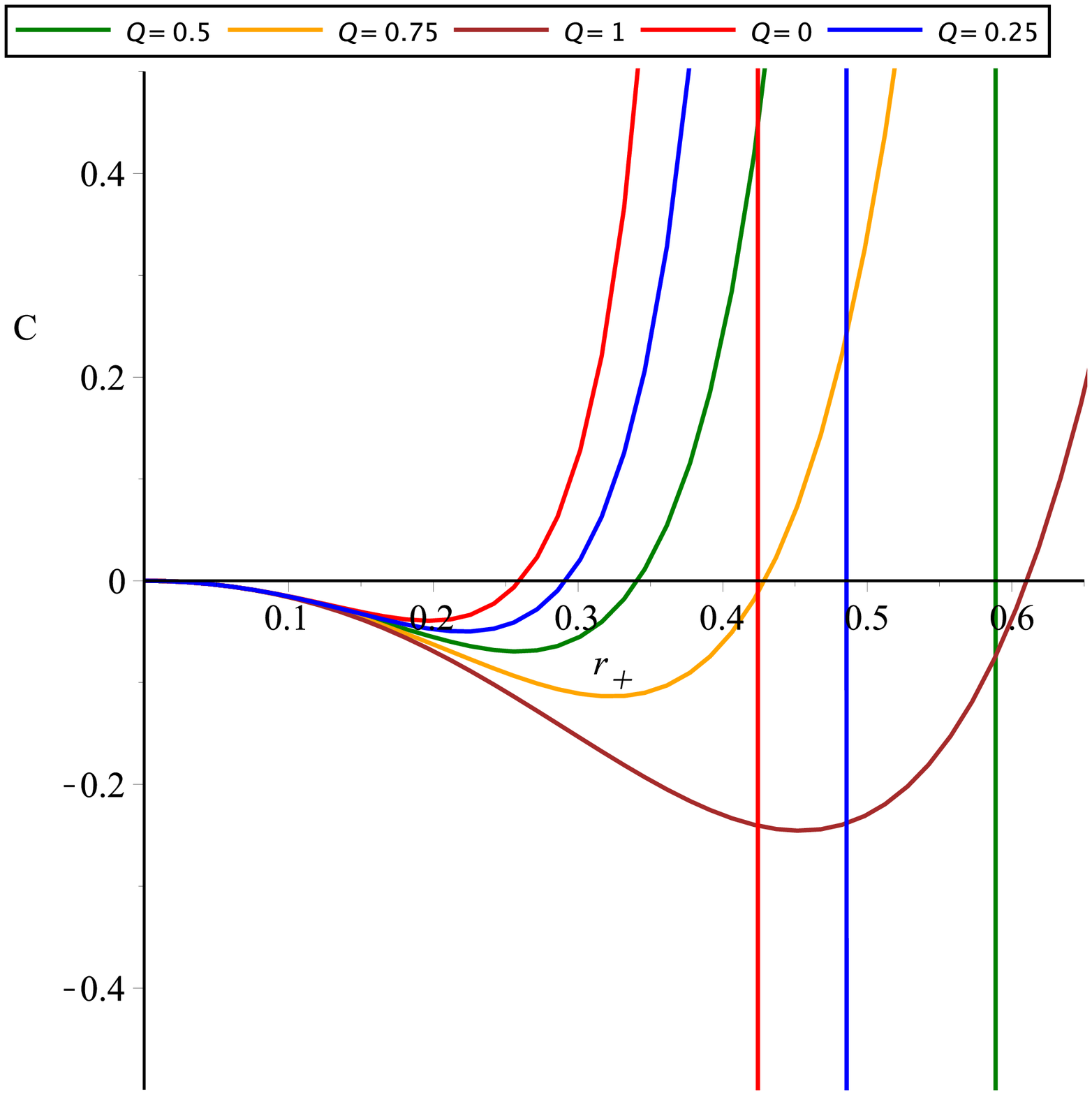}
	}
	\subfigure[$ Q=0.5 $, $a=0.15 $]{
		\includegraphics[width=0.3\textwidth]{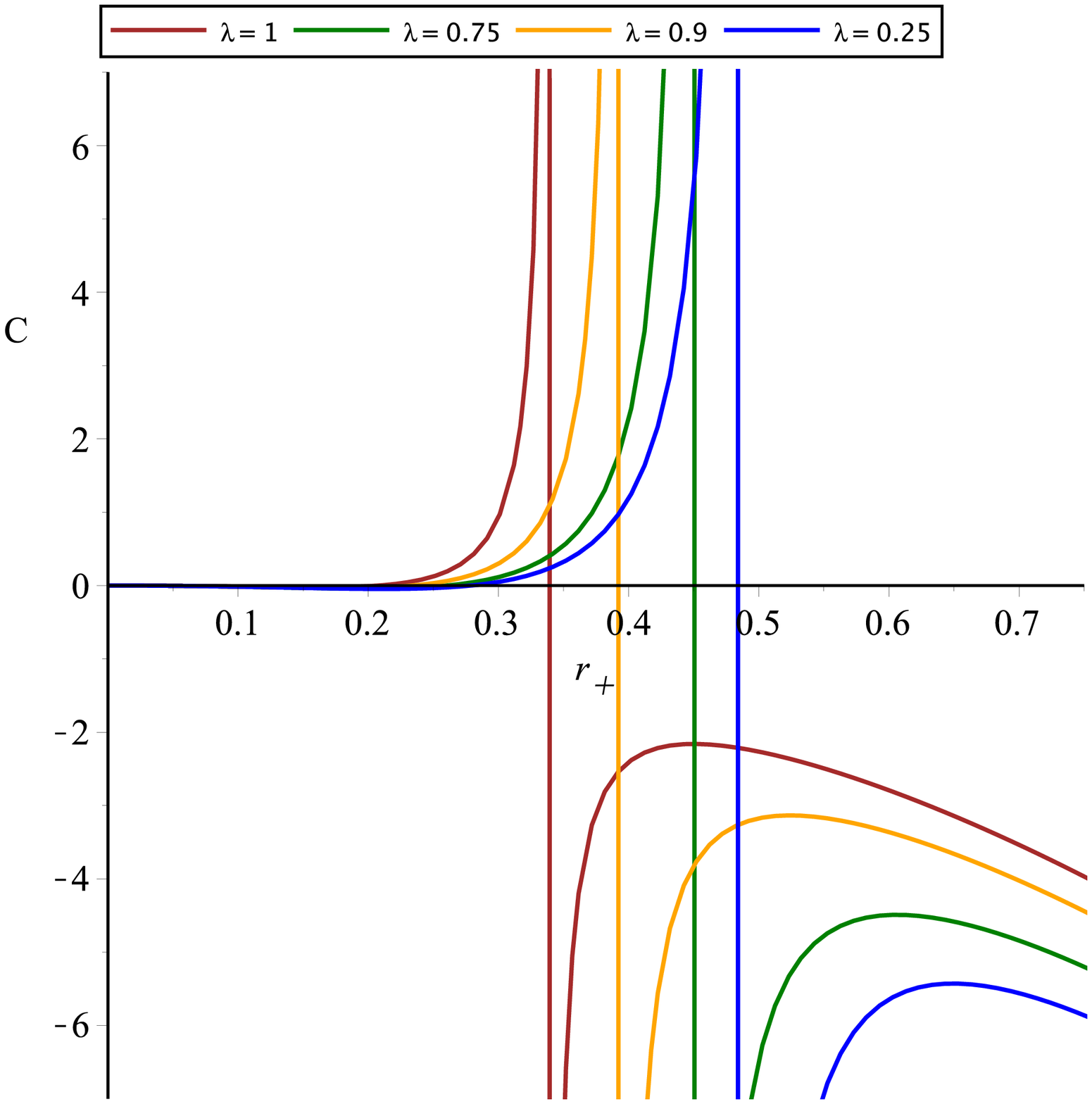}
	}
	\subfigure[closeup of figure (e) ]{
		\includegraphics[width=0.3\textwidth]{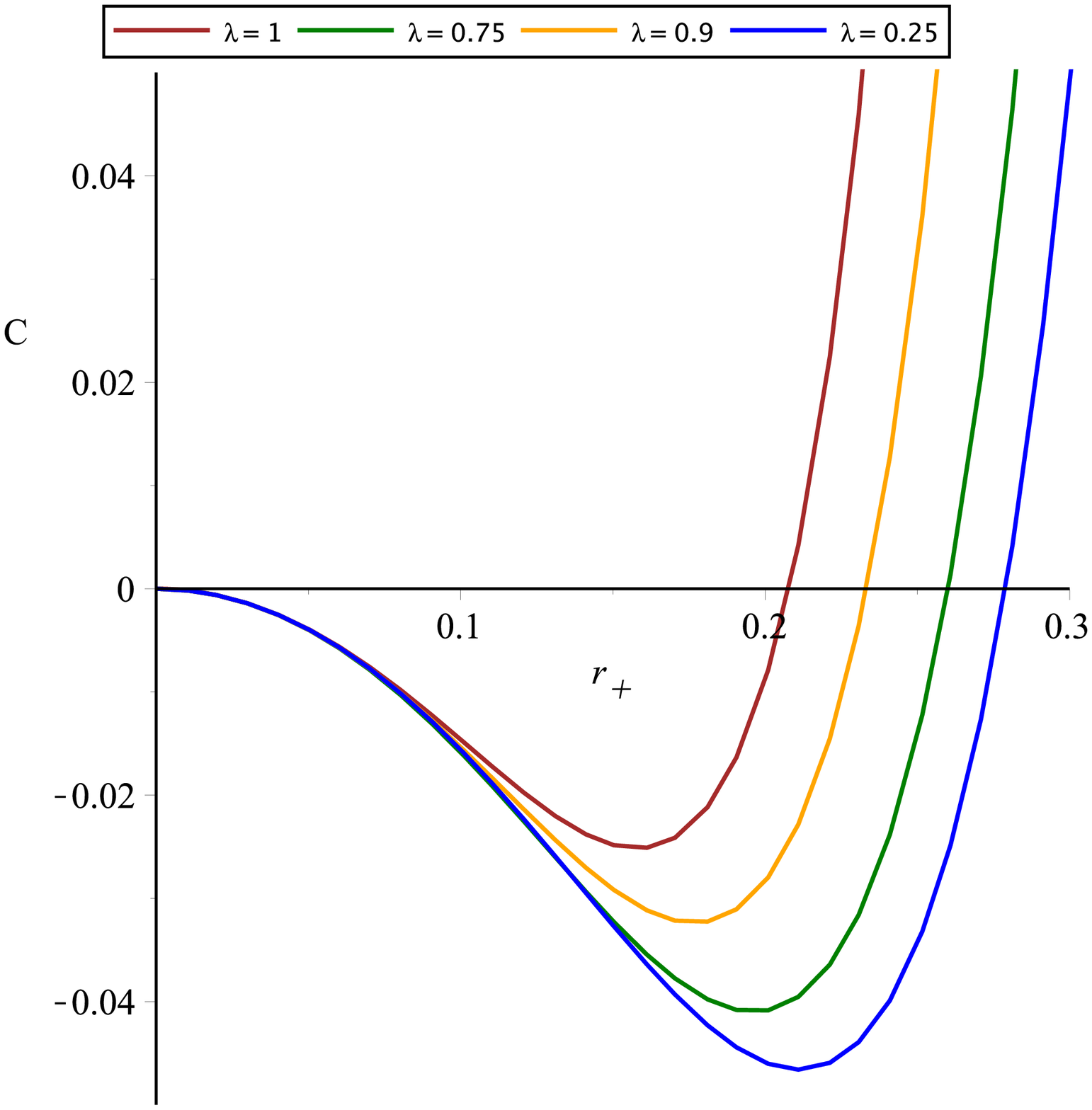}
	}
\subfigure[$Q=0.25$, $\lambda=0.25 $]{
	\includegraphics[width=0.38\textwidth]{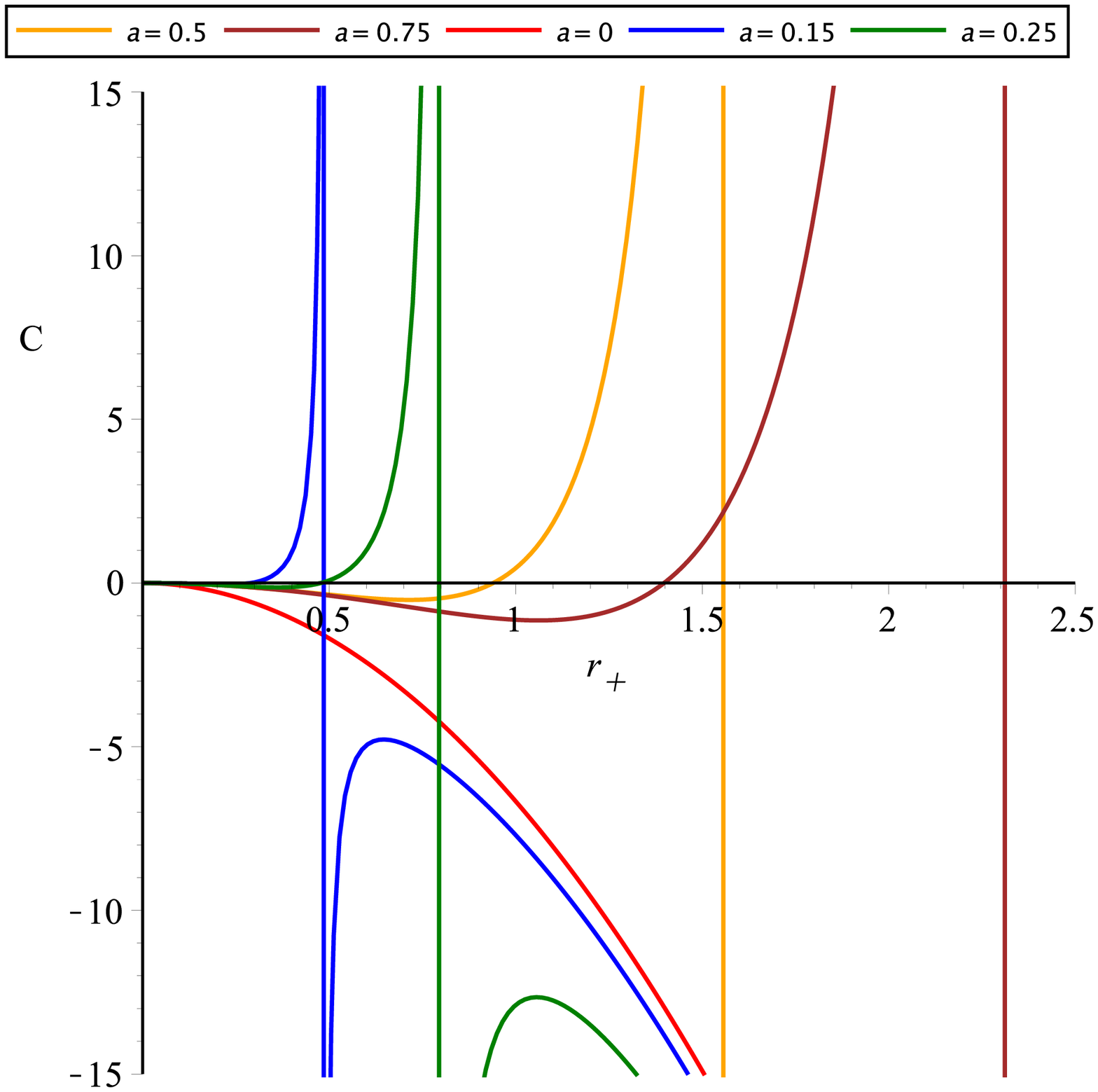}
}
	\caption{{These figures are based on  the evolution of the  heat capacity in terms of horizon radius $ r_{+}$, for a rotating regular black hole in the presence of the CMG corrections. One can consider figures (a) and (b) as general rotating solutions, with fixed values of $Q$, $\lambda$, and $a$. In addition to this, to observe the effects of variation of the parameter $Q$ in such a solution we have plotted figures (c) and (d). The plots (e) and (f) follow such a procedure but here the parameter $\lambda$ is varied rather than parameters $Q$ and $a$. Finally, figure (g) investigates the effect of  changes in the rotation parameter $a$.}}
	\label{pic:CR}
\end{figure}

\textcolor[rgb]{0.00,0.00,1.00}{\subsection{Thermodynamics}\label{subr1}}
{Now we are in a situation that studying of thermodynamics of a rotating charged regular black hole with CMG corrections is achievable. }In doing so at first one has to obtain the mass function in such a configuration. To determine the mass function one can consider the condition  $\Delta(r_{+})=0$, and express it in terms of entropy $S$. By using the relation between entropy  and event horizon radius, i.e. $ S=\pi (r^{2}_{+}+a^2)$, one gets
\begin{equation}\label{MR}
M(S,Q,a)=\dfrac{1}{2}\dfrac{Q \pi^{-1+\lambda}(-a \pi+\sqrt{\pi S})^{2-\lambda}+2a\sqrt{\pi S}-2a^2 \pi-S}{\sqrt{\pi}(a\sqrt{\pi}-\sqrt{S})}\,,
\end{equation}
and therefore Hawking temperature  $ T={\partial M}/{\partial S}$, reads

\begin{equation}\label{TR}
T=\dfrac{1}{4}\dfrac{(-a \pi+\sqrt{\pi S})^{-\lambda}\mathcal{F}-2a\sqrt{\pi S}+S}{\sqrt{\pi S}(a\sqrt{\pi}-\sqrt{S})^2}\,,
\end{equation}
where
\begin{equation*}
\mathcal{F}=Qa^2\pi^{1+\lambda}(\lambda-1)-2QaS^\frac{1}{2}\pi^{\frac{1}{2}+\lambda}(\lambda-1)+QS\pi^\lambda(\lambda-1)\,.
\end{equation*}
Then heat capacity $C=T{\partial S}/{\partial T}$ can be expressed as follows
\begin{equation}\label{CR}
C=2S^{\frac{3}{2}}(a\sqrt{\pi}-\sqrt{S})(\frac{C_1}{C_2})\,,
\end{equation}
where
\begin{eqnarray*}
C_1&&=-\pi^{1+\lambda}Qa^2(\lambda-1)+2\pi^{\frac{1}{2}+\lambda}QaS^\frac{1}{2}(\lambda-1)\\
&&+(-\pi a+\sqrt{\pi S})^\lambda (2 a \sqrt{\pi S}-S)-\pi^\lambda QS (\lambda-1)\;,\\
C_2&&=(-\pi a+\sqrt{\pi S})^\lambda (3aS^\frac{3}{2} \sqrt{\pi}-S^2)-\pi^{\lambda} QS (\lambda^{2}-1)\\
&&+\pi^{\frac{1}{2}+\lambda}QaS^\frac{3}{2}(2\lambda^2+\lambda-3)+\pi^{1+\lambda}Q a^2 S(2\lambda^2+\lambda-3)\\
&&+\pi^{\frac{3}{2}+\lambda}Q a^3 S^\frac{2}{2}(\lambda-1)\;.
\end{eqnarray*}
To observe the physical properties of charged spinning black hole with CMG corrections, one can utilize the results of the important thermodynamic parameters like $M$, $T$ and $C$. In doing so we plot them versus horizon radius $r_+$, which are appeared in Figs.~\ref{pic:MR}, \ref{pic:TR} and \ref{pic:CR} respectively.
{The behavior of mass function for different values of $Q$, $\lambda$, and $ a $ is illustrated in Fig.~\ref{pic:MR}. From this figure, one can see that the mass function of such a black hole contains a minimum point which for a Kerr like solution it takes the largest value then by increasing the amount of charge, this turning point takes smaller values. Moreover, as can be seen from Figs.~\ref{pic:MR} (a), and (b) by increasing the value of $Q$, $\lambda$, for a fixed $a$, the values of this minimum point shows a decreasing behavior. Contrasted to these two cases, Figs.~\ref{pic:MR} (a), and (b) , in  Fig.~\ref{pic:MR} (c), one can interestingly see  that by increasing the value of $a$, and for fixed $Q$ and $\lambda$, this minimum point behaves as increasing function. It may goes back to the effect of the increasing of the angular momentum on the behavior of the black hole. In these cases, compared to static black holes in last sections, almost diagrams are in a positive region, and interestingly for larger black holes mass function converges to non-rotating cases with $a=0$, see Fig.~\ref{pic:MR} (c).}
%%%%%%%%%%%%%%%%%%%%%%
{Another important thermodynamic parameter of  black hole investigations  one can consider is temperature, the plot of which function versus  horizon radius is plotted in  Fig.~\ref{pic:TR}. From this figure, it is observed that the plot of the temperature is in the negative regions at a particular range of $r_+$, then it tends to zero. Next, the temperature will get the positive values and increase up to a maximum point. After that, it starts decreasing with increasing in $r_+$.
Moreover, Figs.~\ref{pic:TR} (a) and (c), show that as long as the values of $Q$, and $a$ are increasing, the peak value of temperature gets smaller values. Meanwhile, Fig.~\ref{pic:TR} (b) shows that the peak value of temperature obtains larger values, comparing to the two aforementioned cases, by increasing the value of $\lambda$. In addition to the above properties, when black hole tends to be larger its temperature decreases and it almost converge to zero.}
%%%%%%%%%%%%%%%%%%%%%%%%%%%%%%%%%%%%%%%%%%
{To investigate the stability of a black hole, static or stationary, one can consider  heat capacity as one of the  important thermodynamic parameters, to figure out its physical results one can see Fig.~\ref{pic:CR}.
According to this figure, obviously heat capacity has a root besides a divergence point that correspond to the physical limitation point and the phase transition critical point respectively. From Figs.~\ref{pic:CR} (a) and (b), immediately it is figured out that for the interval  $r_{1} \simeq 0.05<r_{+}<r_{2}\simeq 0.29$, heat capacity gets negative values and therefore black hole is in an unstable phase of its evolution. Then for $r_{2} <r_{+} $, it enters a positive region which means that the black hole will be in a stable phase. Additionally, the phase transition critical point is located at $r_+\simeq 0.49$. In a similar procedure, one can study the effects of varying of parameter $Q$ on the evolution of heat capacity when $\lambda$ and $a$ are considered to be fixed,  in Figs.~\ref{pic:CR} (c)-(d).  These plots show that, by increasing the values of charge the physical limitation point and the phase transition critical point  shifted to the larger values of the event horizon radii, i.e. for larger black holes.
As a next step,  from Figs.~\ref{pic:CR} (e)-(f), one can observe that by considering the evolution of parameter $\lambda$, and  fixing parameters $Q$ and $ a $, the location of phase transition points of the system is shifting along horizon radii as well. Ultimately, seeing the impact of the variation of parameter $a$ on the evolution of the black hole one can consider the plot Fig.~\ref{pic:CR} (g).
}

%%%%%%%%%%%%%%%%%%%%%%%%%%%%%%%%%%%%%%%%%%%%%
%%%%%%%%%%%%%%%%%%%%%%%%%%%%%%%%%%%%%%%%%%%%%%%%%%

\begin{figure}[]
	\centering
	\subfigure[]{
		\includegraphics[width=0.4\textwidth]{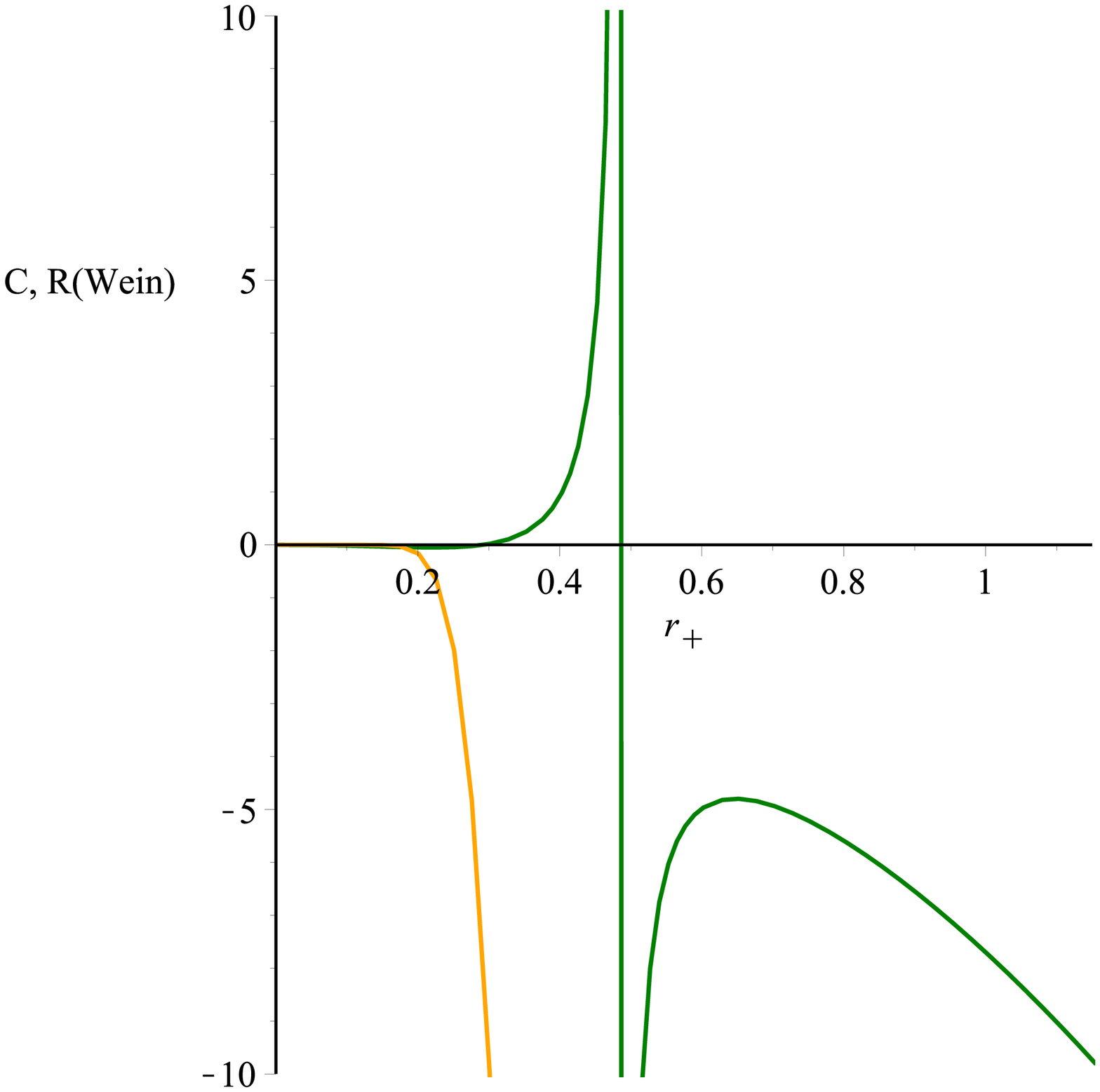}
	}
	\subfigure[]{
		\includegraphics[width=0.4\textwidth]{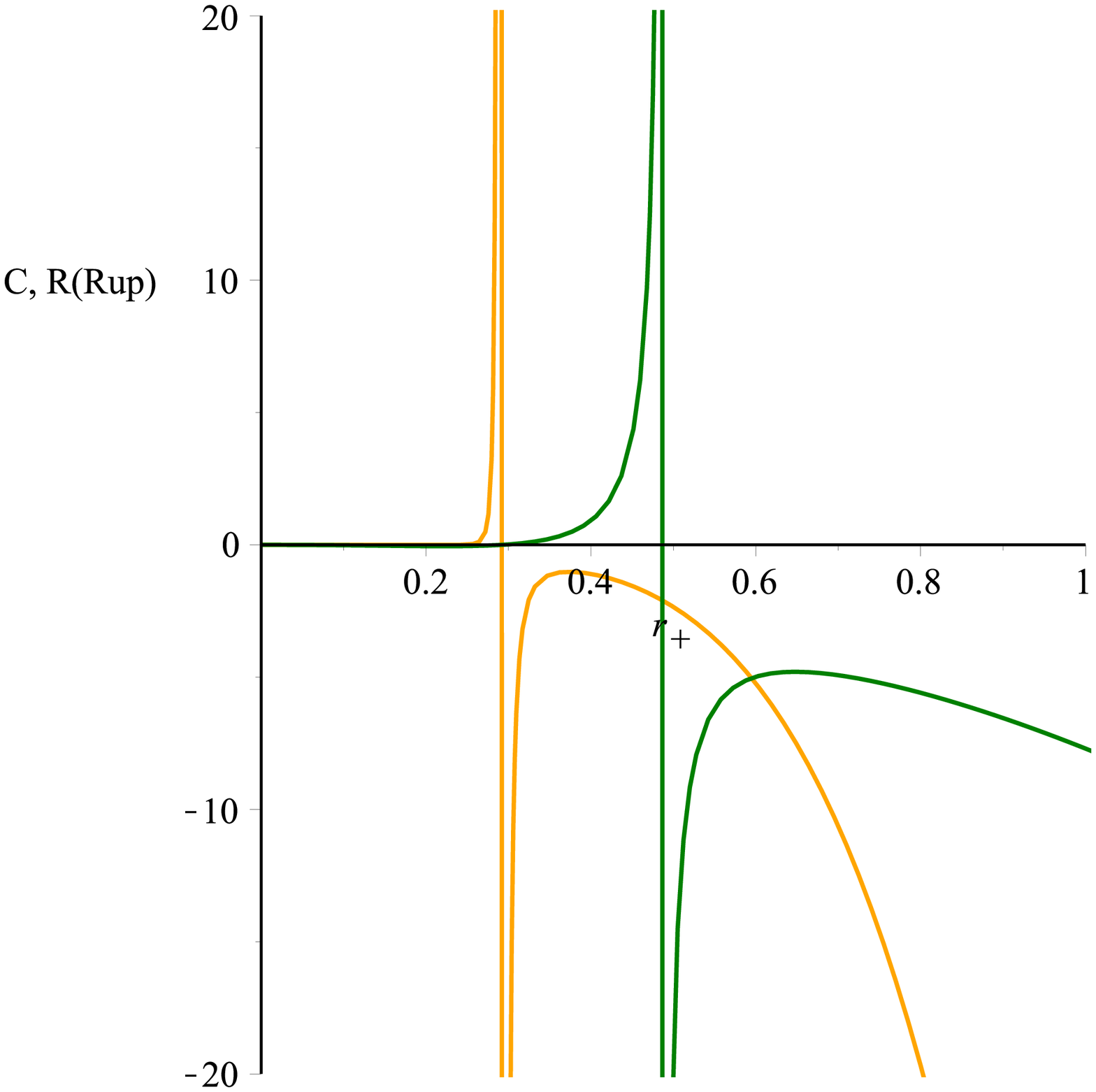}
	}
	\subfigure[A close-up view of figure (b)]{
		\includegraphics[width=0.4\textwidth]{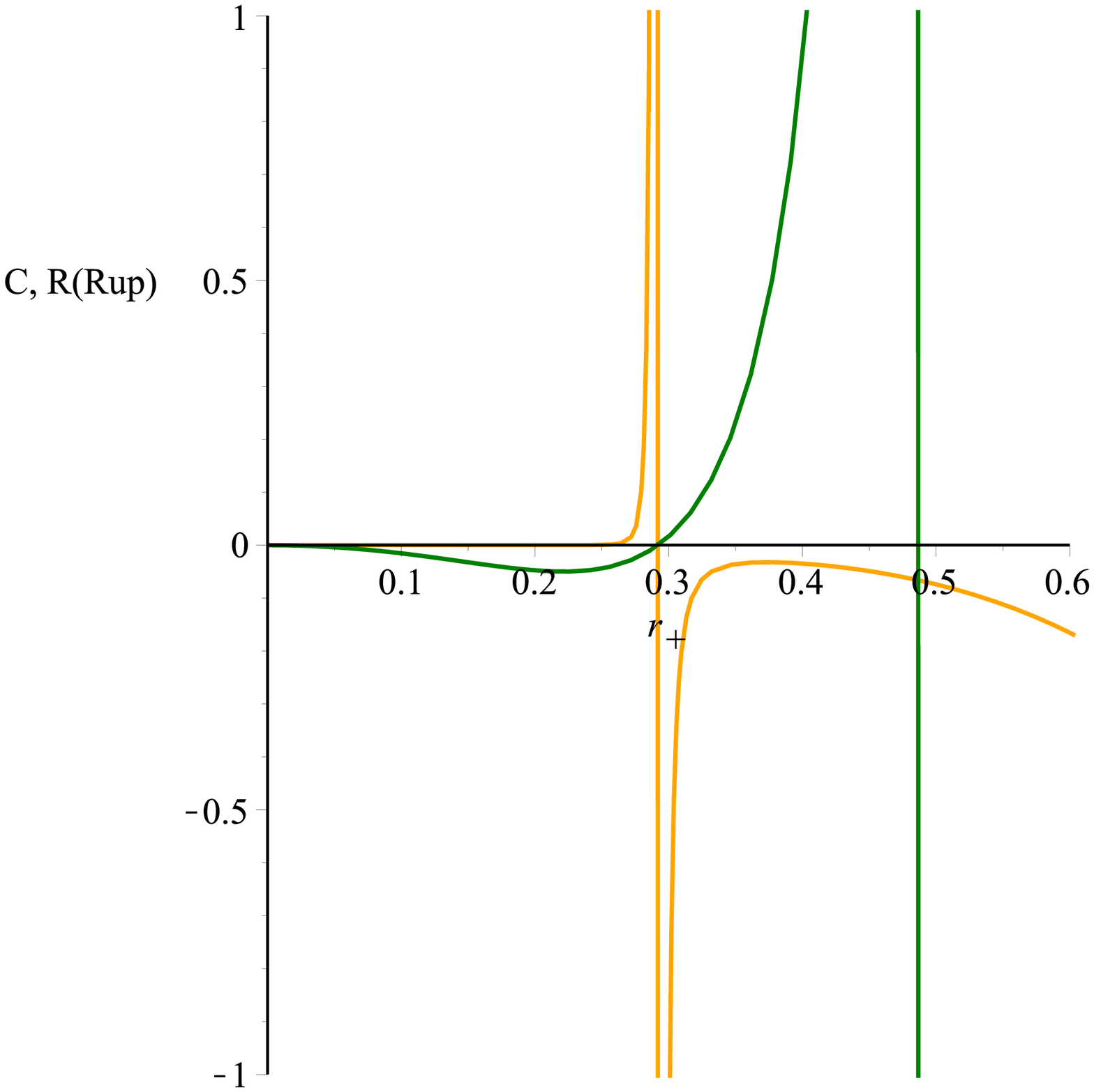}
	}
	\subfigure[]{
		\includegraphics[width=0.4\textwidth]{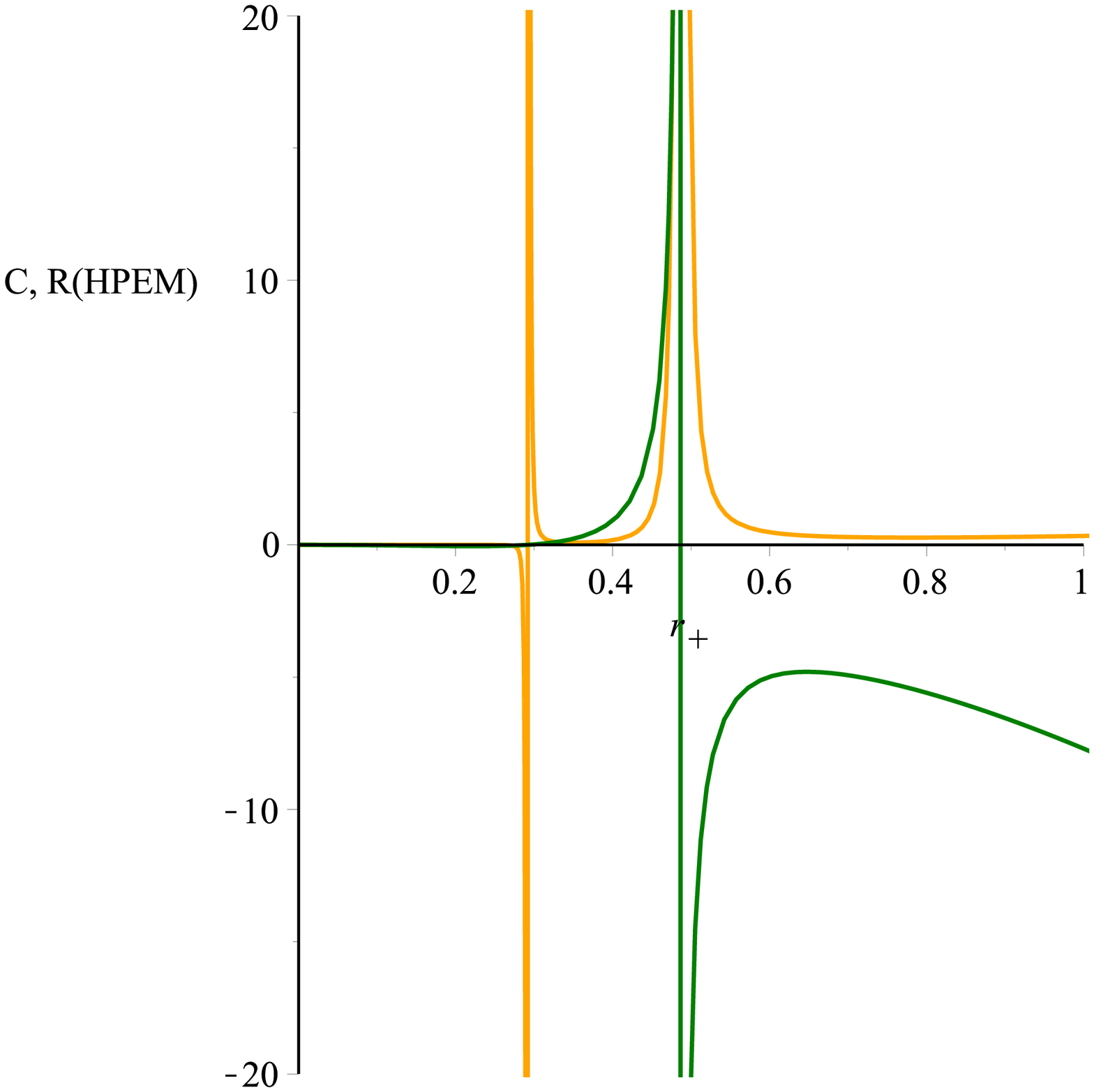}
	}
	\subfigure[A close-up  view of figure (d)]{
		\includegraphics[width=0.4\textwidth]{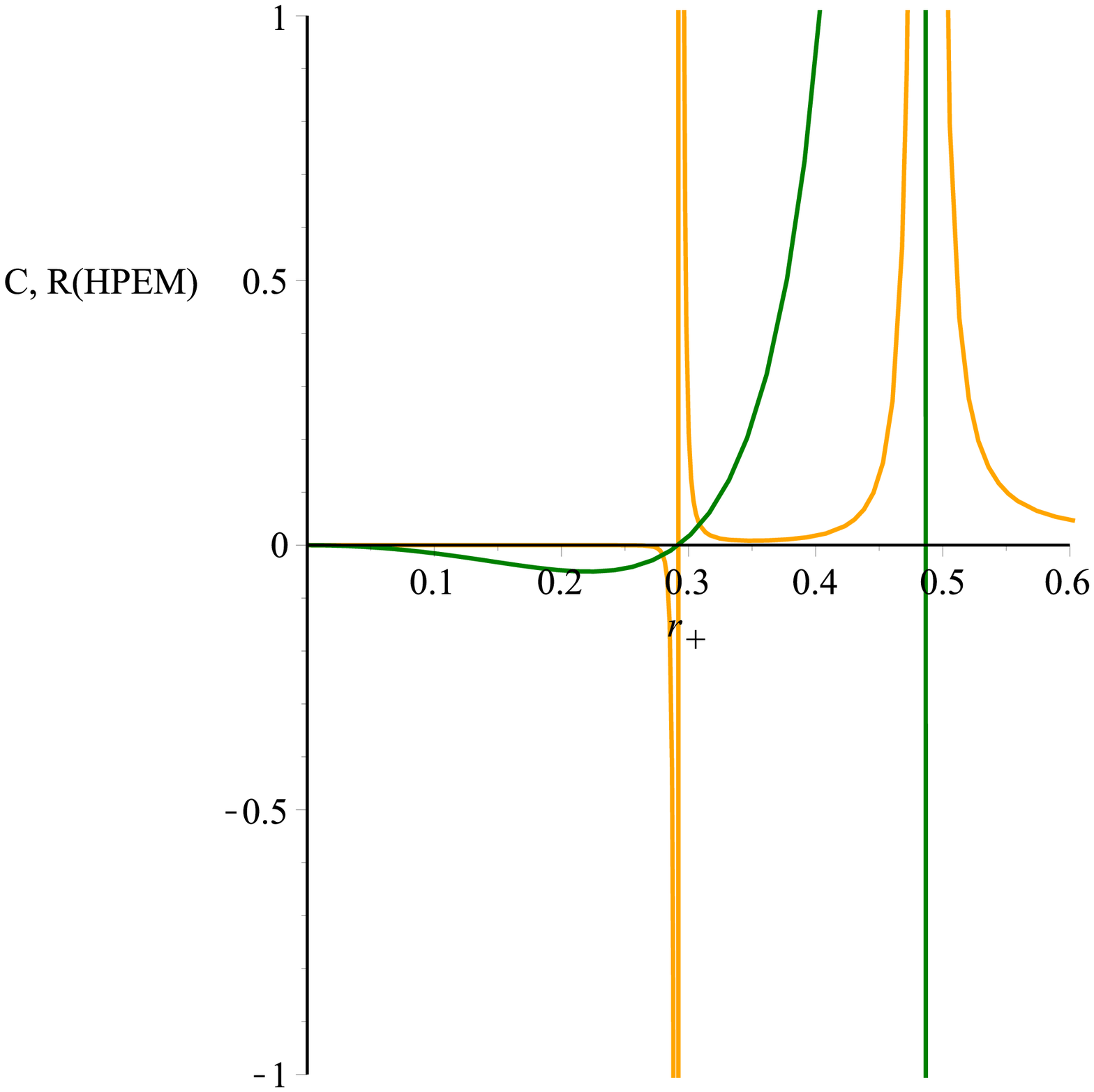}
	}
	\caption{{ }{In these figures the curvature scalar variation of Weinhold, Ruppeiner and HPEM metrics, indicated by orange lines, in addition to heat capacity variations, distinguished by Green lines, versus event horizon radii  are plotted. They have been illustrated for $ Q=\lambda=0.25 $, and $a=0.15 $, considering a rotating regular charged black hole in the presence of a CMG correction. Based on figures (d) and (e), one sees that the divergence points of the Ricci scalar of the HPEM metric coincide with both the zero and divergence points of the heat capacity, physical limitation and phase transition critical points.}}
	\label{pic:CRWeinQueHPEM}
\end{figure}

\textcolor[rgb]{0.00,0.00,1.00}{\subsection{Thermodynamic geometry}\label{subr2}}
In this section, we would like to study the thermodynamic geometry of a regular rotating black hole in the presence of CMG corrections. {But before going further it is perhaps worth going a little bit more through  the Weinhold, and Ruppeiner metrics. As discussed already, In the Weinhold metric usually in the Hessian definition energy as a function of entropy and other extensive variables can appear. Such a definition from the mathematical point of view will be natural, albeit if there be an well defined  space coordinatized affine structure, which means metric should preserve its form under affine coordinate transformations, see \cite{Aman:2015wsa}. To define the Ruppeiner metric, one might consider a Gibbs surface of a thermodynamic system utilizing a relation between entropy and extensive variables of the system including its energy \cite{Ruppeiner:1995zz}. We know that  a black hole is a thermodynamic system or in other words is an equilibrium state of general relativity with the total mass $M$, which one can find a relation between the area of its event horizon and the entropy. Obviously this entropy function has dependency on  conserved parameters like mass, angular momentum and charge as well and therefore an affine structure can be exist. This entropy does not behave like a concave function and therefore the Ruppeiner metric for instance is not positive-definite anymore. Technically it means that, the density of states quickly are growing up with energy and therefore the canonical ensembles do not behave satisfactorily \cite{Hawking:1976de}. }
Obviously here, the thermodynamical space will change from M(S,Q) which appeared in previous section, say static black hole, to $M(S,Q,a)$, for a rotating charged black hole. Therefore, the Weinhold, Ruppeiner, Quevedo and HPEM metrics can be expressed, collectively, as follows \cite{Hendi:2015rja,Hendi:2015xya,EslamPanah:2018ums,Soroushfar:2019ihn,Soroushfar:2020wch,Pourhassan:2021mhb}
\begin{equation}
ds^{2}=
\left\{\begin{array}{ll}
 \quad \quad \quad Mg_{ab}^{W}dX^{a}dX^{b}\,,  & \text{Weinhold} \\
\\
\quad \quad  -\dfrac{1}{T}Mg_{ab}^{R}dX^{a}dX^{b}\,,  & \text{Ruppeiner} \\
\\
(SM_{S}+aM_{a}+QM_{Q})(-M_{SS}dS^{2}+M_{aa}da^{2}+M_{QQ}dQ^{2})\,,   & \text{Quevedo Case I}   \\
\\
\quad SM_{S}(-M_{SS}dS^{2}+M_{aa}da^{2}+M_{QQ}dQ^{2})\,,    &\text{Quevedo~ Case~ II}    \\
\\
\quad \quad \frac{SM_{S}}{\left(M_{aa}M_{QQ} \right)^{3}}\left(-M_{SS}dS^{2}+M_{aa}da^{2}+M_{QQ}dQ^{2}\right)\,. & \text{HPEM}
\end{array}\right.
\end{equation}
These metrics have the following denominator for their Ricci scalars \cite{Hendi:2015rja,Hendi:2015xya,EslamPanah:2018ums,Soroushfar:2019ihn,Soroushfar:2020wch,Pourhassan:2021mhb}:
\begin{equation}
{denom}(R)=
\left\{\begin{array}{ll}
-2M^{3}(M_{Qa}^{2}M_{SS}+M_{SQ}^{2}M_{aa}+M_{Sa}^{2}M_{QQ}-\\
M_{SS}M_{aa}M_{QQ}-2M_{SQ}M_{Sa}M_{Qa})^{2}   &\text{Weinhold}\,,\\
\\

-2M^{3}T^{3}(M_{Qa}^{2}M_{SS}+M_{SQ}^{2}M_{aa}+M_{Sa}^{2}M_{QQ}-\\
M_{SS}M_{aa}M_{QQ}-2M_{SQ}M_{Sa}M_{Qa})^{2}& \text{Ruppeiner}\,,\\
\\
2M_{SS}^{2}M_{aa}^{2}M_{QQ}^{2}(SM_{S}+aM_{a}+QM_{Q})^{3}\,,  & \text{Quevedo~ Case~ I}   \\
\\
\quad \quad 2S^{3}M_{SS}^{2}M_{aa}^{2}M_{QQ}^{2}M_{S}^{3}\,,  & \text{Quevedo ~Case ~II}  \\
\\
\quad \quad \quad 2S^{3}M_{SS}^{2}M_{S}^{3}\,. & \text{HPEM}
\end{array}\right.
\end{equation}
{Examining the above equations shows that the Quevedo formalism does not contain any specific physical information about the system. For the remnant  formalisms, the resulting curvature scalars are plotted according to the horizon radius aiming to investigate the thermodynamic phase transition, see Fig.~\ref{pic:CRWeinQueHPEM}. It can be seen  from Fig.~\ref{pic:CRWeinQueHPEM} (a), that the singular point of the curvature scalar of the Weinhold metric does not coincide with any of zeros and divergence points of the heat capacity. But for the other two methods, the situation is much better. Figs.~\ref{pic:CRWeinQueHPEM} (b), (c), show that the singular point of the curvature scalar of the Ruppeiner metric are compatible with the root of the heat capacity. However, there is no correspondence between divergence points of the Ruppeiner metric and the heat capacity. Moreover, according to Figs.~\ref{pic:CRWeinQueHPEM} (d), (e), we see that divergence points of the Ricci scalar of the HPEM metric coincide with both zero and divergence points of the heat capacity, physical limitation and phase transition critical points. So, clearly one can get more information for the thermodynamic phase transition from the HPEM metric compared to the Ruppeiner metric and the other methods mentioned in this paper.}

%%%%%%%%%%%%%%%%%%%%%%%%%%%%%%%%%%%%%%%%%%%%

\begin{figure}[]
	\centering
\subfigure[$\lambda=+0.25$, $a=0.15$]{\includegraphics[width=5cm]{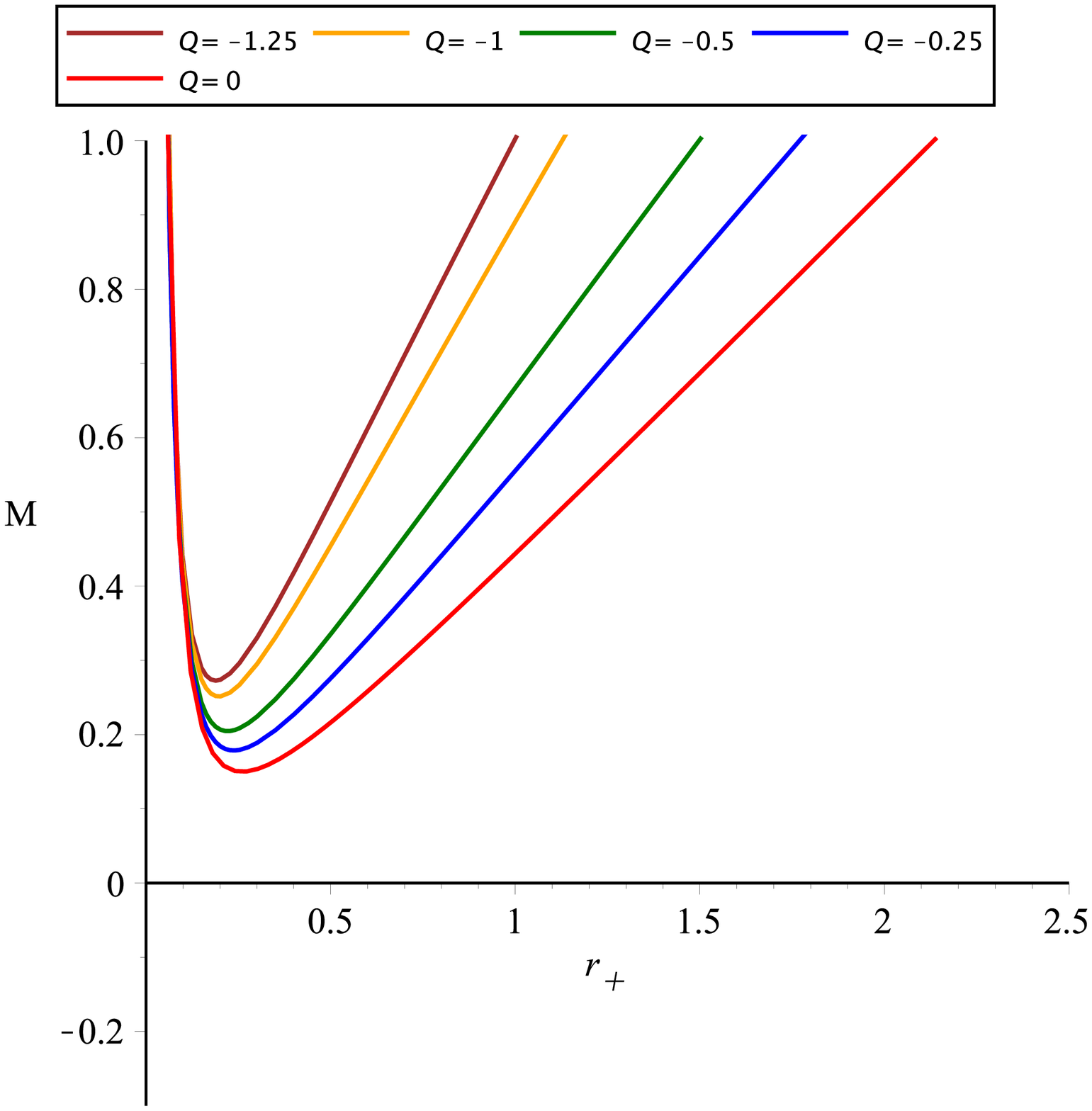}}
\subfigure[$Q=+0.25$, $a=0.15$]{	\includegraphics[width=5cm]{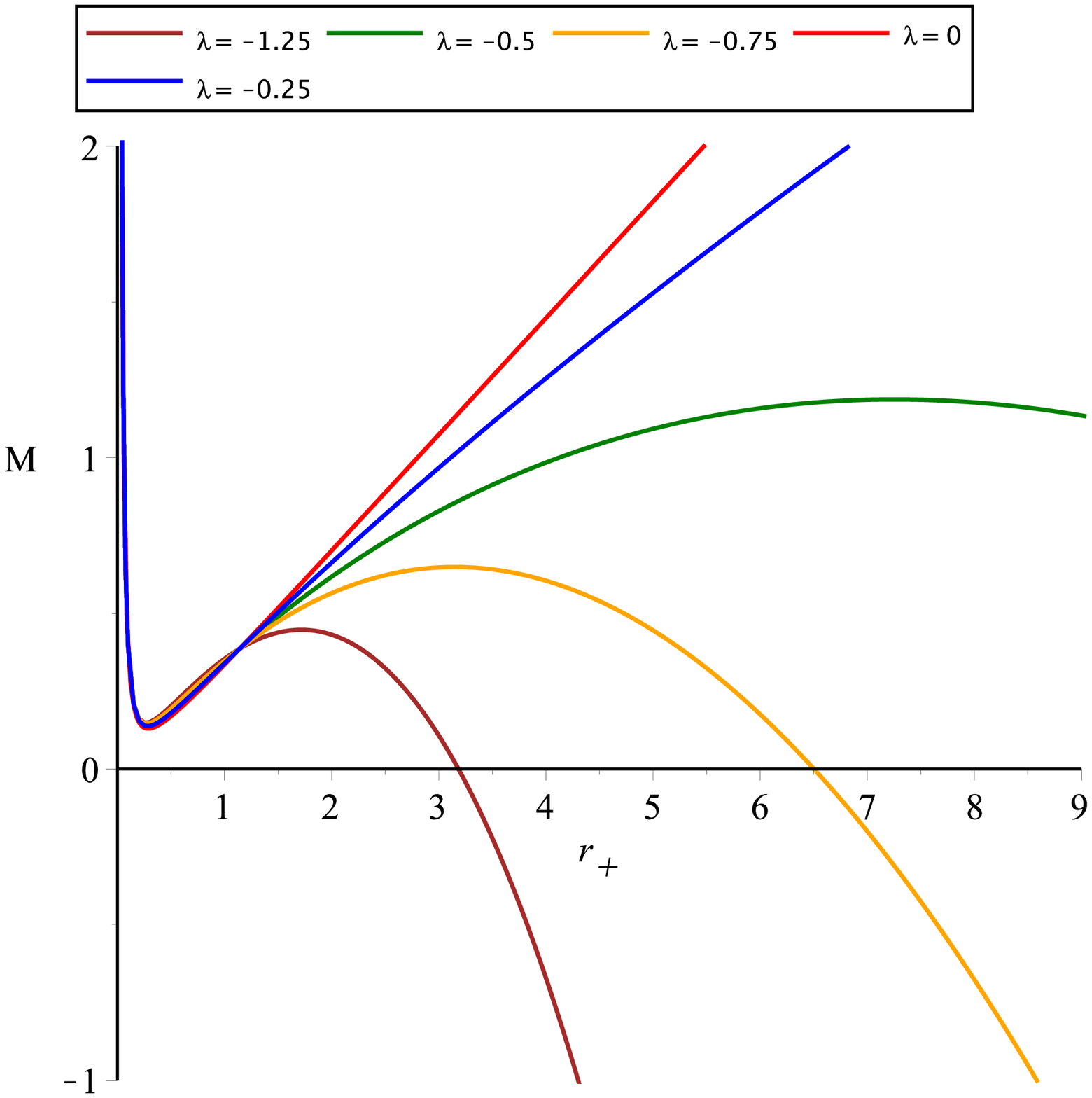}}
\subfigure[$\lambda=-0.25$, $a=0.15$]{\includegraphics[width=5cm]{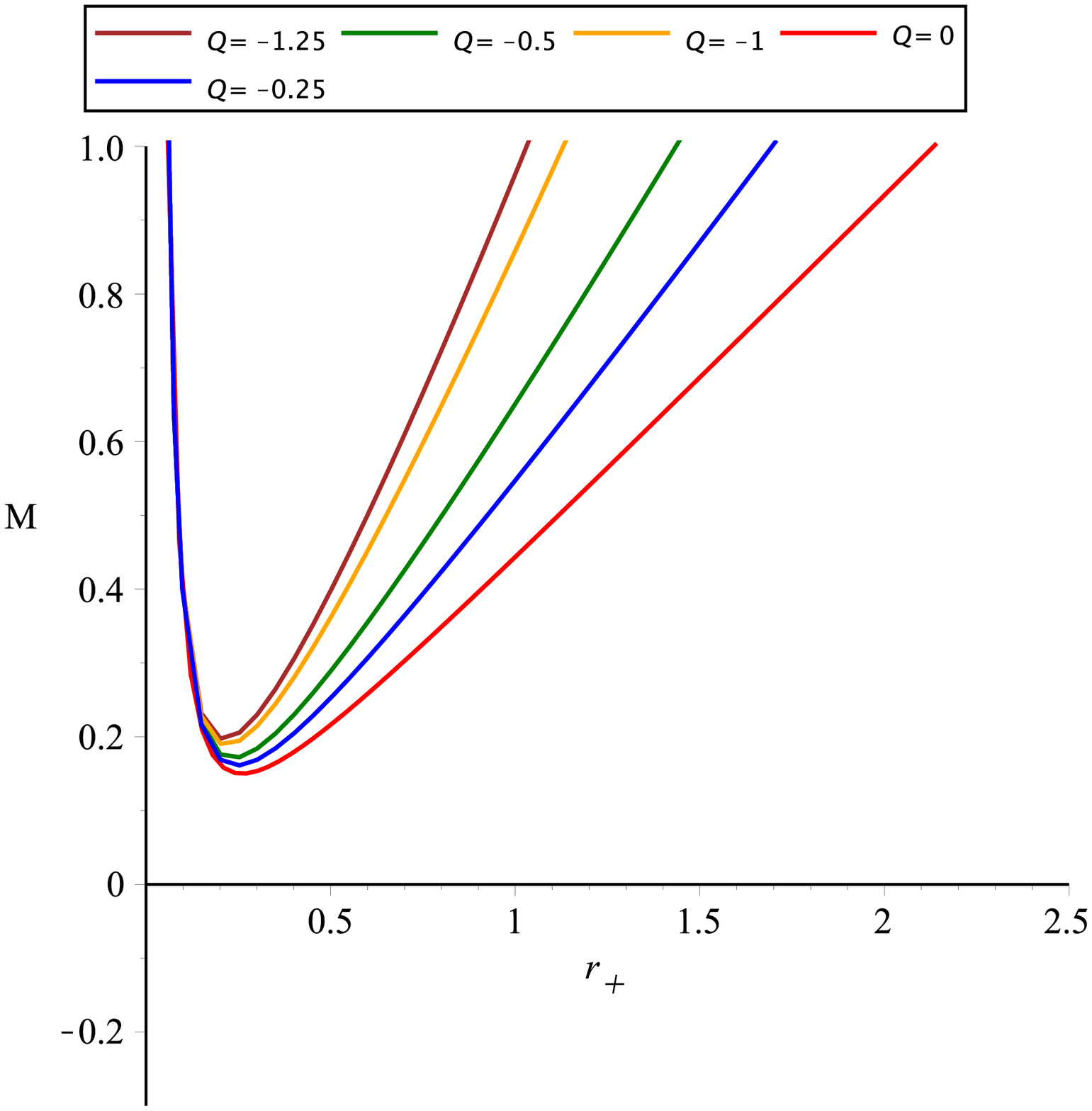}}
\subfigure[$Q=-0.25$, $a=0.15$]{	\includegraphics[width=5cm]{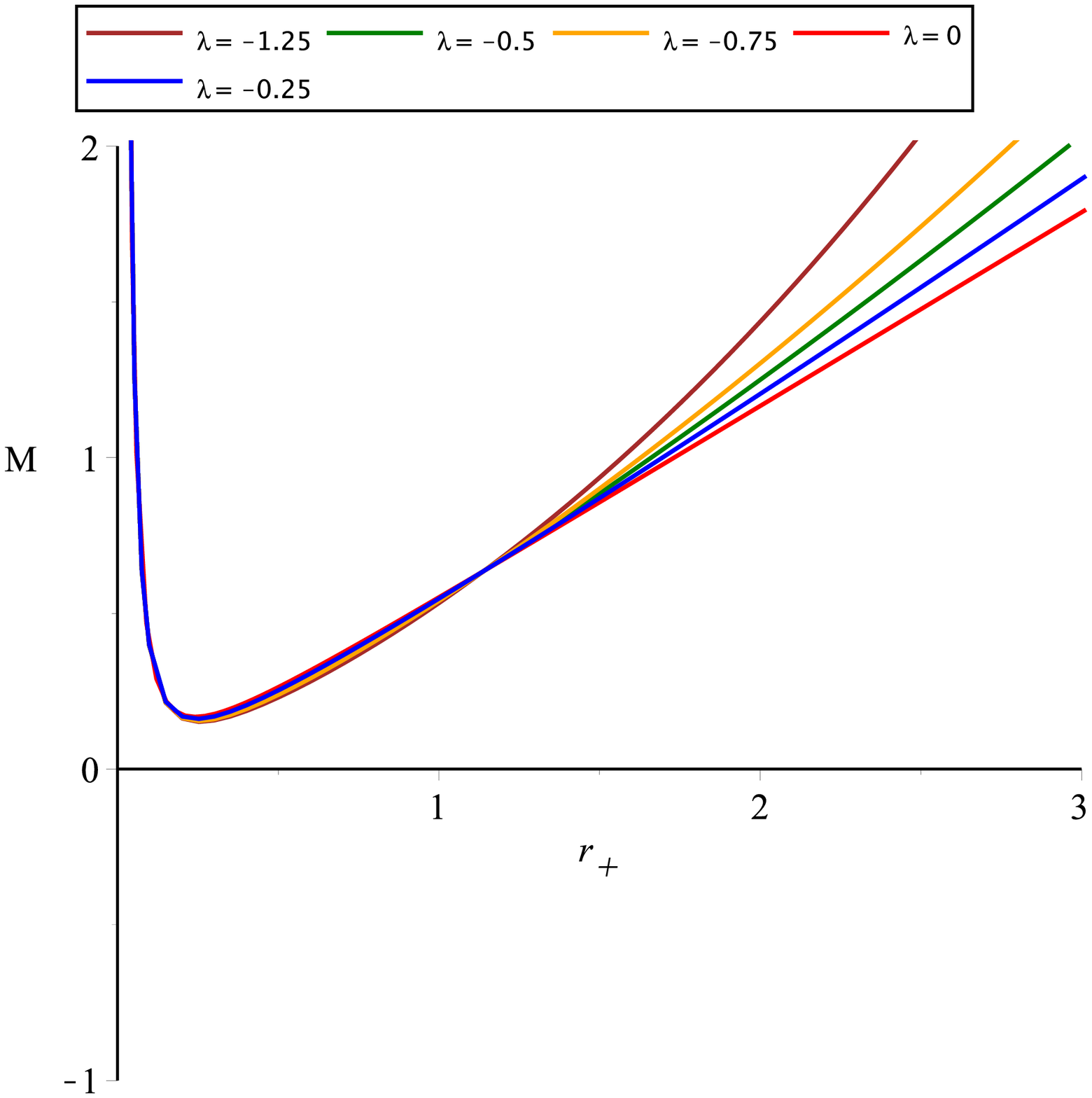}}
	\caption{ {The Variation of mass function  in terms of horizon radius $ r_{+}$, for a rotating regular black hole in the presence of  CMG corrections is illustrated. In plot (a) the focus is on the changes in amount of $Q<0$ when the parameter $\lambda$ gets a fixed positive amount. In figure (b)  the changes are related to the parameter  $\lambda <0 $ when  $Q$ gets a fixed positive constant. Then in figure (c),  mass function varies for $Q<0$ when $\lambda$ gets a fixed negative value is depicted. And finally figure (d) shows the behaviour of the mass function $\lambda<0$ when $Q$ is a fixed negative constant. For figures (a), (c) and (d) mass function gets only positive values and only plot (b) in some cases shows negative values for mass function. }  }
	\label{pic:MN}
\end{figure}

\begin{figure}[]
	\centering
	\subfigure[$Q=+0.25$, $a=0.15$]{
		\includegraphics[width=0.38\textwidth]{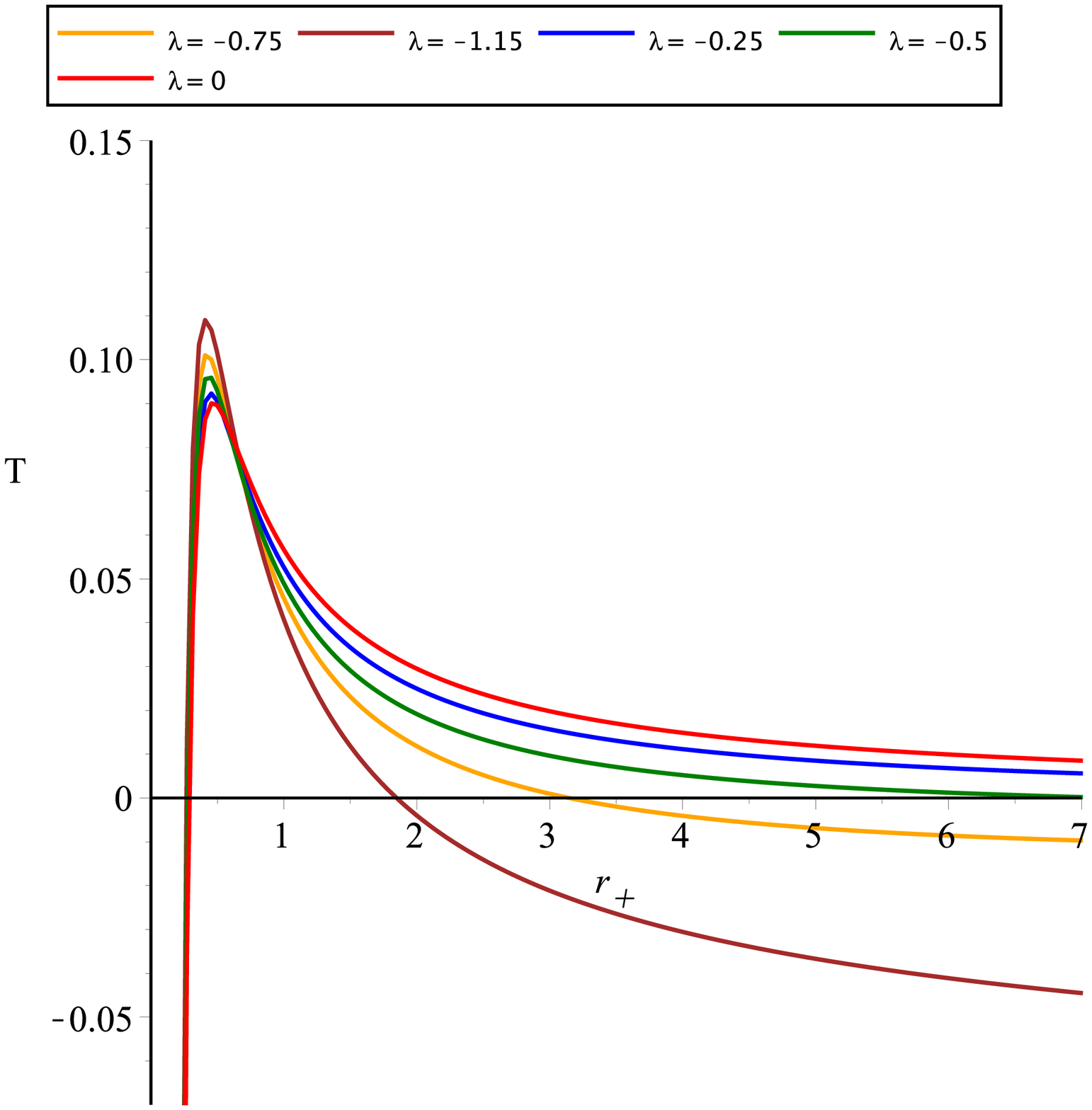}
	}
\subfigure[$\lambda=+0.25$, $a=0.15$]{
	\includegraphics[width=0.38\textwidth]{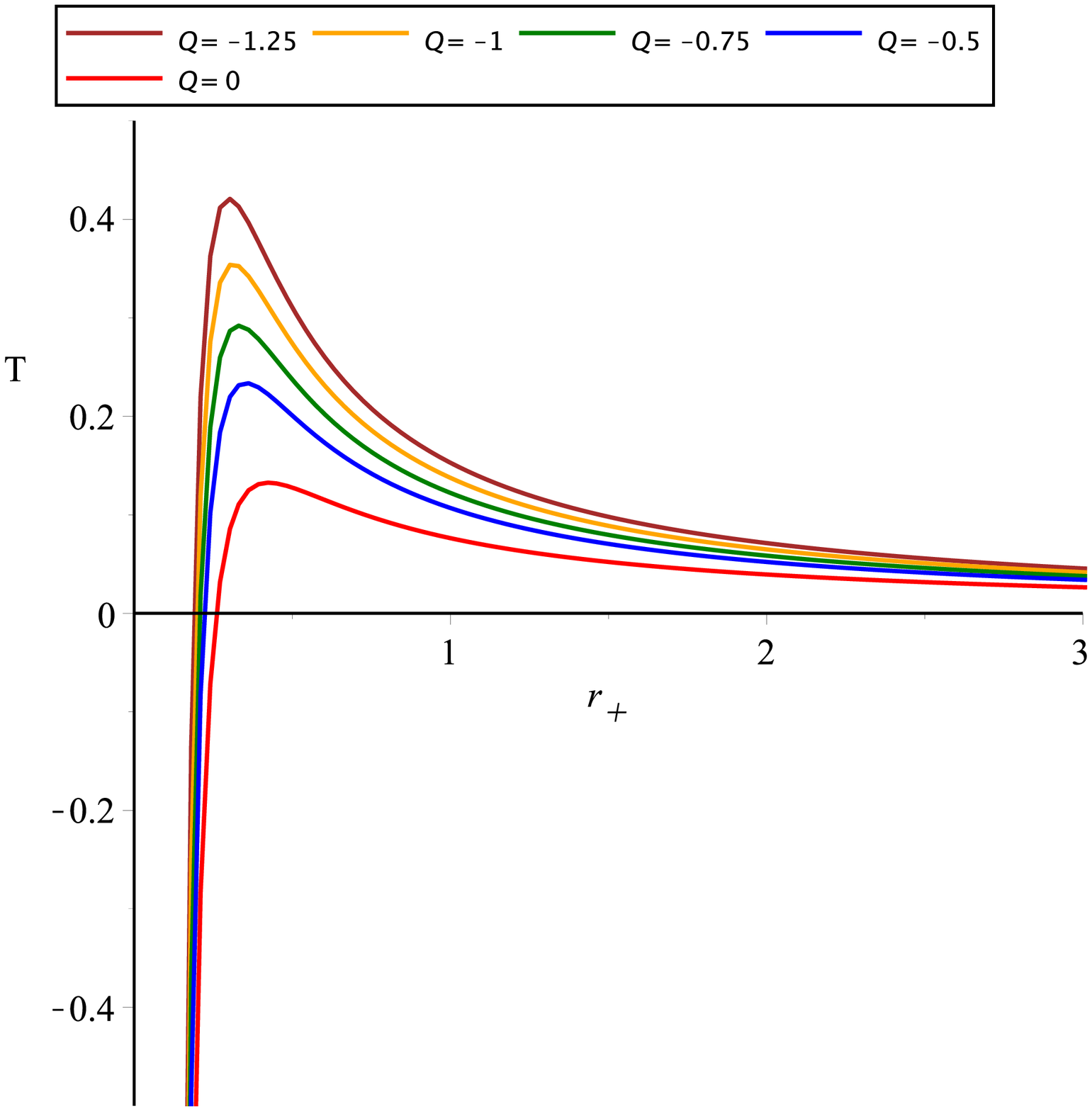}
}
	\subfigure[$Q=-0.25$, $a=0.15$]{
		\includegraphics[width=0.38\textwidth]{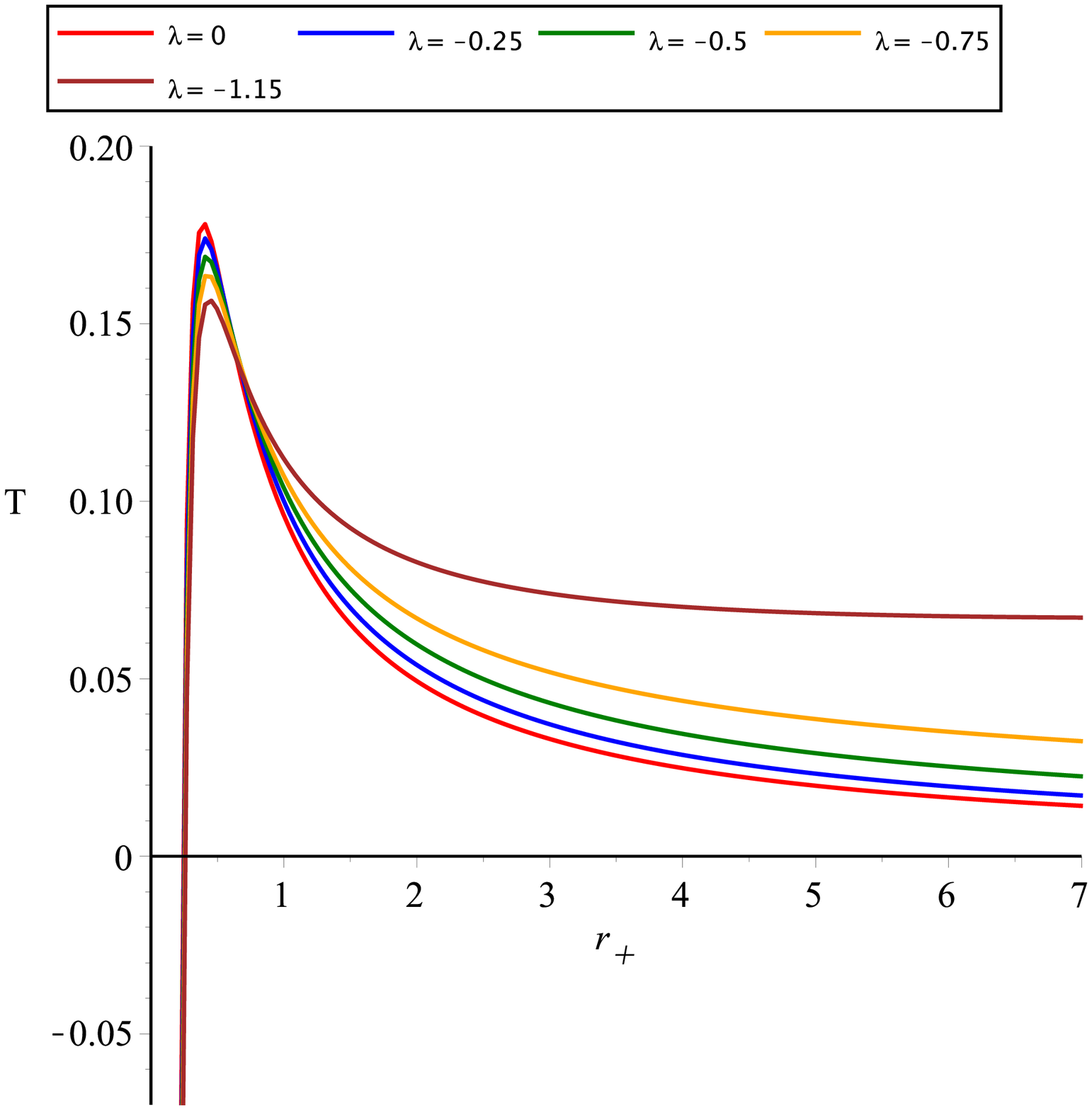}
	}
	\subfigure[$\lambda=-0.25$, $a=0.15$]{
		\includegraphics[width=0.38\textwidth]{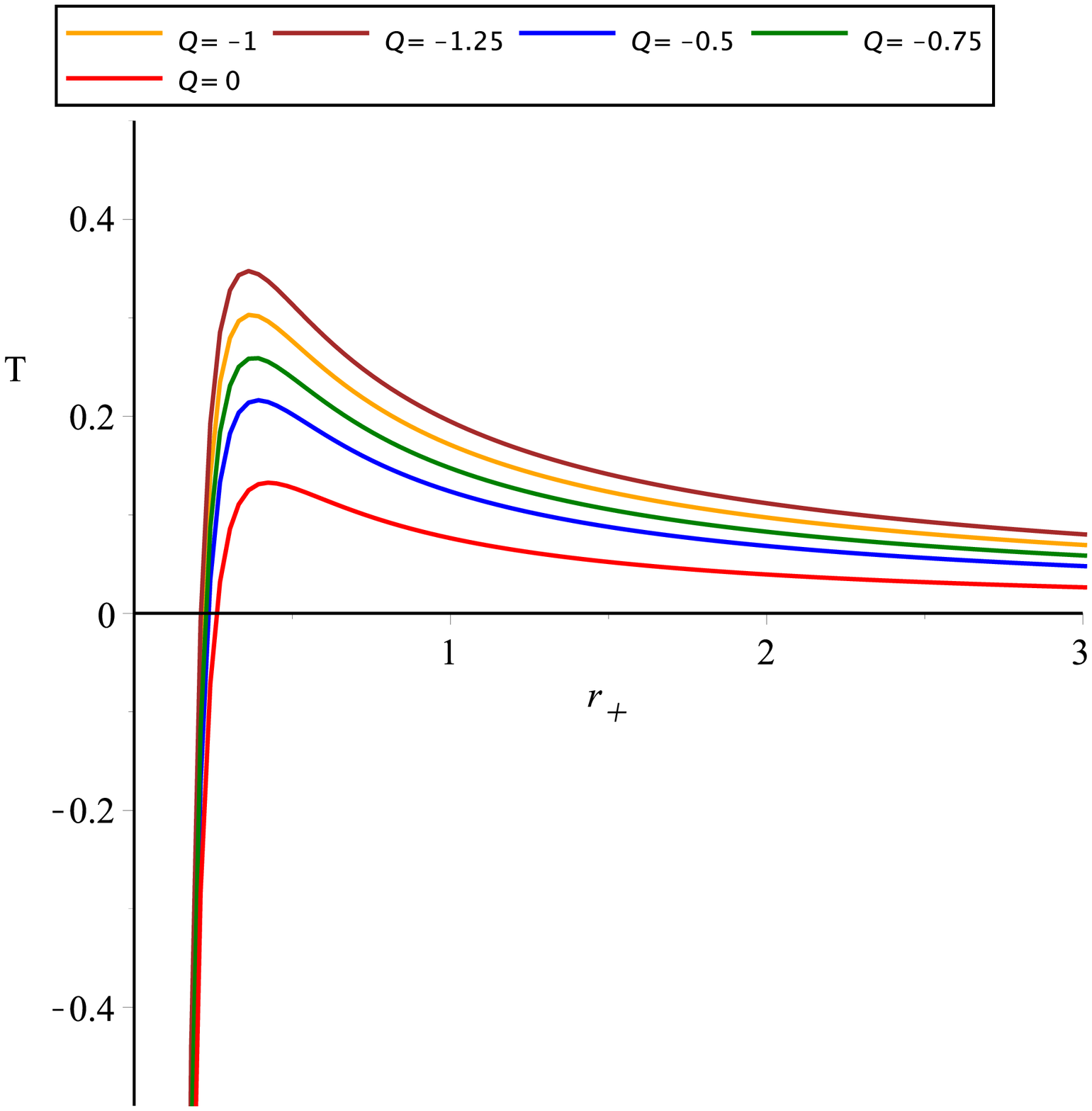}
	}
	\caption{{These figures show the evolution of temperature in terms of horizon radius $ r_{+}$, for a rotating regular black hole in CMG background. In figure (a) the behaviour of temperature when  $\lambda$ gets negative values and $Q>0$ is depicted. The plot (b) is illustrated when $Q<0$ varies and $\lambda $ is a positive constant. In plot (c) and (d) both the parameters $Q$ and $\lambda$ are negative. Interestingly for these plots the non-physical region does not appear.  }}
	\label{pic:TN}
\end{figure}

\begin{figure}[]
	\centering
\subfigure[$\lambda=+0.25$, $a=0.15$ ]{
	\includegraphics[width=5cm]{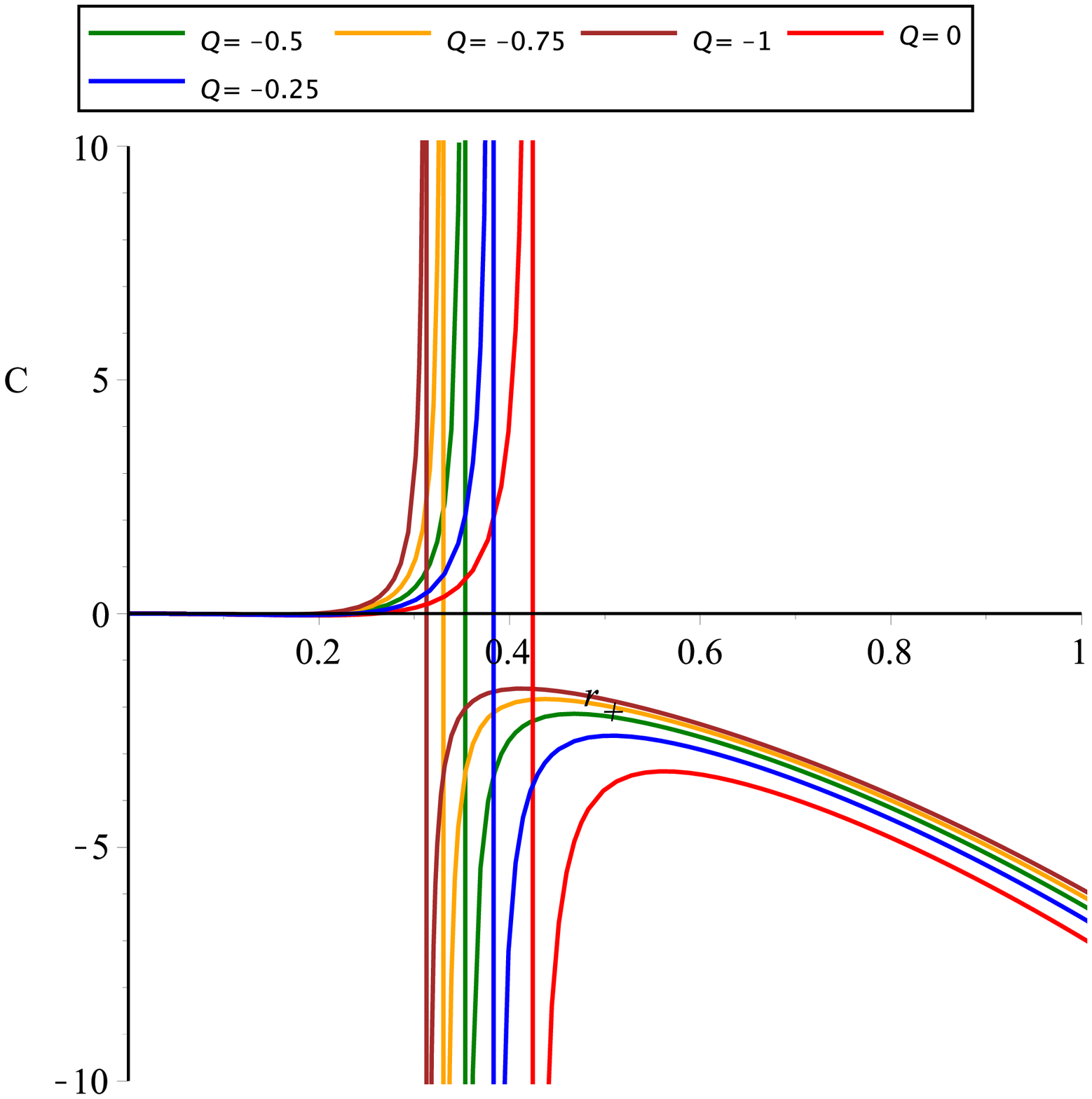}
}
	\subfigure[$Q=+0.25$, $a=0.15$]{
		\includegraphics[width=5cm]{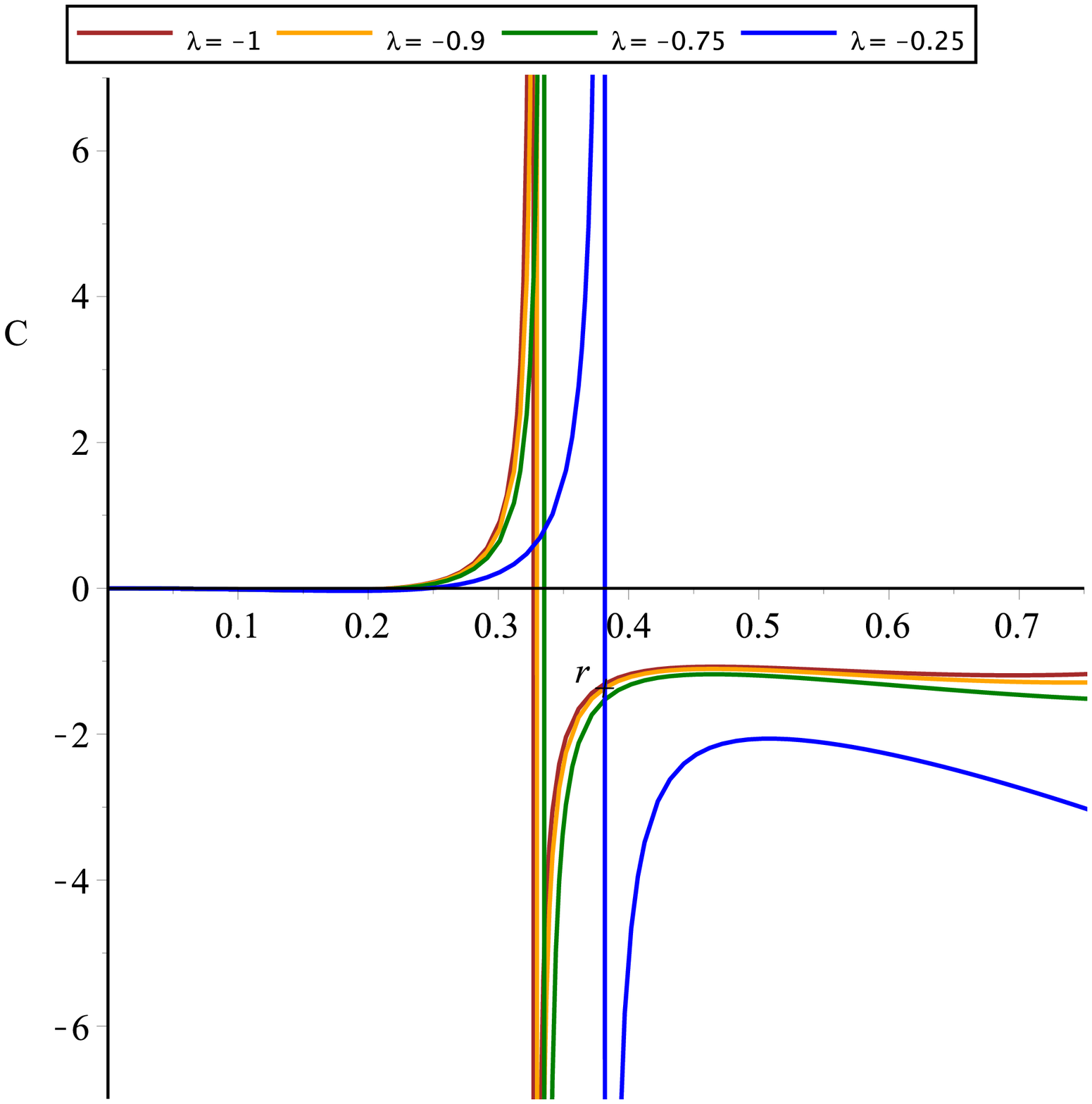}
	}
\subfigure[ $\lambda=-0.25$, $a=0.15$]{
	\includegraphics[width=5cm]{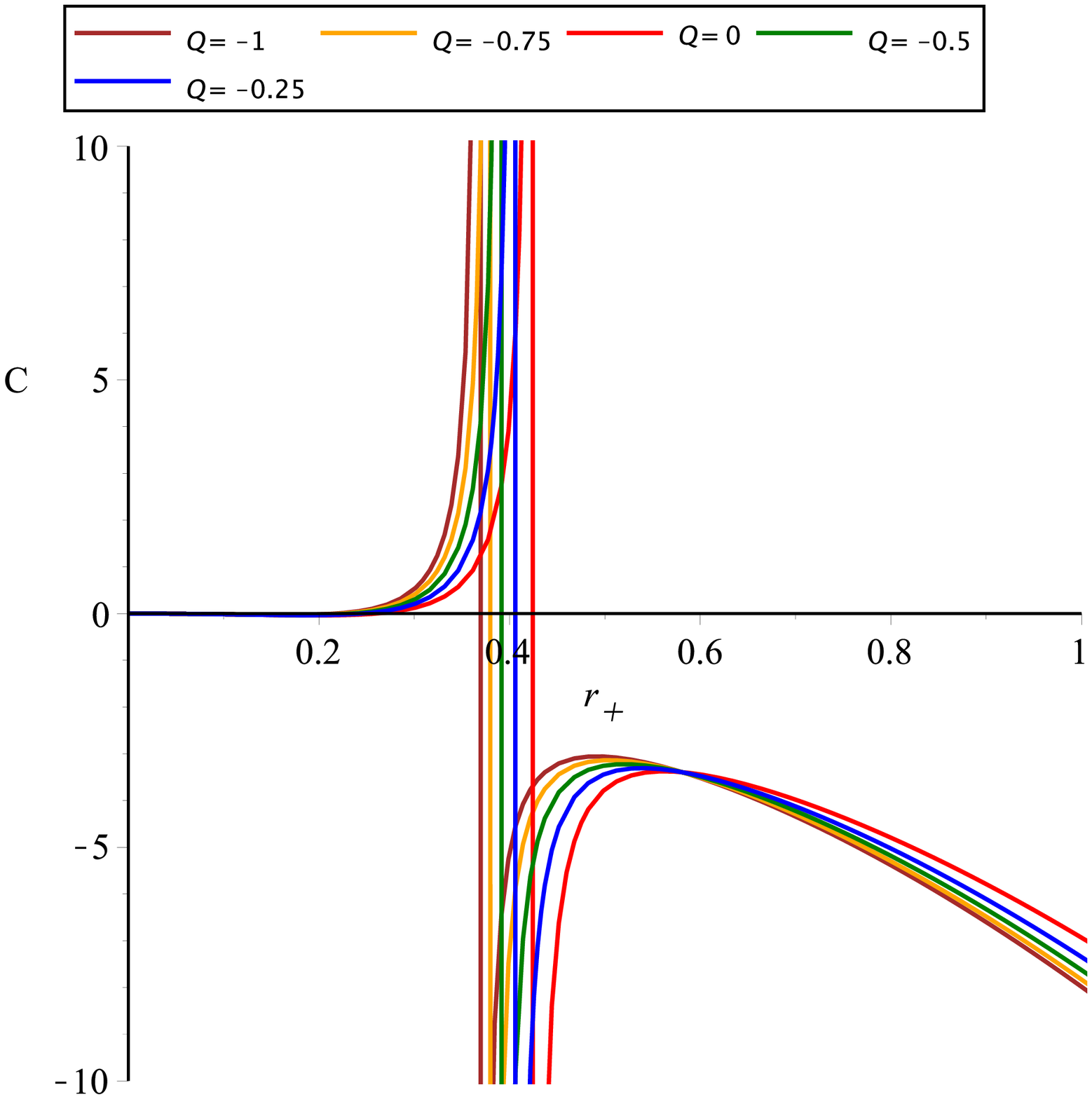}
}
\subfigure[$Q=-0.25$, $a=0.15$]{
	\includegraphics[width=5cm]{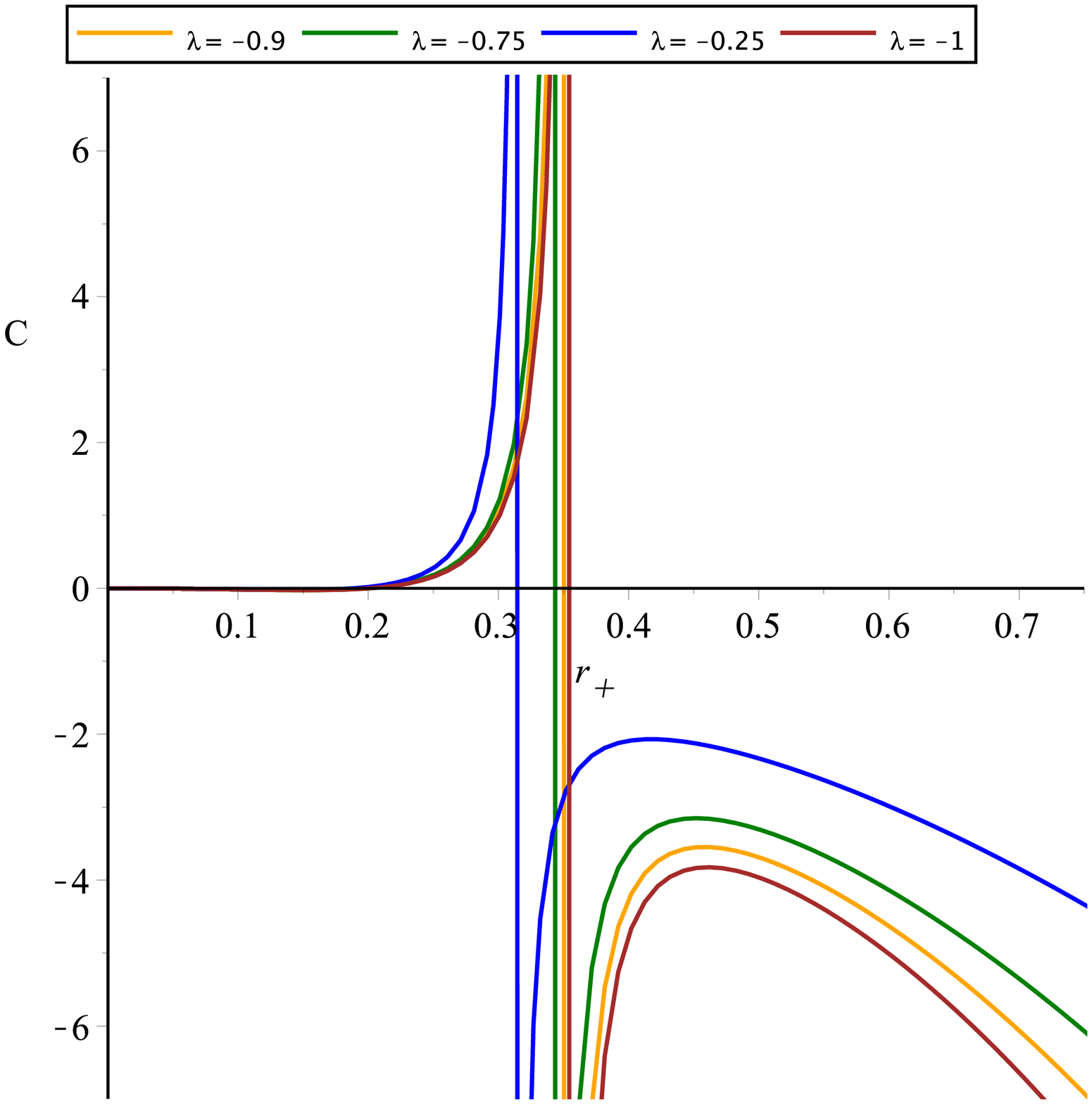}
}
	\subfigure[close up of diagrams (a)]{
		\includegraphics[width=5cm]{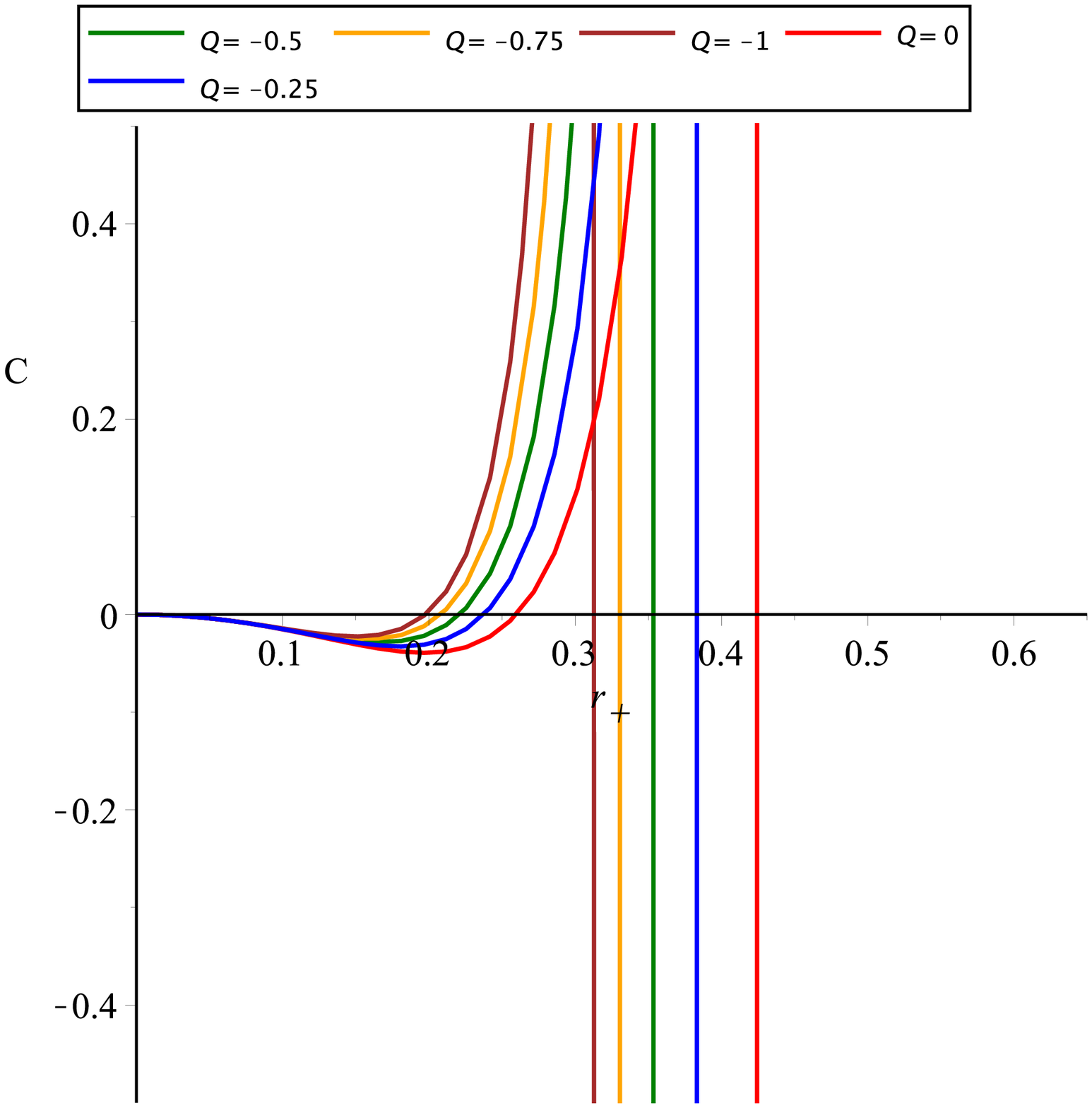}
	}
\subfigure[close up of diagrams (c)]{
	\includegraphics[width=5cm]{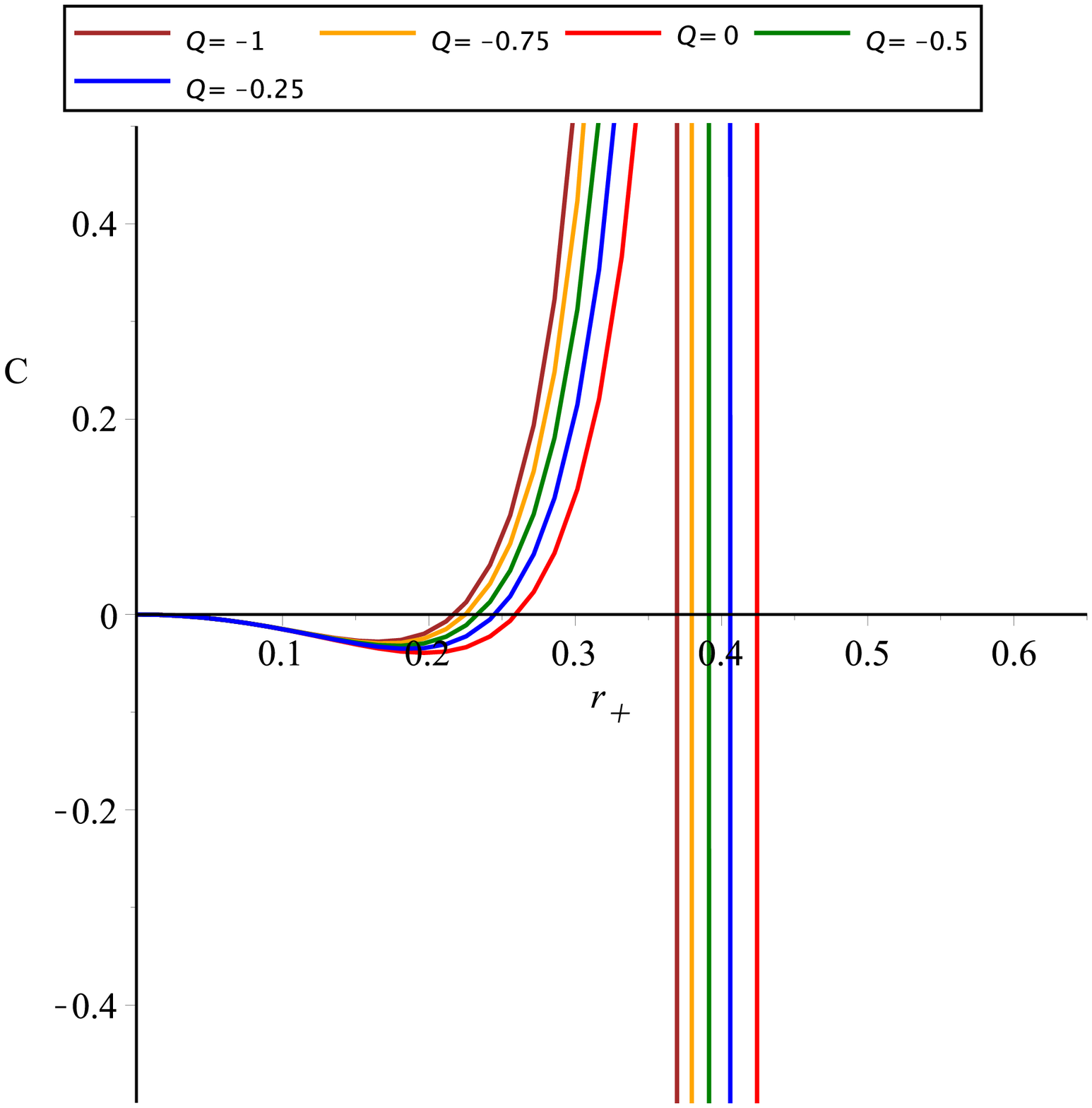}
}
\\
	\caption{{These figures indicate the variations of the heat capacity in terms of horizon radius $ r_{+}$, for a rotating regular black hole in the presence of the CMG corrections. In plot (a) the behaviour of heat capacity when  $Q$ gets negative values and $\lambda>0$, as a constant,  is demonstrated. The plot (b) is illustrated when $\lambda<0$ varies and $Q$  is a positive constant. In plot (c) and (d) both the parameters $Q$ and $\lambda$ are negative in which for (c) parameter $Q$ varies and $\lambda$ is a negative constant and vice versa for the diagram (d). Plots (e) and (f) are devoted to demonstrate a close up of diagrams (a) and (c) respectively. }}
	\label{pic:CN}
\end{figure}

\textcolor[rgb]{0.00,0.00,1.00}{\subsection{Investigating the rotating black hole for $Q<0$ and $\lambda<0$}\label{subrot2}}
{Here, we want to examine the effects of $Q<0$ and  $\lambda<0$ on the evolution of the parameters  mass, temperature and heat capacity for a rotating black hole in the presence of CMG corrections.
In doing so, the behaviour of the mass function is illustrated in Fig.~\eqref{pic:MN} and  it can be understood that for plots (a), (c) and (d) mass function gets only positive values and plot (b) in some cases can get negative values. Besides these, the behaviour of temperature and heat capacity are illustrated in Fig.~\eqref{pic:TN} and Fig.~\eqref{pic:CN}. In all plots (a) to (d) of  Fig.\eqref{pic:TN}, temperature behaviour is as same as the cases where $Q$ and $\lambda$ are positive. For a rotating regular black hole in the presence of CMG correction when one of the parameters $Q$ and $\lambda$ or both of them get negative values heat capacity shows phase transition interestingly.}

% \clearpage

\textcolor[rgb]{0.00,0.00,1.00}{\section{Discussion and Conclusions}\label{section4}}
{ }{In this manuscript, we studied the geometrothermodynamical properties of both static and rotating regular charged blackholes in the presence of the CMG corrections. In both cases, we calculated the thermodynamic quantities, mass $M$, the Hawking temperature $T$ and the heat capacity $C$, and the associated thermodynamic geometries of the black holes in the Weinhold, Ruppeiner, Quevedo and HPEM metric formalisms. The resulting quantities were plotted against the event horizon radii for chosen values of the defining model parameters.
In the case of the static black hole solutions of the CMG theory, $ M $, $ T $ and $ C $ of the system can be expressed in terms of the extensive parameters entropy, $S$, and  scalar charge, $Q$, and the hair parameter  $\lambda$. Figs.~\ref{pic:M} (a) and (b) showed that while the black hole mass has a minimum point and is an increasing function of \rt, the minimum point decreases with increasing $Q$ and $\lambda$. }

{In Fig.~\ref{pic:M} (a),  when $Q=0$, Schwarzschild-like solution, mass function gets only positive values, but by increasing the size of the black hole and for R-N black holes in the presence of the CMG corrections mass function accepts smaller negative values then enters to the positive regions. For Fig.~\ref{pic:M} (b), when hair parameter, $\lambda=0$  and $Q\neq0$,  mass function in the presence of a CMG correction shows a same behaviour, when $Q$ tends zero as same as case (a). Interestingly, it was found out that for a R-N black hole with a fixed amount of $Q$, when hair parameter varies then mass function is almost independent of variation of parameter $\lambda$ at $r_{+}\approx 1.$ {There was also  a physical interpretation for negative mass appeared in this model utilizing the quantum gravity. For instance in \cite{Mann:1997jb} the author showed that by means of quantum fluctuations some regions with negative energy can exist that under some certain conditions will undergo gravitational collapse to form a black hole. The exterior of such a black hole definitely has negative mass with a non-trivial topology.} }

{Figs.~\ref{pic:T}, (a), (b) and (c), on the other hand, showed that  Hawking temperature starts off as negative for a particular range of \rt, increases to a maximum positive value and eventually decreases with \rt. Increasing $Q$ and $\lambda$ decreases the peak values of $T$. }
{In Fig.~\ref{pic:T} (b), temperature was illustrated when $\lambda$ gets a fixed value when$Q$ can vary. When the case was a Schwarzschild-like black hole with CMG corrections, there is no any phase transition for such a configuration. By considering nonzero values for $Q$ , the behavior of temperature for larger black holes interestingly  showed a convergence to almost zero temperature which means as long as black holes become larger then they lost their temperature as well. Then by considering a fixed vale for the parameter $Q$, the impacts of the variation of the hair parameter, $\lambda$ on the evolution of temperature in Fig.~\ref{pic:T} (c) were examined . Interestingly, when $\lambda=0,$ the behavior of temperature was as same as Fig.~\ref{pic:T} (b), when metric is Schwarzschild-like, say when $Q=0$ in the presence of the CMG corrections. It obviously understood that for some particular cases, hair parameter, $\lambda,$ could act like parameter $Q$ . Some other interesting physical properties can be realized from Fig.~\ref{pic:T} (c), they are related to the peak of the Hawking temperature and the converging point of the temperature for larger black holes. By comparing Figs.~\ref{pic:T} (b) and (c) one immediately realizes that by decreasing the values of parameter $\lambda$ maximum points get larger amounts and also for larger black holes temperature will be larger than cases appeared in Fig.~\ref{pic:T} (b) . In addition to the above mentioned properties,  these plots showed that as long as the values of $Q$, and $\lambda$ are  increasing the peak value of temperature becomes smaller in its value.}

{In Figs.~\ref{pic:C}, (a-(f), it is shown that $C$ has one zero point and one divergent point corresponding to physical limitation and phase transition critical points, respectively. The negative and positive regions of $C$ shown correspond, respectively, to the unstable and stable phases of the black hole system. Changes in $Q$ and $\lambda$ shift the transition points along \rt, i.e. for larger black holes.
}
{In other words, for ${r_1} \simeq 0.01\leq{r_ + }\leq{r_2} \simeq 0.06$ heat capacity got negative values and it indicated that the black hole will be in an unstable phase, and then for ${r_ + }> 0.06$  it was entered the positive regions and the black hole will be in its stable phase. {And for its second phase transition one could look at Fig.~\ref{pic:C} (a),  in which it appears  at ${r_ + }\simeq 0.14$ indicating the phase transition critical point of the black hole.} Considering the evolution of heat capacity by emphasising on the effects of the  parameters $Q$ and $\lambda$ when they were varied separately, in Figs.~\ref{pic:C} (c), (d), immediately one can conclude that, when you have a Schwarzschild-like black hole, i.e. when $Q=0$, whereas heat capacity only gets negative values then it  completely is in an unstable phase.   By introducing nonzero values of $Q$ to the in question problem it behaves as same as Figs.~\ref{pic:C} (a), (b). In a similar procedure when $\lambda$ tends zero, Figs.~\ref{pic:C} (e), (f) followed  the discussed behaviour of Figs.~\ref{pic:C} (c), (d). In addition to these, from whole the Figs. \ref{pic:C} (c)-(f), it could be seen that by increasing the values of $Q$, and $\lambda$, the phase transition points of the system are shifting along for larger black holes.}

{Figs. \ref{pic:CRupHPEM} (a)-(d) depicted the curvature scalar $R$ and $C$ variations versus \rt. Whereas the divergence of $R$ in the Ruppeiner metric coincided only with zero points of the heat capacity, physical limitation point, and had no point corresponding to divergence point of $C$, phase transition critical points. But, as can be seen from Figs.~\ref{pic:CRupHPEM} (c), (d), intriguingly  the divergence points of the Ricci scalar of the HPEM metric, is in concordance with both the zero point and the divergence point  of heat capacity. In other words, the divergence points of the Ricci scalar of the HPEM metric capable with both the first and the second phase transitions. Therefore, it was figured out  that one can achieve more physical information considering HPEM formalism instead of Ruppeiner or other aforementioned metrics.}
{
As a next step by considering a regular rotating charge black hole, a same investigation was demonstrated. Obviously in this in question black hole, the thermodynamical quantities depend not only on $S$, $Q$, and $\lambda$, but also on the spin parameter $a$. }

{The behavior of mass function for different values of $Q$, $\lambda$, and $ a $ was appeared in Fig.~\ref{pic:MR}. From this figure, it has been shawn that the mass function of such a black hole contains a minimum point which for a Kerr like case it took the largest value then by increasing the values of parameter charge, this turning point got smaller values. Moreover, as can be seen from Figs.~\ref{pic:MR} (a), and (b) by increasing the value of $Q$, $\lambda$, for a fixed spin parameter, the values of this minimum point showed a decreasing behavior. Contrasted to these two cases, Figs.~\ref{pic:MR} (a), and (b), in  Fig.~\ref{pic:MR} (c), one could realize that by increasing the values of $a$, and for fixed $Q$ and $\lambda$, this minimum point behaved as an increasing function. It may went back to the effect of the increasing of the angular momentum on the behavior of the black hole. In these cases, compared to static black holes in primary sections, almost diagrams are in a positive region, and interestingly for larger black holes mass function converges to non-rotating cases with $a=0$, for more detail see Fig.~\ref{pic:MR} (c).}

{ Figs.~\ref{pic:TR} (a, b), and (c) showed that $T$ started off as negative for a particular range of \rt, increases to a maximum positive value and eventually decreases with \rt.  From Figs.~\ref{pic:TR} (a) and (c), it was realized  that as long as the values of $Q$, and $a$ are increasing, the peak value of temperature got smaller values. But, Fig.~\ref{pic:TR} (b) showed that the peak value of temperature obtains larger values, comparing to the two aforementioned cases, by increasing the value of $\lambda$. In addition to the above properties, the black hole temperature decreased for larger black holes and it almost converge to zero temperature.}

{  In Fig.~\ref{pic:CR},  (a)-(g) it is shown that $C$ has one zero point and one divergent point corresponding to physical limitation and phase transition critical points, respectively. The negative and positive regions of $C$ shown correspond, respectively, to the unstable and stable phases of the black hole system. From Figs.~\ref{pic:CR} (a) and (b), immediately it was figured out that for the interval  $r_{1} \simeq 0.05<r_{+}<r_{2}\simeq 0.29$, heat capacity got negative values and therefore black hole was in an unstable phase. Then for $r_{2} <r_{+} $, it entered a positive region which means that the black hole will be in a stable phase. Additionally, the phase transition critical point was located at $r_+\simeq 0.49$. Then, one could study the effects of varying of parameter $Q$ on the evolution of heat capacity when $\lambda$ and $a$ are considered to be fixed in Figs.~\ref{pic:CR} (c)-(d).  These plots showed that, by increasing the amount of parameter $Q$ the physical limitation point and the phase transition critical point  shifted to the larger values of the event horizon radii, say for larger black holes.
 From Figs.~\ref{pic:CR} (e)-(f), one observed that by considering the evolution of parameter $\lambda$, and  fixing parameters $Q$ and $ a $, the location of phase transition points of the system is shifting along horizon radii. Ultimately, seeing the impact of the variation of parameter $a$ on the evolution of the black hole one can consider the plot Fig.~\ref{pic:CR} (g).}

{ Fig.~\ref{pic:CRWeinQueHPEM} (a), indicated that the singular point of the curvature scalar of the Weinhold metric does not coincide with any of zeros and divergence points of the heat capacity. Figs.~\ref{pic:CRWeinQueHPEM} (b), (c), showed  that the singular point of the curvature scalar of the Ruppeiner metric were exactly compatible with the zero, root, of the heat capacity. However, there was no any correspondence between divergence points of the Ruppeiner metric and the heat capacity. Nevertheless, according to Figs.~\ref{pic:CRWeinQueHPEM} (d), (e), one saw that divergence points of the Ricci scalar of the HPEM metric, oddly coincided with both zero and divergence points of the heat capacity. Therefore, clearly more information for the thermodynamic phase transition from the HPEM metric compared to the Ruppeiner metric and the other methods mentioned in this paper could be achieved.}\\

{Since there is not necessarily a positive constraint on the parameters $Q$ and $\lambda$, we examined the cases in which these parameters also accept negative values. The result of these studies appeared for static black holes in Figs. \eqref{pic:MNS} to  \eqref{pic:CNS} for the function of mass, temperature and heat capacity. We have also tried such studies for the rotating black hole and the results are drawn in the  Figs. \eqref{pic:MN} to  \eqref{pic:CN}. As mentioned, in this case, for static black holes, temperature and mass have interesting physical behavior, and for rotating black holes, mass and heat capacity have an interesting behavior, compared to the case where the scalar charge sign and the hair parameter are positive.}

{The effects of the presence of a nonminimal coupling between the scalar curvature and scalar field may contain some interesting physical results {which can be the subject of future work}.}\\
\\

\textcolor[rgb]{0.00,0.00,1.00}{\section*{Acknowledgments}}
{The authors need to express their appreciation of the critical review made by the anonymous reviewer which for sure improved the clarity and quality of this work. H.S. is grateful to Y. Sobouti who basically taught him the physics of the black holes. H.S. also thanks T. Harko and  H. Firouzjahi for constructive discussions on black holes and perturbations.}
AA acknowledges that this work is based on the research supported in part by the NRF of South Africa with grant number 112131.\\


\begin{thebibliography}{}
%LIGO-EHT

%\cite{Barish:1999vh}
\bibitem{Barish:1999vh}
B.~C.~Barish and R.~Weiss,
``LIGO and the detection of gravitational waves,''
Phys. Today \textbf{52N10}, 44-50 (1999)
doi:10.1063/1.882861
%39 citations counted in INSPIRE as of 08 Jul 2020


%\cite{TheLIGOScientific:2016agk}
\bibitem{TheLIGOScientific:2016agk}
B.~P.~Abbott \textit{et al.} [LIGO Scientific and Virgo],
``GW150914: The Advanced LIGO Detectors in the Era of First Discoveries,''
Phys. Rev. Lett. \textbf{116}, no.13, 131103 (2016)
doi:10.1103/PhysRevLett.116.131103
[arXiv:1602.03838 [gr-qc]].
%298 citations counted in INSPIRE as of 08 Jul 2020


%\cite{Abbott:2016izl}
\bibitem{Abbott:2016izl}
T.~D.~Abbott \textit{et al.} [LIGO Scientific and Virgo],
``Improved analysis of GW150914 using a fully spin-precessing waveform Model,''
Phys. Rev. X \textbf{6}, no.4, 041014 (2016)
doi:10.1103/PhysRevX.6.041014
[arXiv:1606.01210 [gr-qc]].
%102 citations counted in INSPIRE as of 08 Jul 2020


%\cite{Lovelace:2016uwp}
\bibitem{Lovelace:2016uwp}
G.~Lovelace, C.~O.~Lousto, J.~Healy, M.~A.~Scheel, A.~Garcia, R.~O'Shaughnessy, M.~Boyle, M.~Campanelli, D.~A.~Hemberger, L.~E.~Kidder, H.~P.~Pfeiffer, B.~Szilagyi, S.~A.~Teukolsky and Y.~Zlochower,
``Modeling the source of GW150914 with targeted numerical-relativity simulations,''
Class. Quant. Grav. \textbf{33}, no.24, 244002 (2016)
doi:10.1088/0264-9381/33/24/244002
[arXiv:1607.05377 [gr-qc]].
%44 citations counted in INSPIRE as of 08 Jul 2020


%\cite{Calloni:2017whl}
\bibitem{Calloni:2017whl}
E.~Calloni [LIGO Scientific and Virgo],
``Introduction to gravitational wave detection and Advanced Virgo Status and perspectives,''
Nucl. Part. Phys. Proc. \textbf{291-293}, 127-133 (2017)
doi:10.1016/j.nuclphysbps.2017.06.026
%0 citations counted in INSPIRE as of 08 Jul 2020


%EHT


%\cite{Akiyama:2019cqa}
\bibitem{Akiyama:2019cqa}
K.~Akiyama \textit{et al.} [Event Horizon Telescope],
``First M87 Event Horizon Telescope Results. I. The Shadow of the Supermassive Black Hole,''
Astrophys. J. \textbf{875}, no.1, L1 (2019)
doi:10.3847/2041-8213/ab0ec7
[arXiv:1906.11238 [astro-ph.GA]].
%530 citations counted in INSPIRE as of 08 Jul 2020



%\cite{Akiyama:2019brx}
\bibitem{Akiyama:2019brx}
K.~Akiyama \textit{et al.} [Event Horizon Telescope],
``First M87 Event Horizon Telescope Results. II. Array and Instrumentation,''
Astrophys. J. Lett. \textbf{875}, no.1, L2 (2019)
doi:10.3847/2041-8213/ab0c96
[arXiv:1906.11239 [astro-ph.IM]].
%139 citations counted in INSPIRE as of 08 Jul 2020



%\cite{Akiyama:2019sww}
\bibitem{Akiyama:2019sww}
K.~Akiyama \textit{et al.} [Event Horizon Telescope],
``First M87 Event Horizon Telescope Results. III. Data Processing and Calibration,''
Astrophys. J. Lett. \textbf{875}, no.1, L3 (2019)
doi:10.3847/2041-8213/ab0c57
[arXiv:1906.11240 [astro-ph.GA]].
%130 citations counted in INSPIRE as of 08 Jul 2020



%\cite{Akiyama:2019bqs}
\bibitem{Akiyama:2019bqs}
K.~Akiyama \textit{et al.} [Event Horizon Telescope],
``First M87 Event Horizon Telescope Results. IV. Imaging the Central Supermassive Black Hole,''
Astrophys. J. Lett. \textbf{875}, no.1, L4 (2019)
doi:10.3847/2041-8213/ab0e85
[arXiv:1906.11241 [astro-ph.GA]].
%174 citations counted in INSPIRE as of 08 Jul 2020



%\cite{Akiyama:2019fyp}
\bibitem{Akiyama:2019fyp}
K.~Akiyama \textit{et al.} [Event Horizon Telescope],
``First M87 Event Horizon Telescope Results. V. Physical Origin of the Asymmetric Ring,''
Astrophys. J. Lett. \textbf{875}, no.1, L5 (2019)
doi:10.3847/2041-8213/ab0f43
[arXiv:1906.11242 [astro-ph.GA]].
%216 citations counted in INSPIRE as of 08 Jul 2020

%\cite{Akiyama:2019eap}
\bibitem{Akiyama:2019eap}
K.~Akiyama \textit{et al.} [Event Horizon Telescope],
``First M87 Event Horizon Telescope Results. VI. The Shadow and Mass of the Central Black Hole,''
Astrophys. J. Lett. \textbf{875}, no.1, L6 (2019)
doi:10.3847/2041-8213/ab1141
[arXiv:1906.11243 [astro-ph.GA]].
%237 citations counted in INSPIRE as of 08 Jul 2020





%%%%%%%%%%%%%%%%%%%%%%%%%%%%%%%%%%%%
%%%%%%%%%%%%%%%%%%%%%%%%%%%%%%%%%%%%


%{%\cite{Michell:1784xqa}
%\bibitem{Michell:1784xqa}
%J.~Michell,``On the Means of Discovering the Distance, Magnitude, \&c. of the Fixed Stars, in Consequence of the Diminution of the Velocity of Their Light, in Case Such a Diminution Should be Found to Take Place in any of Them, and Such Other Data Should be Procured from Observations, as Would be Farther Necessary for That Purpose.,''
%Phil. Trans. Roy. Soc. Lond. \textbf{74}, 35-57 (1784)
%doi:10.1098/rstl.1784.0008 %78 citations counted in INSPIRE as of 08 Jul 2020}



%Alber Einstein Papers

\bibitem{1912a} A. Einstein, ‘Lichtgeschwindigkeit und Statik des Gravitationsfeldes’. Annalen der Physik, \textbf{38},
355–369 [CPAE 4, 129–145] ( 1912).
\bibitem{1912b} A. Einstein,  ‘Zur Theorie des statischen Gravitationsfeldes’. ibid., \textbf{38}, 443–458
[CPAE 4, 146–164] ( 1912).
\bibitem{1914a} A. Einstein,  ‘Die formale Grundlage der allgemeinen Relativit\"{a}tstheorie’. Koniglich
Preußische Akademie der Wissenschaften (Berlin). Sitzungsberichte, 1030–1085
[CPAE 6, 72–130] ( 1914).
\bibitem{1915a} A. Einstein,  ‘Zur allgemeinen Relativit\"{a}tstheorie’. ibid., 778–786 [CPAE 6, 214–
224] ( 1915).
\bibitem{1915b} A. Einstein,  ‘Zur allgemeinen Relativit\"{a}tstheorie (Nachtrag)’. ibid., 799–801
[CPAE 6, 225–229] (1915).
\bibitem{1915c} A. Einstein,  ‘Erkl\"{a}rung der Perihelbewegung des Merkur aus der allgemeinen
Relativit\"{a}tstheorie’. ibid., 831–839 [CPAE 6, 233–243] ( 1915).
\bibitem{1915d} A. Einstein,  ‘Die Feldgleichungen der Gravitation’. ibid., 844–847 [CPAE 6, 244–
249] ( 1915).
\bibitem{1916a} A. Einstein, ‘Die Grundlage der allgemeinen Relativit\"{a}tstheorie’. Annalen der
Physik, \textbf{49}, 769–822 (also published separately as Leipzig: Teubner) [CPAE 6,
283–339] ( 1916).

%Schwarzschild, K

\bibitem{Schw1916}K. Schwarzschild, "Über das Gravitationsfeld eines Massenpunktes nach der Einsteinschen Theorie". Sitzungsberichte der Königlich Preussischen Akademie der Wissenschaften. \textbf{7}: 189–196 (1916).
%%%%%%%%%%%%%%%%%%%%%%%%%%%%%%%%%%%%%%%%%%%%%%%%%%%%%%%%%%%%%%%%%%%%%%%%%%%%%%%%%%%%%%%%%%%%%%%%%%%%%%%%%%%%%%
%%%%%%%%%%%%%%%%%%%%%%%%%%%%%%%%%%%%%%%%%%%%%%%%%%%%%%%%%%%%%%%%%%%%%%%%%%%%%%%%%%%%%%%%%%%%%%%%%%%%%%%%%%%%%%

\bibitem{Johannes1917} J. Droste,  "On the field of a single centre in Einstein's theory of gravitation, and the motion of a particle in that field", Proceedings Royal Academy Amsterdam.\textbf{ 19} (1): 197–215 (1917).
\bibitem{Edington1920}  A. S. Eddington, "The Internal Constitution of the Stars", \textbf{52}, Issue 1341, pp. 233-240 DOI: 10.1126/science.52.1341.233, (1920).
\bibitem{Edington1924} A. Eddington, "A Comparison of Whitehead's and Einstein's Formulæ", Nature \textbf{113}, 192, https://doi.org/10.1038/113192a0,  (1924).
\bibitem{tHoof2009} G. 't Hooft,   "Introduction to the Theory of Black Holes",  Institute for Theoretical Physics / Spinoza Institute: 47–48. Archived from the original on 21 May 2009, (2009).
\bibitem{Throne1994} K. Thorne, "Black Holes and Time Warps: Einstein's Outrageous Legacy", W. W. Norton \& Company, ISBN $978-0-393-31276-8$, (1994).

 \bibitem{Reissner(1916)} H.  Reissner, (1916). "\"{U}ber die Eigengravitation des elektrischen Feldes nach der Einsteinschen Theorie". Annalen der Physik (in German). 50 (9): 106–120, doi:10.1002/andp.19163550905, (1916).
 \bibitem{Weyl(1917)} H. Weyl,  (1917). "Zur Gravitationstheorie". Annalen der Physik (in German). 54 (18): 117–145, doi:10.1002/andp.19173591804, (1917).
 \bibitem{Nordstron(1918)} G. Nordstr\"{o}m, (1918). "On the Energy of the Gravitational Field in Einstein's Theory". Verhandl. Koninkl. Ned. Akad. Wetenschap., Afdel. Natuurk., Amsterdam. 26: 1201–1208, (1918)
\bibitem{Jeffery(1921)} G. B.  Jeffery,  "The field of an electron on Einstein's theory of gravitation". Proc. Roy. Soc. Lond. A. 99 (697): 123–134, doi:10.1098/rspa.1921.0028, (1921).




%\cite{Kerr:1963ud}
\bibitem{Kerr:1963ud}
R.~P.~Kerr,
``Gravitational field of a spinning mass as an example of algebraically special metrics,''
Phys. Rev. Lett. \textbf{11}, 237-238 (1963)
doi:10.1103/PhysRevLett.11.237
%1516 citations counted in INSPIRE as of 08 Jul 2020


%\cite{Newman:1965my}
\bibitem{Newman:1965my}
E.~T.~Newman, R.~Couch, K.~Chinnapared, A.~Exton, A.~Prakash and R.~Torrence,
``Metric of a Rotating, Charged Mass,''
J. Math. Phys. \textbf{6}, 918-919 (1965)
doi:10.1063/1.1704351
%577 citations counted in INSPIRE as of 08 Jul 2020





%\cite{Israel:1967wq}
\bibitem{Israel:1967wq}
W.~Israel,
``Event horizons in static vacuum space-times,''
Phys. Rev. \textbf{164}, 1776-1779 (1967)
doi:10.1103/PhysRev.164.1776
%753 citations counted in INSPIRE as of 09 Jul 2020

%\cite{Carter:1971zc}
\bibitem{Carter:1971zc}
B.~Carter,
``Axisymmetric Black Hole Has Only Two Degrees of Freedom,''
Phys. Rev. Lett. \textbf{26}, 331-333 (1971)
doi:10.1103/PhysRevLett.26.331
%739 citations counted in INSPIRE as of 09 Jul 2020



%\cite{Robinson:1975bv}
\bibitem{Robinson:1975bv}
D.~C.~Robinson,
``Uniqueness of the Kerr black hole,''
Phys. Rev. Lett. \textbf{34}, 905-906 (1975)
doi:10.1103/PhysRevLett.34.905
%528 citations counted in INSPIRE as of 09 Jul 2020



%\cite{Chrusciel:2012jk}
\bibitem{Chrusciel:2012jk}
P.~T.~Chrusciel, J.~Lopes Costa and M.~Heusler,
``Stationary Black Holes: Uniqueness and Beyond,''
Living Rev. Rel. \textbf{15}, 7 (2012)
doi:10.12942/lrr-2012-7
[arXiv:1205.6112 [gr-qc]].
%239 citations counted in INSPIRE as of 09 Jul 2020



\bibitem{Belinskii1969} V.A. Belinskii, I. M. Khalatnikov, "On the Nature of the Singularities in the General Solution of the Gravitational Equations"- Sov. Phys. JETP, \textbf{29}, 911 (1969).

%\cite{Belinsky:1970ew}
\bibitem{Belinsky:1970ew}
V.~A.~Belinsky, I.~M.~Khalatnikov and E.~M.~Lifshitz,
``Oscillatory approach to a singular point in the relativistic cosmology,''
Adv. Phys. \textbf{19}, 525-573 (1970)
doi:10.1080/00018737000101171
%876 citations counted in INSPIRE as of 09 Jul 2020




%\cite{Penrose:1964wq}
\bibitem{Penrose:1964wq}
R.~Penrose,
``Gravitational collapse and space-time singularities,''
Phys. Rev. Lett. \textbf{14}, 57-59 (1965)
doi:10.1103/PhysRevLett.14.57
%1009 citations counted in INSPIRE as of 09 Jul 2020



\bibitem{Hawking1973} S. W. Hawking,  and G. F. R.  Ellis, The large scale structure of space-time, Cambridge University Press, Cambridge, UK,  (1973).

%\cite{Ford:2003qt}
\bibitem{Ford:2003qt}
L.~H.~Ford,
``The Classical singularity theorems and their quantum loop holes,''
Int. J. Theor. Phys. \textbf{42}, 1219-1227 (2003)
doi:10.1023/A:1025754515197
[arXiv:gr-qc/0301045 [gr-qc]].
%18 citations counted in INSPIRE as of 09 Jul 2020



%\cite{Hawking:1969sw}
\bibitem{Hawking:1969sw}
S.~W.~Hawking and R.~Penrose,
``The Singularities of gravitational collapse and cosmology,''
Proc. Roy. Soc. Lond. A \textbf{A314}, 529-548 (1970)
doi:10.1098/rspa.1970.0021
%832 citations counted in INSPIRE as of 09 Jul 2020










%\cite{Penrose:1969pc}
\bibitem{Penrose:1969pc}
R.~Penrose,
``Gravitational collapse: The role of general relativity,''
Riv. Nuovo Cim. \textbf{1}, 252-276 (1969)
doi:10.1023/A:1016578408204
%1117 citations counted in INSPIRE as of 09 Jul 2020


\bibitem{Christodoulou1999} D. Christodoulou, "The Instability of Naked Singularities in the Gravitational Collapse of a Scalar Field." Annals of Mathematics, Second Series, 149, no. 1, 183-217 (1999) doi:10.2307/121023.

\bibitem{Bekenstein:1972tm}
J.~D.~Bekenstein,
``Black holes and the second law,''
Lett. Nuovo Cim. \textbf{4}, 737-740 (1972)
doi:10.1007/BF02757029
%1032 citations counted in INSPIRE as of 09 Jul 2020

%\cite{Bardeen:1973gs}
\bibitem{Bardeen:1973gs}
J.~M.~Bardeen, B.~Carter and S.~W.~Hawking,
``The Four laws of black hole mechanics,''
Commun. Math. Phys. \textbf{31}, 161-170 (1973)
doi:10.1007/BF01645742
%2189 citations counted in INSPIRE as of 09 Jul 2020



%\cite{Hawking:1974rv}
\bibitem{Hawking:1974rv}
S.~W.~Hawking,
``Black hole explosions,''
Nature \textbf{248}, 30-31 (1974)
doi:10.1038/248030a0
%3170 citations counted in INSPIRE as of 09 Jul 2020

%\cite{Davies:1978zz}
\bibitem{Davies:1978zz}
P.~C.~W.~Davies,
``Thermodynamics of black holes,''
Rept. Prog. Phys. \textbf{41}, 1313-1355 (1978)
doi:10.1088/0034-4885/41/8/004
%115 citations counted in INSPIRE as of 09 Jul 2020


\bibitem{Hut:1977zx} P. Hut, "Charged black holes and phase transitions," Monthly Notices of the Royal Astronomical Society,  \textbf{180},  Issue 3, 379–389 (1977), doi: 10.1093/mnras/180.3.379


%\cite{Sokolowski:1980uva}
\bibitem{Sokolowski:1980uva}
L.~M.~Sokolowski and P.~Mazur,
``Second-order phase transitions in black-hole thermodynamics,''
J. Phys. \textbf{13}, A1113-1120 (1980)
doi:10.1088/0305-4470/13/3/043
%7 citations counted in INSPIRE as of 09 Jul 2020


%\cite{Cai:1997cs}
\bibitem{Cai:1997cs}
R.~G.~Cai,
``Effective spatial dimension of extremal nondilatonic black p-branes and the description of entropy on the world volume,''
Phys. Rev. Lett. \textbf{78}, 2531-2534 (1997)
doi:10.1103/PhysRevLett.78.2531
[arXiv:hep-th/9702142 [hep-th]].
%8 citations counted in INSPIRE as of 09 Jul 2020



%\cite{Maldacena:1997re}
\bibitem{Maldacena:1997re}
J.~M.~Maldacena,
``The Large N limit of superconformal field theories and supergravity,''
Int. J. Theor. Phys. \textbf{38}, 1113-1133 (1999)
doi:10.1023/A:1026654312961
[arXiv:hep-th/9711200 [hep-th]].
%15780 citations counted in INSPIRE as of 09 Jul 2020





%\cite{Maldacena:1997zz}
\bibitem{Maldacena:1997zz}
J.~M.~Maldacena,
``The Large N limit of superconformal field theories and supergravity,''
AIP Conf. Proc. \textbf{484}, no.1, 51 (1999)
doi:10.1063/1.59653
%33 citations counted in INSPIRE as of 09 Jul 2020




%\cite{Aharony:1999ti}
\bibitem{Aharony:1999ti}
O.~Aharony, S.~S.~Gubser, J.~M.~Maldacena, H.~Ooguri and Y.~Oz,
``Large N field theories, string theory and gravity,''
Phys. Rept. \textbf{323}, 183-386 (2000)
doi:10.1016/S0370-1573(99)00083-6
[arXiv:hep-th/9905111 [hep-th]].
%4805 citations counted in INSPIRE as of 10 Jul 2020

%\cite{Shen:2005nu}
\bibitem{Shen:2005nu}
J.~y.~Shen, R.~G.~Cai, B.~Wang and R.~K.~Su,
``Thermodynamic geometry and critical behavior of black holes,''
Int. J. Mod. Phys. A \textbf{22}, 11-27 (2007)
doi:10.1142/S0217751X07034064
[arXiv:gr-qc/0512035 [gr-qc]].
%133 citations counted in INSPIRE as of 10 Jul 2020



%\cite{Witten:1998zw}
\bibitem{Witten:1998zw}
E.~Witten,
``Anti-de Sitter space, thermal phase transition, and confinement in gauge theories,''
Adv. Theor. Math. Phys. \textbf{2}, 505-532 (1998)
doi:10.4310/ATMP.1998.v2.n3.a3
[arXiv:hep-th/9803131 [hep-th]].
%3089 citations counted in INSPIRE as of 10 Jul 2020

%\cite{Hawking:1982dh}
\bibitem{Hawking:1982dh}
S.~W.~Hawking and D.~N.~Page,
``Thermodynamics of Black Holes in anti-De Sitter Space,''
Commun. Math. Phys. \textbf{87}, 577 (1983)
doi:10.1007/BF01208266
%1777 citations counted in INSPIRE as of 10 Jul 2020


%\cite{Chamblin:1999tk}
\bibitem{Chamblin:1999tk}
A.~Chamblin, R.~Emparan, C.~V.~Johnson and R.~C.~Myers,
``Charged AdS black holes and catastrophic holography,''
Phys. Rev. D \textbf{60}, 064018 (1999)
doi:10.1103/PhysRevD.60.064018
[arXiv:hep-th/9902170 [hep-th]].
%934 citations counted in INSPIRE as of 10 Jul 2020


%\cite{tHooft:1993dmi}
\bibitem{tHooft:1993dmi}
G.~'t Hooft,
``Dimensional reduction in quantum gravity,''
Conf. Proc. C \textbf{930308}, 284-296 (1993)
[arXiv:gr-qc/9310026 [gr-qc]].
%2366 citations counted in INSPIRE as of 10 Jul 2020


%\cite{Susskind:1994vu}
\bibitem{Susskind:1994vu}
L.~Susskind,
``The World as a hologram,''
J. Math. Phys. \textbf{36}, 6377-6396 (1995)
doi:10.1063/1.531249
[arXiv:hep-th/9409089 [hep-th]].
%2900 citations counted in INSPIRE as of 10 Jul 2020

%\cite{Sheikhahmadi:2021jkz}
\bibitem{Sheikhahmadi:2021jkz}
H.~Sheikhahmadi,
``Is the Universe actually holographic?,''
[arXiv:2109.15022 [hep-th]].
%0 citations counted in INSPIRE as of 02 Oct 2021



%\cite{Ruppeiner:1995zz}
\bibitem{Ruppeiner:1995zz}
G.~Ruppeiner,
``Riemannian geometry in thermodynamic fluctuation theory,''
Rev. Mod. Phys. \textbf{67}, 605-659 (1995)
doi:10.1103/RevModPhys.67.605
%289 citations counted in INSPIRE as of 10 Jul 2020


%\cite{Ruppeiner:1983zz}
\bibitem{Ruppeiner:1983zz}
G.~Ruppeiner,
``Thermodynamic Critical Fluctuation Theory?,''
Phys. Rev. Lett. \textbf{50}, 287-290 (1983)
doi:10.1103/PhysRevLett.50.287
%34 citations counted in INSPIRE as of 10 Jul 2020



\bibitem{Weinhold} F. Weinhold, "Metric geometry of equilibrium thermodynamics,"
J. Chem. Phys. \textbf{63}, 2479 (1975); https://doi.org/10.1063/1.431689


\bibitem{Weinhold1975zx} F. Weinhold, "Metric geometry of equilibrium thermodynamics. II. Scaling, homogeneity, and generalized Gibbs–Duhem relations,"
J. Chem. Phys. \textbf{63}, 2484 (1975); https://doi.org/10.1063/1.431635

\bibitem{Weinhold1975zya} F. Weinhold, "Metric geometry of equilibrium thermodynamics. III. Elementary formal structure of a vector‐algebraic representation of equilibrium thermodynamics,"
J. Chem. Phys. \textbf{63}, 2488 (1975); https://doi.org/10.1063/1.431636




\bibitem{Weinhold1976zxt} F. Weinhold, "Thermodynamics and geometry"
Physics Today \textbf{29}, 23 (1976); https://doi.org/10.1063/1.3023366


\bibitem{Ruppeiner} G. Ruppeiner, "Thermodynamics: A Riemannian geometric model," Phys. Rev. A \textbf{20}, 1608 (1979)  https://doi.org/10.1103/PhysRevA.20.1608


%\cite{Quevedo:2006xk}
\bibitem{Quevedo:2006xk}
H.~Quevedo,
``Geometrothermodynamics,''
J. Math. Phys. \textbf{48}, 013506 (2007)
doi:10.1063/1.2409524
[arXiv:physics/0604164 [physics]].
%166 citations counted in INSPIRE as of 10 Jul 2020

\bibitem{Hendi:2015rja}
S.~H.~Hendi, S.~Panahiyan, B.~Eslam Panah and M.~Momennia,
``A new approach toward geometrical concept of black hole thermodynamics,''
Eur. Phys. J. C \textbf{75}, no.10, 507 (2015)
doi:10.1140/epjc/s10052-015-3701-5
[arXiv:1506.08092 [gr-qc]].
%46 citations counted in INSPIRE as of 10 Jul 2020



%\cite{Fierz:1939ix}
\bibitem{Fierz:1939ix}
M.~Fierz and W.~Pauli,
``On relativistic wave equations for particles of arbitrary spin in an electromagnetic field,''
Proc. Roy. Soc. Lond. A \textbf{A173}, 211-232 (1939)
doi:10.1098/rspa.1939.0140
%1362 citations counted in INSPIRE as of 10 Jul 2020

%\cite{Pauli:1939xp}
\bibitem{Pauli:1939xp}
W.~Pauli and M.~Fierz,
``On Relativistic Field Equations of Particles With Arbitrary Spin in an Electromagnetic Field,''
Helv. Phys. Acta \textbf{12}, 297-300 (1939)
%117 citations counted in INSPIRE as of 10 Jul 2020


%\cite{vanDam:1970vg}
\bibitem{vanDam:1970vg}
H.~van Dam and M.~J.~G.~Veltman,
``Massive and massless Yang-Mills and gravitational fields,''
Nucl. Phys. B \textbf{22}, 397-411 (1970)
doi:10.1016/0550-3213(70)90416-5
%1036 citations counted in INSPIRE as of 10 Jul 2020



%\cite{Zakharov:1970cc}
\bibitem{Zakharov:1970cc}
V.~I.~Zakharov,
``Linearized gravitation theory and the graviton mass,''
JETP Lett. \textbf{12}, 312 (1970)
%755 citations counted in INSPIRE as of 10 Jul 2020



%\cite{Vainshtein:1972sx}
\bibitem{Vainshtein:1972sx}
A.~I.~Vainshtein,
``To the problem of nonvanishing gravitation mass,''
Phys. Lett. B \textbf{39}, 393-394 (1972)
doi:10.1016/0370-2693(72)90147-5
%1250 citations counted in INSPIRE as of 10 Jul 2020


%\cite{Boulware:1973my}
\bibitem{Boulware:1973my}
D.~G.~Boulware and S.~Deser,
``Can gravitation have a finite range?,''
Phys. Rev. D \textbf{6}, 3368-3382 (1972)
doi:10.1103/PhysRevD.6.3368
%988 citations counted in INSPIRE as of 10 Jul 2020






%\cite{deRham:2010ik}
\bibitem{deRham:2010ik}
C.~de Rham and G.~Gabadadze,
``Generalization of the Fierz-Pauli Action,''
Phys. Rev. D \textbf{82}, 044020 (2010)
doi:10.1103/PhysRevD.82.044020
[arXiv:1007.0443 [hep-th]].
%928 citations counted in INSPIRE as of 10 Jul 2020


%\cite{deRham:2010kj}
\bibitem{deRham:2010kj}
C.~de Rham, G.~Gabadadze and A.~J.~Tolley,
``Resummation of Massive Gravity,''
Phys. Rev. Lett. \textbf{106}, 231101 (2011)
doi:10.1103/PhysRevLett.106.231101
[arXiv:1011.1232 [hep-th]].
%1301 citations counted in INSPIRE as of 10 Jul 2020

%\cite{Deser:2013eua}
\bibitem{Deser:2013eua}
S.~Deser, K.~Izumi, Y.~C.~Ong and A.~Waldron,
``Massive Gravity Acausality Redux,''
Phys. Lett. B \textbf{726}, 544-548 (2013)
doi:10.1016/j.physletb.2013.09.001
[arXiv:1306.5457 [hep-th]].
%78 citations counted in INSPIRE as of 10 Jul 2020



%\cite{tHooft:2011aa}
\bibitem{tHooft:2011aa}
G.~'t Hooft,
``A class of elementary particle models without any adjustable real parameters,''
Found. Phys. \textbf{41}, 1829-1856 (2011)
doi:10.1007/s10701-011-9586-8
[arXiv:1104.4543 [gr-qc]].
%110 citations counted in INSPIRE as of 10 Jul 2020


%\cite{Faria:2013hxa}
\bibitem{Faria:2013hxa}
F.~F.~Faria,
``Massive conformal gravity,''
Adv. High Energy Phys. \textbf{2014}, 520259 (2014)
doi:10.1155/2014/520259
[arXiv:1312.5553 [gr-qc]].
%17 citations counted in INSPIRE as of 10 Jul 2020


%\cite{Bebronne:2009mz}
\bibitem{Bebronne:2009mz}
M.~V.~Bebronne and P.~G.~Tinyakov,
``Black hole solutions in massive gravity,''
JHEP \textbf{04}, 100 (2009)
doi:10.1007/ Erratum-ibid 06(2011)018
[arXiv:0902.3899 [gr-qc]].
%34 citations counted in INSPIRE as of 10 Jul 2020



%\cite{Bebronne:2007qh}
\bibitem{Bebronne:2007qh}
M.~V.~Bebronne and P.~G.~Tinyakov,
``Massive gravity and structure formation,''
Phys. Rev. D \textbf{76}, 084011 (2007)
doi:10.1103/PhysRevD.76.084011
[arXiv:0705.1301 [astro-ph]].
%35 citations counted in INSPIRE as of 10 Jul 2020



%\cite{Capela:2011mh}
\bibitem{Capela:2011mh}
F.~Capela and P.~G.~Tinyakov,
``Black Hole Thermodynamics and Massive Gravity,''
JHEP \textbf{04}, 042 (2011)
doi:10.1007/JHEP04(2011)042
[arXiv:1102.0479 [gr-qc]].
%26 citations counted in INSPIRE as of 10 Jul 2020


%\cite{Faria:2020kbv}
\bibitem{Faria:2020kbv}
F.~F.~Faria,
``Gravitational waves in massive conformal gravity,''
[arXiv:2007.03637 [gr-qc]].
%0 citations counted in INSPIRE as of 10 Jul 2020

%\cite{deRham:2020yet}
\bibitem{deRham:2020yet}
C.~de Rham and V.~Pozsgay,
``Proca-Nuevo,''
[arXiv:2003.13773 [hep-th]].
%0 citations counted in INSPIRE as of 10 Jul 2020



%\cite{Kenna-Allison:2019tbu}
\bibitem{Kenna-Allison:2019tbu}
M.~Kenna-Allison, A.~E.~Gümrükçüoglu and K.~Koyama,
``Stable cosmology in generalized massive gravity,''
Phys. Rev. D \textbf{101}, no.8, 084014 (2020)
doi:10.1103/PhysRevD.101.084014
[arXiv:1912.08560 [hep-th]].
%1 citations counted in INSPIRE as of 10 Jul 2020


%\cite{Cayuso:2019ieu}
\bibitem{Cayuso:2019ieu}
R.~Cayuso, O.~J.~C.~Dias, F.~Gray, D.~Kubizňák, A.~Margalit, J.~E.~Santos, R.~Gomes Souza and L.~Thiele,
``Massive vector fields in Kerr-Newman and Kerr-Sen black hole spacetimes,''
JHEP \textbf{04}, 159 (2020)
doi:10.1007/JHEP04(2020)159
[arXiv:1912.08224 [hep-th]].
%4 citations counted in INSPIRE as of 10 Jul 2020

%%%%%%%%%%%%%%%%%%%%%%%%%%%%%%%%%%%%%%%%%%%%%%%%%%%%%%%%%%%%%%%%%%%%%%%%%%%%%%%%%%%%%%%%%%%%%%%
%%%%%%%%%%%%%%%%%%%%%%%%%%%%%%%%%%%%%%%%%%%%%%%%%%%%%%%%%%%%%%%%%%%%%%%%%%%%%%%%%%%%%%%%%%%%%%%
%%%%%%%%%%%%%%%%%%%%%%%%%%%%%%%%%%%%%%%%%%%%%%%%%%%%%%%%%%%%%%%%%%%%%%%%%%%%%%%%%%%%%%%%%%%%%%%
%%%%%%%%%%%%%%%%%%%%%%%%%%%%%%%%%%%%%%%%%%%%%%%%%%%%%%%%%%%%%%%%%%%%%%%%%%%%%%%%%%%%%%%%%%%%%%%



%\bibitem{Dubovsky:2004}
%\cite{Dubovsky:2004sg}
%\bibitem{Dubovsky:2004sg}
\bibitem{Dubovsky:2004}
S.~L.~Dubovsky,
``Phases of massive gravity,''
JHEP \textbf{10}, 076 (2004)
doi:10.1088/1126-6708/2004/10/076
[arXiv:hep-th/0409124 [hep-th]].
%252 citations counted in INSPIRE as of 10 Jul 2020

%\cite{Jusufi:2019caq}
\bibitem{Jusufi:2019caq}
K.~Jusufi, M.~Jamil, H.~Chakrabarty, Q.~Wu, C.~Bambi and A.~Wang,
``Rotating regular black holes in conformal massive gravity,''
Phys. Rev. D \textbf{101}, no.4, 044035 (2020)
doi:10.1103/PhysRevD.101.044035
[arXiv:1911.07520 [gr-qc]].
%12 citations counted in INSPIRE as of 21 Jun 2020



%\cite{Pacilio:2018gom}
\bibitem{Pacilio:2018gom}
C.~Pacilio,
``Scalar charge of black holes in Einstein-Maxwell-dilaton theory,''
Phys. Rev. D \textbf{98}, no.6, 064055 (2018)
doi:10.1103/PhysRevD.98.064055
[arXiv:1806.10238 [gr-qc]].
%9 citations counted in INSPIRE as of 16 Nov 2021



%\cite{EslamPanah:2018ums}
\bibitem{EslamPanah:2018ums}
B.~Eslam Panah,
``Effects of energy dependent spacetime on geometrical thermodynamics and heat engine of black holes: gravity's rainbow,''
Phys. Lett. B \textbf{787}, 45-55 (2018)
doi:10.1016/j.physletb.2018.10.042
[arXiv:1805.03014 [hep-th]].
%25 citations counted in INSPIRE as of 10 Jul 2020
%$Kumara:2019xgt,AhmedRizwan:2019yxk,Gunasekaran:2012dq



%\cite{Kumara:2019xgt}
\bibitem{Kumara:2019xgt}
A.~N.~Kumara, C.~L.~A.~Rizwan, D.~Vaid and K.~M.~Ajith,
``Critical Behaviour and Microscopic Structure of Charged AdS Black Hole with a Global Monopole in Extended and Alternate Phase Spaces,''
[arXiv:1906.11550 [gr-qc]].
%7 citations counted in INSPIRE as of 10 Jul 2020


%\cite{Rizwan:2018mpy}
%\bibitem{Rizwan:2018mpy}
\bibitem{AhmedRizwan:2019yxk}
A.~Rizwan C.L., N.~Kumara A., D.~Vaid and K.~M.~Ajith,
``Joule-Thomson expansion in AdS black hole with a global monopole,''
Int. J. Mod. Phys. A \textbf{33}, no.35, 1850210 (2019)
doi:10.1142/S0217751X1850210X
[arXiv:1805.11053 [gr-qc]].
%18 citations counted in INSPIRE as of 10 Jul 2020



%\cite{Gunasekaran:2012dq}
\bibitem{Gunasekaran:2012dq}
S.~Gunasekaran, R.~B.~Mann and D.~Kubiznak,
``Extended phase space thermodynamics for charged and rotating black holes and Born-Infeld vacuum polarization,''
JHEP \textbf{11}, 110 (2012)
doi:10.1007/JHEP11(2012)110
[arXiv:1208.6251 [hep-th]].
%362 citations counted in INSPIRE as of 10 Jul 2020

%\cite{Hoseini:2016nzw}
\bibitem{Hoseini:2016nzw}
B.~Hoseini, R.~Saffari, S.~Soroushfar, J.~Kunz and S.~Grunau,
``Analytic treatment of complete geodesics in a static cylindrically symmetric conformal spacetime,''
Phys. Rev. D \textbf{94} (2016) no.4, 044021
doi:10.1103/PhysRevD.94.044021
[arXiv:1602.03898 [gr-qc]].
%8 citations counted in INSPIRE as of 11 May 2021

%\cite{Hoseini:2016tvu}
\bibitem{Hoseini:2016tvu}
B.~Hoseini, R.~Saffari and S.~Soroushfar,
``Study of the geodesic equations of a spherical symmetric spacetime in conformal Weyl gravity,''
Class. Quant. Grav. \textbf{34} (2017) no.5, 055004
doi:10.1088/1361-6382/aa5a63
[arXiv:1606.06558 [gr-qc]].
%3 citations counted in INSPIRE as of 11 May 2021




%\cite{Hoseini:2016ztk}
\bibitem{Hoseini:2016ztk}
B.~Hoseini, R.~Saffari and S.~Soroushfar,
``Geodesic Motion in the Spacetime Of a SU(2)-Colored (A)dS Black Hole in Conformal Gravity,''
Eur. Phys. J. Plus \textbf{136} (2021) no.5, 489
doi:10.1140/epjp/s13360-021-01511-y
[arXiv:1606.06545 [gr-qc]].
%1 citations counted in INSPIRE as of 11 May 2021

%\cite{Mann:1997jb}
\bibitem{Mann:1997jb}
R.~B.~Mann,
``Black holes of negative mass,''
Class. Quant. Grav. \textbf{14}, 2927-2930 (1997)
doi:10.1088/0264-9381/14/10/018
[arXiv:gr-qc/9705007 [gr-qc]].
%100 citations counted in INSPIRE as of 07 Aug 2021






%\cite{Bondi:1957zz}
\bibitem{Bondi:1957zz}
H.~Bondi,
``Negative Mass in General Relativity,''
Rev. Mod. Phys. \textbf{29}, 423-428 (1957)
doi:10.1103/RevModPhys.29.423
%156 citations counted in INSPIRE as of 19 Nov 2021

%\cite{Belletete:2013nqa}
\bibitem{Belletete:2013nqa}
J.~Bellet\^ete and M.~B.~Paranjape,
``On negative mass,''
Int. J. Mod. Phys. D \textbf{22}, 1341017 (2013)
doi:10.1142/S0218271813410174
[arXiv:1304.1566 [gr-qc]].
%11 citations counted in INSPIRE as of 19 Nov 2021



\bibitem{Castelvecchi:2016gbf}
Davide Castelvecchi,
"Artificial black hole creates its own version of Hawking radiation,"
Nature  \textbf{18};536 (7616) :258-9 (2016).
doi: 10.1038/536258a.





\bibitem{Khamehchi:2017xzz}
M. A. Khamhchi, et al.
"Negative-Mass Hydrodynamics in a Spin-Orbit–Coupled Bose-Einstein Condensate,"
Phys. Rev. Lett. \textbf{118}, 155301 (2017)
doi:10.1103/PhysRevLett.118.155301,
[arXiv:1612.04055 [cond-mat.quant-gas]].


%\cite{Farnes:2017gbf}
\bibitem{Farnes:2017gbf}
J.~S.~Farnes,
``A unifying theory of dark energy and dark matter: Negative masses and matter creation within a modified $\Lambda$CDM framework,''
Astron. Astrophys. \textbf{620}, A92 (2018)
doi:10.1051/0004-6361/201832898
[arXiv:1712.07962 [physics.gen-ph]].
%35 citations counted in INSPIRE as of 19 Nov 2021








%\cite{Sanchez:2016ger}
\bibitem{Sanchez:2016ger}
A.~Sanchez,
``Geometrothermodynamics of black holes in Lorentz noninvariant massive gravity,''
Phys. Rev. D \textbf{94}, no.2, 024037 (2016)
doi:10.1103/PhysRevD.94.024037
[arXiv:1603.02259 [gr-qc]].
%5 citations counted in INSPIRE as of 14 Aug 2021

%\cite{Fernando:2016qhq}
\bibitem{Fernando:2016qhq}
S.~Fernando,
``Phase transitions of black holes in massive gravity,''
Mod. Phys. Lett. A \textbf{31}, no.16, 1650096 (2016)
doi:10.1142/S0217732316500966
[arXiv:1605.04860 [gr-qc]].
%11 citations counted in INSPIRE as of 14 Aug 2021


%\cite{CruzNeto:2020lqb}
\bibitem{CruzNeto:2020lqb}
F.~A.~Cruz Neto, A.~G.~Jir\'on Vicente and L.~B.~Castro,
``Comment on \textquotedblleft{}Dirac fermions in Som\textendash{}Raychaudhuri space-time with scalar and vector potential and the energy momentum distributions [Eur. Phys. J. C (2019) 79:541]\textquotedblright{},''
Eur. Phys. J. C \textbf{80}, no.4, 348 (2020)
doi:10.1140/epjc/s10052-020-7914-x
[arXiv:2001.00911 [hep-th]].
%1 citations counted in INSPIRE as of 14 Aug 2021






%\cite{Soroushfar:2016nbu}
\bibitem{Soroushfar:2016nbu}
S.~Soroushfar, R.~Saffari and N.~Kamvar,
``Thermodynamic geometry of black holes in f(R) gravity,''
Eur. Phys. J. C \textbf{76}, no.9, 476 (2016)
doi:10.1140/epjc/s10052-016-4311-6
[arXiv:1605.00767 [gr-qc]].
%15 citations counted in INSPIRE as of 10 Jul 2020






\bibitem{Salamon} P. Salamon, J. D. Nulton and E. Ihrig, "Thermodynamic lengths and intrinsic time scales in molecular relaxation,"
 J. Chem. Phys, 80,436 (1984) https://doi.org/10.1063/1.449666

\bibitem{Mrugala:1984} R.~Mrugala, "On equivalence of two metrics in classical thermodynamics," Physica A: Statistical Mechanics and its Applications, { 125}, 631 (1984)
https://doi.org/10.1016/0378-4371(84)90074-8
%\cite{Quevedo:2006xk}

%\cite{Hendi:2015xya}
\bibitem{Hendi:2015xya}
S.~H.~Hendi, A.~Sheykhi, S.~Panahiyan and B.~Eslam Panah,
``Phase transition and thermodynamic geometry of Einstein-Maxwell-dilaton black holes,''
Phys. Rev. D \textbf{92}, no.6, 064028 (2015)
doi:10.1103/PhysRevD.92.064028
[arXiv:1509.08593 [hep-th]].
%71 citations counted in INSPIRE as of 10 Jul 2020.


%\cite{Soroushfar:2019ihn}
\bibitem{Soroushfar:2019ihn}
S.~Soroushfar, R.~Saffari and S.~Upadhyay,
``Thermodynamic geometry of a black hole surrounded by perfect fluid in Rastall theory,''
Gen. Rel. Grav. \textbf{51}, no.10, 130 (2019)
doi:10.1007/s10714-019-2614-2
[arXiv:1908.02133 [gr-qc]].
%8 citations counted in INSPIRE as of 10 Jul 2020

%\cite{Aman:2015wsa}
\bibitem{Aman:2015wsa}
J.~\r{A}man, I.~Bengtsson and N.~Pidokrajt,
``Thermodynamic Metrics and Black Hole Physics,''
Entropy \textbf{17}, 6503-6518 (2015)
doi:10.3390/e17096503
[arXiv:1507.06097 [gr-qc]].
%5 citations counted in INSPIRE as of 08 Aug 2021

%\cite{Hawking:1976de}
\bibitem{Hawking:1976de}
S.~W.~Hawking,
``Black Holes and Thermodynamics,''
Phys. Rev. D \textbf{13}, 191-197 (1976)
doi:10.1103/PhysRevD.13.191
%1049 citations counted in INSPIRE as of 08 Aug 2021




%\cite{Soroushfar:2020wch}
\bibitem{Soroushfar:2020wch}
S.~Soroushfar and S.~Upadhyay,
``Phase transition of a charged AdS black hole with a global monopole through geometrical thermodynamics,''
Phys. Lett. B \textbf{804} (2020), 135360
doi:10.1016/j.physletb.2020.135360
[arXiv:2003.06714 [gr-qc]].
%4 citations counted in INSPIRE as of 11 May 2021

%\cite{Pourhassan:2021mhb}
\bibitem{Pourhassan:2021mhb}
B.~Pourhassan, S.~S.~Wani, S.~Soroushfar and M.~Faizal,
``Quantum Work and Information Geometry of a Quantum Myers-Perry Black Hole,''
[arXiv:2102.03296 [hep-th]].
%0 citations counted in INSPIRE as of 11 May 2021
%%%%%%%%%%%%%%%%%%%%%%%%%%%%%%%%%%%%%%%%%%%%%%%%%%%%%%%%%%%%


%%%%%%%%%%%%%%%%%%%%%%%%%%%%%%%%%%%%%%%%%%%%%%%%%%%%%%%%%%%%

\end{thebibliography}
\end{document}